\pdfoutput=1
\RequirePackage{./StyleFilesMacros/lhchiggs}
\documentclass{cernyrep}
\usepackage{etoolbox}
\usepackage[stable]{footmisc}
\usepackage{adforn}
\usepackage{floatrow}
\usepackage[T1]{fontenc}
\usepackage{esvect}

\usepackage[dvipsnames]{xcolor}
\usepackage{graphicx}
\usepackage{tikz}
\usetikzlibrary{arrows,decorations.pathmorphing,backgrounds,positioning,fit,petri,automata,shadows,calendar,mindmap,decorations.markings,calc}
\tikzset{->-/.style={decoration={markings, mark=at position .5 with {\arrow{>}}},postaction={decorate}}}
\tikzset{-<-/.style={decoration={markings, mark=at position .5 with {\arrow{<}}},postaction={decorate}}}
\usepackage{cancel}
\usepackage{pdflscape}
\usepackage{float}
\usepackage{./StyleFilesMacros/axodraw}
\usepackage{lineno}
\usepackage{enumitem}
\setlist{nolistsep} 
\usepackage{./StyleFilesMacros/lhchiggs}
\usepackage{heppennames2}
\usepackage{cernunits}

\usepackage{./StyleFilesMacros/hepparticles}

\usepackage{fancyhdr}
\usepackage{tikz}

\newcommand{\cref}[1]{Chapter~\ref{c.#1}}

\newcommand{\nc}{\newcommand}
\nc{\gev}{\;\mathrm{GeV}}

\usepackage{hypernat}
\RequirePackage{./StyleFilesMacros/relsize}
\usepackage[toc,page]{appendix}
\usepackage{booktabs}
\usepackage{bigdelim}

\def\reffi#1{\mbox{Figure~\ref{#1}}}

\def\refta#1{\mbox{Table~\ref{#1}}}

\def\refse#1{\mbox{Section~\ref{#1}}}

\def\eq#1{{Eq.~(\ref{#1})}}

\def\be{\begin{equation}}
\def\ee{\end{equation}}
\def\bea{\begin{eqnarray}}
\def\eea{\end{eqnarray}}

\newcommand{\mgFull}{\texttt{MadGraph5\_aMC@NLO}}
\newcommand{\mgamc}{\texttt{mg5amc}}

\newcommand{\ms}{\texttt{MadSpin}}
\def\MG5{{\tt MadGraph5\_aMC@NLO}}

\newcommand{\ab}{ab}

\newcommand{\GeV}{\ensuremath{\,\text{GeV}}\xspace}
\newcommand{\TeV}{\ensuremath{\,\text{TeV}}\xspace}

\newcommand{\tev}{{\unskip\,\text{TeV}}}

\newcommand{\pt}{\ensuremath{p_\text{T}}\xspace}

\newcommand{\fr}{\frac}

\newcolumntype{.}{D{.}{.}{-1}}
\newcolumntype{d}[1]{D{.}{.}{#1}}
\newcolumntype{x}[1]{{\centering\hspace{0pt}}p{#1}}
\newcolumntype{C}[1]{>{\centering\arraybackslash}p{#1}}

\newlength{\width}
\newlength{\height}

\newcommand{\crn}{\nonumber \\}
\newcommand{\nnb}{\nonumber}

\newfloatcommand{capbtabbox}{table}[][\FBwidth]

\def\draftdate{\relax}
\def\mda{\relax}
\def\mua{\relax}
\def\mla{\relax}
\def\draft{
\def\thtystars{******************************}
\def\sixtystars{\thtystars\thtystars}
\typeout{}
\typeout{\sixtystars**}
\typeout{* Draft mode!
         For final version remove \protect\draft\space in source file *}
\typeout{\sixtystars**}
\typeout{}
\def\draftdate{\today}
\def\mua{\marginpar[\boldmath\hfil$\uparrow$]%
                   {\boldmath$\uparrow$\hfil}\color{black}%
                    \typeout{marginpar: $\uparrow$}\ignorespaces}
\def\mda{\color{red}\marginpar[\boldmath\hfil$\downarrow$]%
                   {\boldmath$\downarrow$\hfil}%
                    \typeout{marginpar: $\downarrow$}\ignorespaces}
\def\mla{\marginpar[\boldmath\hfil$\rightarrow$]%
                   {\boldmath$\leftarrow $\hfil}%
                    \typeout{marginpar: $\leftrightarrow$}\ignorespaces}
\def\Mua{\marginpar[\boldmath\hfil$\Uparrow$]%
                   {\boldmath$\Uparrow$\hfil}\color{black}%
                    \typeout{marginpar: $\uparrow$}\ignorespaces}
\def\Mda{\color{red}\marginpar[\boldmath\hfil$\Downarrow$]%
                   {\boldmath$\Downarrow$\hfil}%
                    \typeout{marginpar: $\downarrow$}\ignorespaces}
\def\Mla{\marginpar[\boldmath\hfil\textcolor{red}{$\Rightarrow$}]%
                   {\boldmath\textcolor{red}{$\Leftarrow $}\hfil}%
                    \typeout{marginpar: $\leftrightarrow$}\ignorespaces}
\overfullrule 5pt
\oddsidemargin 15mm
\marginparwidth 29mm
}

\usepackage[hyperfootnotes=true,bookmarks=false,hypertexnames=true,pagebackref=true,linktocpage=false]{hyperref}
\hypersetup{
   unicode=true,          	
   pdftoolbar=true,        	
   pdfmenubar=true,        	
   pdffitwindow=false,     	
   pdfstartview={FitH},    	
   pdfnewwindow=true,   	
   colorlinks=true,       	
   linkcolor=MidnightBlue,	
   citecolor=OliveGreen,	
   filecolor=Cyan,          	
   urlcolor=RedViolet       	
}

\renewcommand*\backref[1]{\ifx#1\relax \else (#1) \fi}
\usepackage{chngcntr}
\usepackage[subfigure]{tocloft} 
\usepackage{subfigure} 
\usepackage{wrapfig}
\usepackage{afterpage}

\usepackage[export]{adjustbox} 

	\counterwithin{chapter}{part}	
	\counterwithin{section}{chapter}
	\counterwithin{equation}{chapter}
	\counterwithout{figure}{chapter}
	\counterwithout{table}{chapter}
  \counterwithout{equation}{chapter}
	\counterwithout{footnote}{chapter}
\makeatletter
	\@addtoreset{chapter}{part}
	\@addtoreset{footnote}{part}
\makeatother
\usepackage[export]{adjustbox}

\title{VBSCan Mid-Term Scientific Meeting}

\newcommand{\myabstract}
{
This document summarises the talks and discussions happened during the
VBSCan Mid-Term Scientific Meeting workshop.
The VBSCan COST action is dedicated to the coordinated study of vector boson
scattering (VBS) from the phenomenological and experimental point of view,
for the best exploitation of the data that will be delivered by existing and future particle colliders.
}

\newcommand{\editors}
{
\begin{flushright}
I.~Brivio, C.~Charlot,       \\
R.~Covarelli, R.L.~Delgado, \\
K.~Lohwasser, M.~Pellen,    \\ 
M.~Slawinska, G.~Ortona,    \\ 
K.~Ozdemir, C.~Petridou,    \\ 
I.~Puljak, M.~Zaro          \\ 
\end{flushright}
}

\newcommand{\docID}{VBSCan-PUB-02-20}
\usepackage{layout}
\newcommand{\mydate}{\today}

\begin{document}


\pagestyle{plain}
  

\thispagestyle{empty}
{

\begin{center}

{\fontfamily{qhv}\selectfont

\begin{minipage}[t]{0.25\textwidth}
  \mydate
\end{minipage}
\begin{minipage}[t]{0.4\textwidth}
  \begin{center}
  \textit{VBSCan COST Action report}
  \end{center}
\end{minipage}
\begin{minipage}[t]{0.25\textwidth}
  \begin{flushright}
  \docID\\
  UWThPh 2020-3\\
  IFIRSE-TH-2019-6\\
  DESY~20-026\\
  Cavendish-HEP-20/02
  TIF-UNIMI-2020-13
  \end{flushright}
\end{minipage}


\vspace{5cm}

\makeatletter

{\LARGE\textbf{\@title}}

\makeatother

\vspace {1cm}
\adforn{50}~~\adforn{10}~~\adforn{22}
\vspace {1cm}

\begin{minipage}{0.7\textwidth}

\myabstract
\end{minipage}

\vfill


\begin{minipage}[b]{0.35\textwidth}
  \includegraphics[height=1.1cm,valign=c]{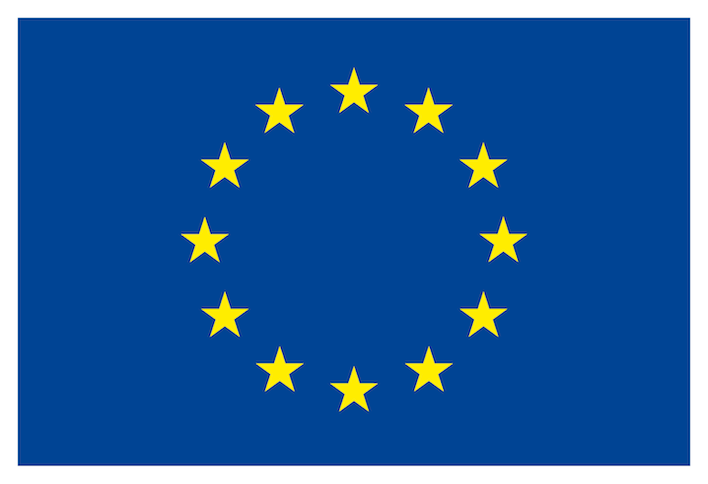}
  \hspace{0.1cm}
  \includegraphics[height=1cm,valign=c]{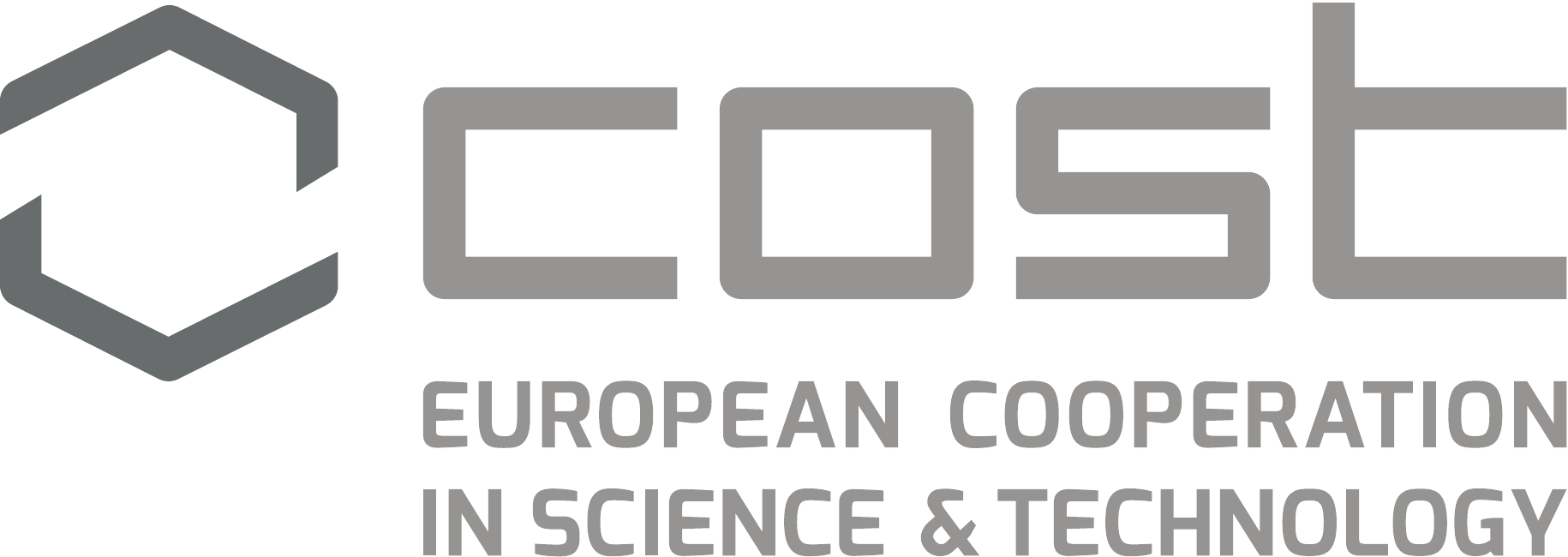}\\
\end{minipage}
\begin{minipage}[t]{0.2\textwidth}
  \begin{center}
  \includegraphics[width=0.8\textwidth]{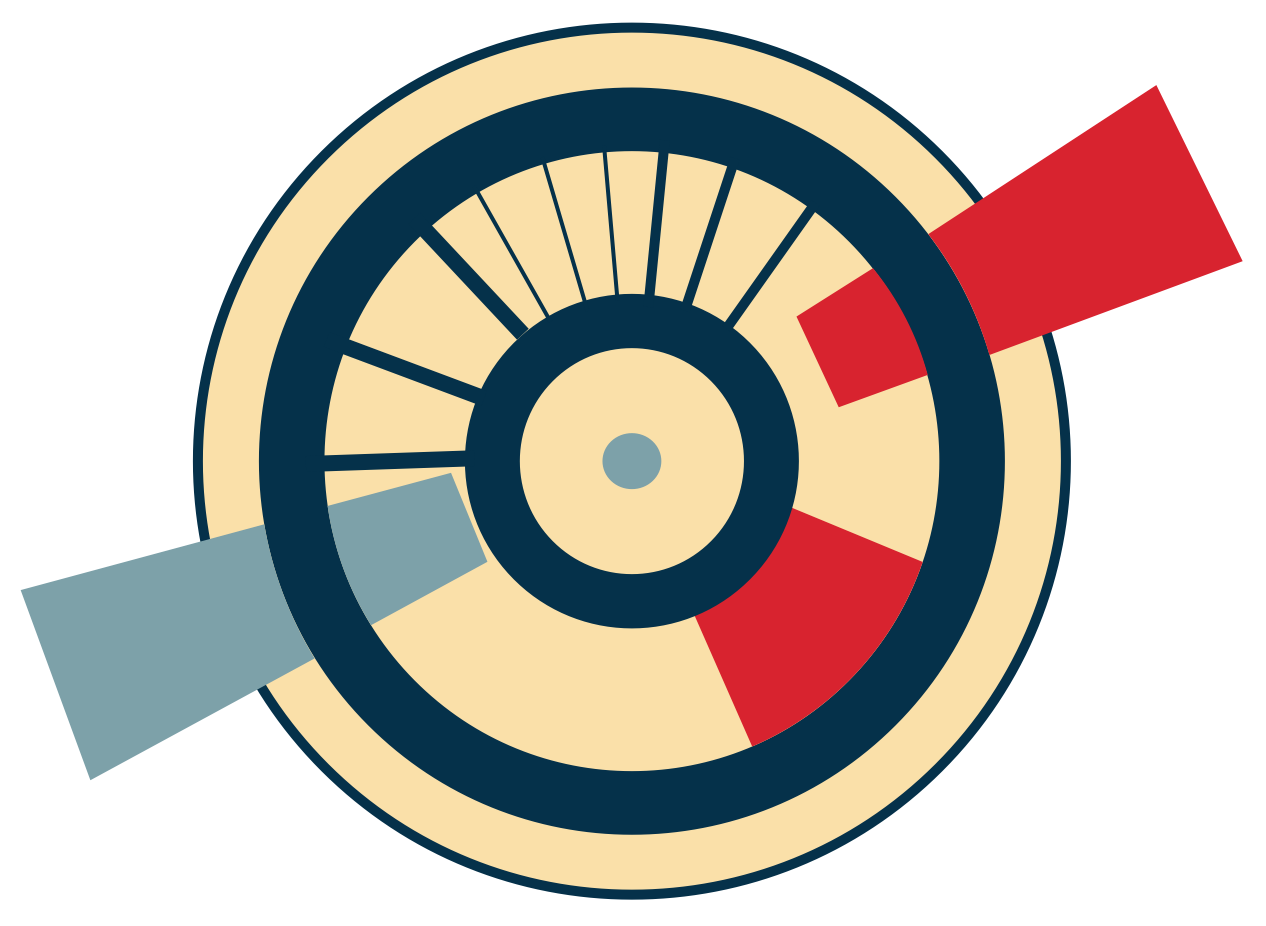}
  \end{center}
\end{minipage}
\begin{minipage}[b]{0.35\textwidth}
  \begin{flushright}
  \textit{Editors:}
  \editors
  \end{flushright}
\end{minipage}

} 
\end{center}

\newpage{}


}


\pagenumbering{roman}


{\pagestyle{plain}
\thispagestyle{plain}

\leftline{\textbf{Authors}}
\begin{flushleft}
Julien Baglio$^{1}$,
Alessandro Ballestrero$^{2}$,
Riccardo Bellan$^{3}$,
Carsten Bittrich$^{4}$,
Simon Bra\ss{}$^{5}$,
Ilaria Brivio$^{6}$,
Diogo Buarque Franzosi$^{7}$,
Claude Charlot$^{8}$,
Roberto Covarelli$^{3}$,
Javier Cuevas$^{9}$,
Michele Gallinaro$^{10}$,
Raquel Gomez-Ambrosio$^{11}$,
Pietro Govoni$^{12}$,
Michele Grossi$^{13}$,
Alexander Karlberg$^{14}$,
Aysel Kayis Topaksu$^{15}$,
Borut Kersevan$^{16}$,
Wolfgang Kilian$^{17}$,
Patrick Kirchgae\ss{}er$^{18}$,
Rafael L. Delgado$^{19}$,
Kristin Lohwasser$^{20}$,
Narei Lorenzo Martinez$^{21}$,
Ezio Maina$^{3}$,
Olivier Mattelaer$^{22}$,
Ankita Mehta$^{23}$,
Predrag Milenovic$^{24}$,
Le Duc Ninh$^{25}$,
Jakob Novak$^{26}$,
Giacomo Ortona$^{27}$,
Kadri \"Ozdemir$^{28}$,
Andreas Papaefstathiou$^{29}$,
Gabriella P\'asztor$^{23}$,
Mathieu Pellen$^{30}$,
Giovanni Pelliccioli$^{3,31}$,
Chara Petridou$^{32}$,
Simon Pl\"atzer$^{33,34}$,
Ivica Puljak$^{35}$,
Daniela Rebuzzi$^{36}$,
J\"urgen Reuter$^{5}$,
Richard Ruiz$^{22}$,
Despoina Sampsonidou$^{32}$,
Emmanuel Sauvan$^{21}$,
Sujay Shil$^{37,38}$,
Magdalena Slawinska$^{39}$,
Philip Sommer$^{20}$,
Micha\l{} Szleper$^{40}$,
Stefanie Todt$^{4}$,
Davide Valsecchi$^{1,41}$,
Dimitris Varouchas$^{27}$,
Pietro Vischia$^{22}$,
Marco Zaro$^{42,12}$
\end{flushleft}

\begin{itemize}
\item[$^{1}$] CERN (CH)
\item[$^{2}$] INFN Torino (IT)
\item[$^{3}$] University and INFN Torino (IT)
\item[$^{4}$] Technische Universitaet Dresden (DE)
\item[$^{5}$] Deutsches Elektronen-Synchrotron, Hamburg (DE)
\item[$^{6}$] Ruprecht-Karls-Universit\"at Heidelberg (DE)
\item[$^{7}$] Chalmers University of Technology, G\"oteborg (SE)
\item[$^{8}$] LLR, \'Ecole Polytechnique, CNRS/IN2P3, Universit\'e Paris-Saclay (FR)
\item[$^{9}$] University of Oviedo (SP)
\item[$^{10}$] LIP Lisbon (PT)
\item[$^{11}$] Institute for Particle Physics Phenomenology, Department of Physics, University of Durham, Durham DH1 3LE (UK)
\item[$^{12}$] University and INFN, Milano-Bicocca (IT)
\item[$^{13}$] University of Pavia and IBM Italy (IT)
\item[$^{14}$] Rudolf Peierls Centre for Theoretical Physics, University of Oxford, Clarendon Laboratory, Parks Road, Oxford OX1 3PU (UK)
\item[$^{15}$] Cukurova University, Science and Art Faculty, Physics Department,  Adana, Turkey
\item[$^{16}$] Faculty of Mathematics and Physics, University of Ljubljana (SI)
\item[$^{17}$] University of Siegen (DE)
\item[$^{18}$] KIT - Karlsruhe Institute of Technology (DE)
\item[$^{19}$] INFN, Firenze (IT)
\item[$^{20}$] University of Sheffield (GB)
\item[$^{21}$] LAPP, Univ. Grenoble Alpes, Univ. Savoie Mont Blanc, CNRS/IN2P3, Annecy (FR)
\item[$^{22}$] Centre for Cosmology, Particle Physics and Phenomenology Universit\'e{} catholique de Louvain, B-1348 Louvain-la-Neuve (BE)
\item[$^{23}$] MTA-ELTE Lend\"ulet CMS Particle and Nuclear Physics Group, E\"otv\"os Lor\'and University, Budapest (HU)
\item[$^{24}$] Faculty of Physics, University of Belgrade (RS)
\item[$^{25}$] Institute For Interdisciplinary Research in Science and Education, ICISE, Quy Nhon, (VN)
\item[$^{26}$] Department of Experimental Particle Physics, Jo\v{z}ef Stefan Institute and Department of Physics, University of Ljubljana (SI)
\item[$^{27}$] LAL, Universit\'e Paris-Sud, CNRS/IN2P3, Universit\'e Paris-Saclay, Orsay (FR)
\item[$^{28}$] P\^ir\^is Reis University, Engineering Faculty, Istanbul (TR)
\item[$^{29}$] Higgs Centre for Theoretical Physics, University of Edinburgh (UK)
\item[$^{30}$] University of Cambridge (UK)
\item[$^{31}$] University of W\"urzburg (DE)
\item[$^{32}$] Aristotle University of Thessalon\'iki (GR)
\item[$^{33}$] Fakult\"at Physik, University of Vienna, Vienna (AT)
\item[$^{34}$] Erwin Schr\"odinger International Institute for Mathematical Physics (ESI), Boltzmanngasse 9, A-1090 Vienna, Austria
\item[$^{35}$] University of Split, FESB (HR)
\item[$^{36}$] Universit\`a di Pavia e INFN, Sezione di Pavia (IT)
\item[$^{37}$] Homi Bhabha National Institute, Training School Complex, Anushakti Nagar, Mumbai (IN)
\item[$^{38}$] Institute of Physics, Sachivalaya Marg, Bhubaneswar, Odisha (IN)
\item[$^{39}$] Polish Academy of Sciences (PL)
\item[$^{40}$] National Center for Nuclear Research, Warsaw (PL)
\item[$^{41}$] Universit\`a degli Studi di Milano (IT)
\item[$^{42}$] Nikhef National institute for subatomic physics (NL)
\end{itemize}

}
\newpage
\tableofcontents
%


\renewcommand{\thechapter}{\arabic{chapter}}
\renewcommand{\thesection}{\arabic{chapter}.\arabic{section}}
\renewcommand{\thesubsection}{\arabic{chapter}.\arabic{section}.\alph{subsection}}
\renewcommand{\thesubsubsection}{\arabic{chapter}.\arabic{section}.\alph{subsection}.\roman{subsubsection}}

\makeatletter
\renewcommand{\thesubfigure}{(\alph{subfigure})}
\renewcommand{\@thesubfigure}{{\subcaplabelfont\thesubfigure}\space}
\let\p@subfigure\thefigure
\makeatother

\pagenumbering{arabic}
\setcounter{chapter}{0}
\setcounter{section}{0}
\setcounter{subsection}{0}
\setcounter{subsubsection}{0}
\setcounter{footnote}{0}

\let\cleardoublepage\clearpage

\cleardoublepage
\phantomsection
\addcontentsline{toc}{part}{Introduction}

\chapter*{Introduction\markboth{Introduction}{Introduction}} 
\label{chapter:intro}
This document summarises the presentations given at the third annual meeting
of the VBSCan COST action, held in Istanbul in July 2019\footnote{https://indico.cern.ch/event/808557/}. The
VBSCan action is funded by the Horizon 2020 Framework of the European Union, and aims at a coordinated
effort in the study of Vector-Boson Scattering (VBS) by involving relevant players
in the high-energy physics community.

The first annual meeting of the action, held in Split in 2017\footnote{https://indico.cern.ch/event/629638/},
set the ground for the various research directions that will be followed during the 
course of the action as it is reported in~\cite{Anders:2018gfr}. The second meeting, 
in Thessaloniki (2018)\footnote{https://indico.cern.ch/event/706178/}
showcased the results obtained during the first grant period of the action, summarised in~\cite{Bellan:2019xpr}. 
The third meeting is devoted to summarise the many goals achieved in the first half of the action, by theory 
and experimental collaborations.

This report follows the structure of the action in different working groups (WGs). Three working groups are devoted
to physics (WG1 ``Theoretical understanding'', WG2 ``Analysis techniques'' and WG3 ``Experimental techniques''),
and two to other matters relevant for the life of the action (WG4 ``Knowledge exchange and cross-activities'' and WG5 ``Inclusiveness policies''). The
progress in WG1, WG2 and WG3 will be reported respectively in Chapter 1, 2 and 3.
Topics treated in WG1 include the study of polarised VBS and the parameterisation of the impact of new physics in the VBS domain
through the effective field theory (EFT). For what concerns WG2, the most recent analysis techniques employed by the ATLAS and CMS collaborations are outlined.
Finally WG3 reports on recent experimental measurements performed by the two collaborations, focusing on their interpretation
using EFT.


\renewcommand*{\thefootnote}{\fnsymbol{footnote}}

\chapter{Theoretical Understanding}
\label{WG1}

\section{Polarized Particles with \mgFull\footnote{speaker: Diogo Buarque Franzosi; authors: Olivier Mattelaer, Richard Ruiz, Sujay Shil}}

Measuring the helicity of Standard-Model (SM) particles is an important programme of high energy physics. 
Present experimental analyses at colliders require the precise simulation of scattering events via Monte Carlo (MC) tools, which should include a consistent description of polarized particles in their frameworks.
In light of the current and anticipated performance of LHC experiments, these predictions should achieve at least up to next-to-leading (NLO) order in QCD (and ideally in electroweak) interactions.
Simulations should also be able to describe unstable polarized particles, with the decay products maintaining the spin correlation of their parent particle's polarized state.
In this report, we present the implementation of polarized parton scattering in the \mgFull~(\mgamc) framework~\cite{Alwall:2014hca,ToAppear}. 
The \mgamc~software suite provides a powerful and flexible framework for simulating fully differential SM and beyond the SM (BSM) processes with (in principle) arbitrary multiplicities in the final state.
Spin-correlated decays of resonant states are propagated through the \texttt{MadSpin} formalism~\cite{Artoisenet:2012st}.
 For tree-induced process, simulations are achievable up to NLO in QCD with parton shower (PS) matching via the {MC@NLO} formalism~\cite{Frixione:2002ik}, and NLO in electroweak (EW)~\cite{Frederix:2018nkq}. For loop-induced processes, leading-order (LO) predictions are automated~\cite{Hirschi:2015iia}.
Multijet matching is also possible through several LO and NLO techniques, as are specialized SCET-based resummation computations for color-singlet processes.
The description of polarized states can thus be combined with any of these features in an automatic fashion\footnote{The notable exception to this is the production of polarized QCD partons or heavy quarks beyond LO, which requires similar extension of the parton distribution function, {MC@NLO}, and PS formalisms to include polarized states.}.

In this report we give a short summary of the \mgamc~syntax to simulate polarized states and briefly discuss two applications.
In our conclusions we briefly comment about the discussions that have taken place in the VBSCan workshop in Istanbul.
The detailed description of the implementation and more applications can be found in Ref.~\cite{BuarqueFranzosi:2019boy}.
The capabilities of handling polarization are available inside \mgFull\ version 2.7.1 or later.

\subsection*{Implementation and syntax}

The syntax $\pt\{X\}$ can be used to specify a polarization $X$ of particle $\pt$ in any process generated with the usual commands of \mgamc. The available values of $X$ depend on the spin of the particle $\pt$ and the mode (explained below) the user wants to run. 
The available values for $X$ are:
\begin{itemize}
\item spin 1/2: $L(R)$ or ${-}({+})$ for left (right) helicity, and available in all modes.
\item spin 1: ${0(T)}$ for longitudinal (transverse) helicities (available in all modes), with $+(-)$ denoting right (left) circular polarization (only available in Mode I, see below). $A$ for auxiliary (only available in Mode II since it vanishes on-shell).
\item spin 3/2: -3, -1, 1, 3. Only available in Mode I.
\item spin 2: -2, -1, 0, 1, 2. Only available in Mode I.
\end{itemize} 
The two available modes are: 
\begin{itemize}
\item Mode I: final state particles. \\
\textit{Only the required polarizations are considered  in the calculation of the summed/averaged matrix element squared. }
	\begin{itemize}
	\item Applicable to particles of spin 1/2, 1, 3/2, and 2.
	\item Examples:
	\begin{verbatim}
	generate p p > t t~{L}
	generate e+{L} e- > w+{0} w-{T}
	\end{verbatim}
	\end{itemize}
\item Mode II: Unstable  particles in the spin-correlated narrow-width approximation. \\
\textit{The propagator is split in the different polarization configurations in the spirit of the implementation in the \texttt{Phantom} code~\cite{Ballestrero:2017bxn}. }
	\begin{itemize}
	\item Can be used via the decay chain syntax, \textit{e.g.}
	\begin{verbatim}
	generate p p > t t~{L}, t~ > b~ w-
	generate e+ e- > w+{0} w-{T}, w+ > e+ ve, w- > e- ve~
	\end{verbatim}
	\item Equivalent result can be attained via \ms, where the events with decayed particles are generated \textit{a posteriori}. E.g.
	\begin{verbatim}
	generate e+ e- > w+{0} w-{T}	
	\end{verbatim}
	and modify \texttt{madspin\_card.dat} with
	\begin{verbatim}
	decay w+ > e+ ve
	decay w- > e- ve~
	\end{verbatim}
	\item \ms~ will automatically recognize that the produced particles are polarized and modify the propagators accordingly.
	\item This mode supports particles of spin 1/2 and 1.
	\end{itemize}
\end{itemize}

For LO event generation the user can choose the rest frame in which the polarizations are defined. This can be done via the new \texttt{me\_frame} parameter in \texttt{run\_card.dat}. This parameter receives a list of integers associated to the process particles that defines the rest frame where the matrix elements are evaluated. For instance 
\begin{verbatim}
[1,2] = me_frame
\end{verbatim}
corresponds to the partonic centre-of-mass frame for the process $p_1 + p_2 \to p_3 + \dots$.
At the moment, it is not possible to define the polarization in the lab frame. 

At NLO in QCD only the polarization of QCD neutral particles can be required. Apart from this restriction, NLO in QCD accuracy can be achieved with the usual syntax with \texttt{[QCD]} appended. \emph{e.g.}
 	\begin{verbatim}
	generate p p > w+ w-{T} [QCD]
	\end{verbatim}

\subsection*{High-Energy Vector-Boson Scattering}

In models where the Higgs boson is a composite state from a dynamical EW symmetry breaking sector, the couplings between the Higgs boson and the EW bosons are modified with respect to the SM. 
This modification leads to gauge miscancellations between diagrams and the consequent growing behavior with $E^2$ of the amplitudes of vector boson scattering (VBS) processes. 
The couplings in this class of models can be parametrized by the Lagrangian
\begin{equation}
\mathcal{L} \supset \left(\frac{m_Z^2}{2} Z_\mu Z^\mu + m_W^2 W^+_\mu W^{-\mu}\right)\left(1+2 a \frac{h}{v}+\cdots\right)\,.
\end{equation}

We illustrate the BSM features of our implementation by looking at high-energy VBS in composite Higgs models. 
Using the described tools, one can enrich specific polarization samples and analyse in detail distributions of each particular polarization configuration in an exclusive fashion.
For that application we make use of the Higgs Characterization UFO model~\cite{Artoisenet:2013puc} with the following syntax
\begin{verbatim}
import model HC_UFO
generate p p > j j w+{X} w-{Y} 
output VBSCH_pp-wpXwmY
generate p p > j j w+ w- 
output VBSCH_pp-wpwm
\end{verbatim}
where we replace $X,\,Y$ by all combinations of $0$  (longitudinal) and $T$ (transverse).
For event generation, we consider one BSM benchmark scenario $a=0.8$ and the SM $a=1$, where $a$ is identified  with the model parameter \texttt{kSM} that can be set in \texttt{param\_card.dat}.
We further apply the following selection cuts to enhance the VBS topology: 
\begin{eqnarray}
p_T(j)>20\GeV,\quad |\eta(j)|<5,\quad m(jj)>250\GeV,\quad \lvert\Delta \eta(jj)\rvert>2.5, \nonumber \\
m(W^+W^-)>300\GeV,\quad p_T(W^\pm)>30\GeV,\quad \lvert\eta(W^\pm)\rvert 2.5\,.
\label{eq:CHcuts}
\end{eqnarray}

In \refta{tab:CHxsPart} we show the effective cross section for each polarization configuration for the composite-Higgs (CH) scenario ($a=0.8$) and the SM. 
The polarization rates and the ratio between CH and SM are also shown.
As expected, the growing behavior in the $(\lambda_{W^+},\lambda_{W^-})=(0,0)$ helicity configuration is a manifestation of the CH interaction, which results in a 30\% increase in the cross section.

\begin{table}[!t]
\begin{center}
\begin{tabular}{ c || c | c | c | c | c }
\hline
\hline
			&  \multicolumn{2}{c|}{SM}	&  \multicolumn{2}{c|}{$a=0.8$}	&  CH/SM		\tabularnewline
Process		&	$\sigma$ [fb]		& $f_{\lambda,\lambda'}$ [\%]	&	$\sigma$ [fb]		& $f_{\lambda,\lambda'}$ [\%]	&	\tabularnewline
\hline
$jjW^+W^-$			& 169.0		   	     & 100 	    &   169.2   & 100       &  1.00 	     \tabularnewline
$jjW_T^+W_T^-$		&  119.2			&	70.5 	&   116.4	& 68.8 		&  0.98  \tabularnewline
$jjW_0^+W_T^-$		& 20.6				& 12.2    	&  21.54	&	12.7  	&  1.05 	     \tabularnewline
$jjW_T^+W_0^-$		&  23.8				&  14.1  	&   24.06	&14.2   	&  1.01 	     \tabularnewline
$jjW_0^+W_0^-$		&  5.45				& 	3.2	    &   7.167	& 4.2  		&  1.31 	     \tabularnewline
	    \hline
\end{tabular}
\caption{
Total effective cross section after cuts of \eq{eq:CHcuts} for the LO process $pp\to jj W^+W^-$. Polarizations defined in the partonic C.M. frame. 
}
\label{tab:CHxsPart}
\end{center}
\end{table}

\begin{figure*}[!t]
\begin{center}
\includegraphics[width=.75\textwidth]{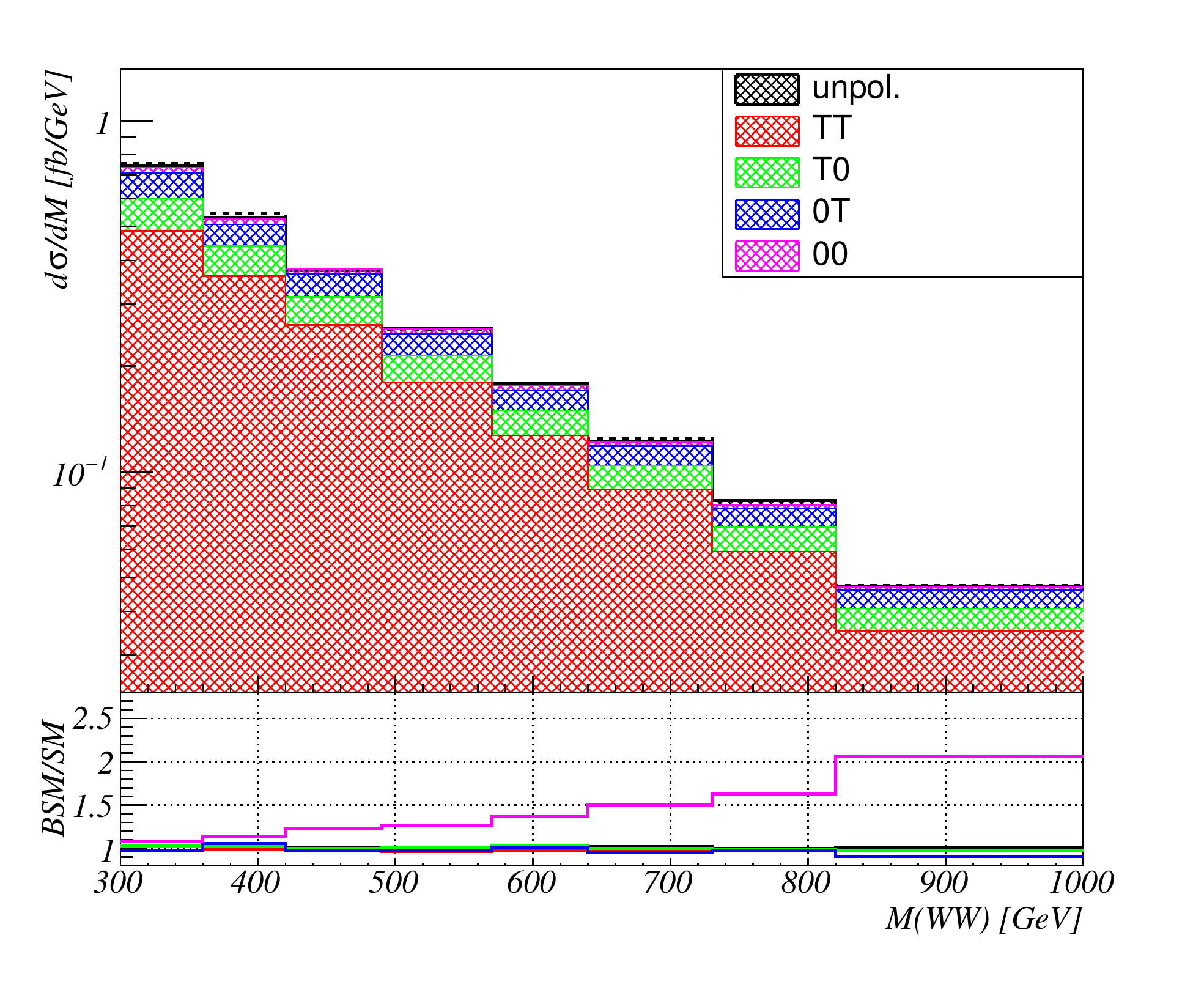}							\label{fig:VBSCH_mww_rf12}							

\end{center}
\caption{
The $WW$ invariant mass spectrum $(d\sigma/dM)$ for the unpolarized, EW process $pp\to jj W^+_\lambda W^-_{\lambda'}$ at LO, in the SM limit $(a=1.0)$.
On the lower panel the ratio $d\sigma^{\rm CH}/dM(WW) ~/~ d\sigma^{\rm SM}/dM(WW)$ of the Composite Higgs scenario with $a=0.8$ w.r.t. the SM.
Helicity polarization $(\lambda,\lambda')$ are defined in the parton c.m.~frame.
}
\label{fig:CHWWdist}
\end{figure*}

The $WW$-system invariant mass $M(WW)$ differential cross section is shown in \reffi{fig:CHWWdist}. The different polarization configurations are stacked on top of each other. We can observe again the growing behavior of the CH case w.r.t. the SM prediction only manifesting in the purely longitudinal scattering.

\subsection*{Polarized di-boson production at NLO in QCD}

To illustrate the use of mode II with decayed polarized particles, the NLO-QCD features and parton shower matching, 
we show cross sections and distributions for the process 
\begin{equation}
q \bar{q'} \to W^\pm Z,\, W^\pm\to \ell^\pm\nu, \, Z\to \tau^+\tau^- \,.
\end{equation}
The syntax for the production of polarized di-bosons at NLO in QCD is 
\begin{verbatim}
import model loop_sm-lepton_masses
define ww = w+ w-
generate p p > ww{X} z{Y} [QCD] 
\end{verbatim}
with $X,Y\in (0,T)$.

The total cross section for different collider energies are reported in \refta{tb:multiboson_xSecWZ}.
In the upper panel, we show the total cross section [pb] at NLO for inclusive, unpolarized $pp\to W^\pm Z$ production, with renormalization ($\mu_r$) and factorization ($\mu_f$) scale variation [\%], and NLO in QCD $K$-factor.
In the subsequent rows, we show the same for individual  $W_\lambda,Z_{\lambda'}$ polarizations with their fractional contribution $f^{\rm NLO~(LO)}_{\lambda,\lambda'}$ [\%] to the total, 
unpolarized rate at NLO~(LO). In the lower panel, we show the same but with $p_T(W),~p_T(Z)>200\GeV$ phase space cuts applied.

\begin{table}[!t]
\begin{center}
\resizebox{\columnwidth}{!}{
\begin{tabular}{ c || c | c | c | c | c | c | c | c | c}
\hline
\hline
$\sqrt{s}$		&  \multicolumn{3}{c|}{$13\TeV$}	&  \multicolumn{3}{c|}{$14\TeV$}	&  \multicolumn{3}{c}{$100\TeV$}	 	\tabularnewline
Process		&	$\sigma^{\rm NLO}$ [pb]	& $K$ &$f^{\rm NLO}_{\lambda,\lambda'}~(f^{\rm LO}_{\lambda,\lambda'})$
			&	$\sigma^{\rm NLO}$ [pb]	& $K$ &$f^{\rm NLO}_{\lambda,\lambda'}~(f^{\rm LO}_{\lambda,\lambda'})$
			&	$\sigma^{\rm NLO}$ [pb]	& $K$ &$f^{\rm NLO}_{\lambda,\lambda'}~(f^{\rm LO}_{\lambda,\lambda'})$					\tabularnewline
\hline
										&  \multicolumn{9}{c}{Region I: Inclusive $pp\to W_\lambda^\pm Z_{\lambda'}$}				\tabularnewline
$WZ$		& $44.7^{+4\%}_{-4\%}$ & $1.59$ & $\dots$    
			& $50.2^{+5\%}_{-5\%}$ & $1.62$ & $\dots$    
			& $571^{+11\%}_{-12\%}$ & $2.08$ & $\dots$ \tabularnewline
$W_TZ_T$	& $32.4^{+4\%}_{-4\%}$ & $1.51$ & $73\%~(76\%)$    
			& $35.8^{+4\%}_{-4\%}$ & $1.52$ & $71\%~(76\%)$    
			& $404^{+10\%}_{-11\%}$ & $1.90$ & $71\%~(77\%)$ \tabularnewline
$W_0Z_T$	& $5.43^{+6\%}_{-6\%}$ & $2.05$ & $12\%~(9\%)$    
			& $6.05^{+6\%}_{-6\%}$ & $2.07$ & $12\%~(9\%)$    
			& $72.0^{+15\%}_{-15\%}$ & $3.00$ & $13\%~(9\%)$ \tabularnewline
$W_TZ_0$	& $5.06^{+6\%}_{-6\%}$ & $2.10$ & $11\%~(9\%)$    
			& $5.63^{+6\%}_{-6\%}$ & $2.12$ & $11\%~(9\%)$    
			& $67.1^{+15\%}_{-15\%}$ & $3.06$ & $12\%~(8\%)$ \tabularnewline
$W_0Z_0$	& $2.31^{+4\%}_{-4\%}$ & $1.34$ & $5\%~(6\%)$    
			& $2.52^{+4\%}_{-4\%}$ & $1.34$ & $5\%~(6\%)$    
			& $23.3^{+7\%}_{-10\%}$ & $1.44$ & $4\%~(6\%)$ \tabularnewline
										&  \multicolumn{9}{c}{Region II: $p_T(W^\pm),~p_T(Z)  >200\GeV$}				\tabularnewline
$WZ$		& $0.531^{+6\%}_{-5\%}$ & $1.55$ & $\dots$    
			& $0.617^{+6\%}_{-5\%}$ & $1.60$ & $\dots$    
			& $16.3^{+7\%}_{-7\%}$ & $2.82$ & $\dots$ \tabularnewline		
$W_TZ_T$	& $0.409^{+8\%}_{-6\%}$ & $1.72$ & $77\%~(70\%)$    
			& $0.475^{+7\%}_{-6\%}$ & $1.76$ & $77\%~(70\%)$    	
			& $13.3^{+8\%}_{-8\%}$ & $3.12$ & $82\%~(74\%)$ \tabularnewline
$W_0Z_T$	& $25.6\times10^{-3}~^{+8\%}_{-6\%}$ & $1.71$ & $5\%~(4\%)$    
			& $29.5\times10^{-3}~^{+8\%}_{-6\%}$ & $1.75$ & $5\%~(4\%)$    		
			& $0.882^{+8\%}_{-8\%}$ & $4.25$ & $5\%~(4\%)$ \tabularnewline
$W_TZ_0$	& $25.3\times10^{-3}~^{+8\%}_{-7\%}$ & $1.84$ & $5\%~(4\%)$    
			& $29.8\times10^{-3}~^{+9\%}_{-7\%}$ & $1.92$ & $5\%~(4\%)$    		
			& $0.902^{+9\%}_{-8\%}$ & $4.69$ & $6\%~(3\%)$ \tabularnewline
$W_0Z_0$	& $74.6\times10^{-3}~^{+0.5\%}_{-<0.5\%}$ & $1.01$ & $14\%~(22\%)$    
			& $83.4\times10^{-3}~^{+<0.5\%}_{-<0.5\%}$ & $1.00$ & $14\%~(22\%)$    	
			& $1.07^{+1.5\%}_{-1.8\%}$ & $1.00$ & $7\%~(19\%)$ \tabularnewline
	    \hline\hline	    
\end{tabular}
} 
\caption{
Upper: Total cross section [pb] at NLO for inclusive, unpolarization $pp\to W^\pm Z$ production, with scale variation [\%], and NLO in QCD $K$-factor,
as well as the same for individual  $W_\lambda,Z_{\lambda'}$ polarizations along with their fractional contribution $f^{\rm NLO~(LO)}_{\lambda,\lambda'}$ [\%] at NLO~(LO).
Lower: Same as upper but with $p_T(W),~p_T(Z)>200\GeV$ phase space cuts applied.
}
\label{tb:multiboson_xSecWZ}
\end{center}
\end{table}

Next we can consider the decay of the bosons via \ms, and define the decay channels in \texttt{madspin\_card.dat} via the commands
\begin{verbatim}
define ww = w+ w-
decay ww > emu vem
decay  z > ta+ ta-
\end{verbatim}

\begin{figure*}[!t]
\begin{center}
\includegraphics[width=.45\textwidth]{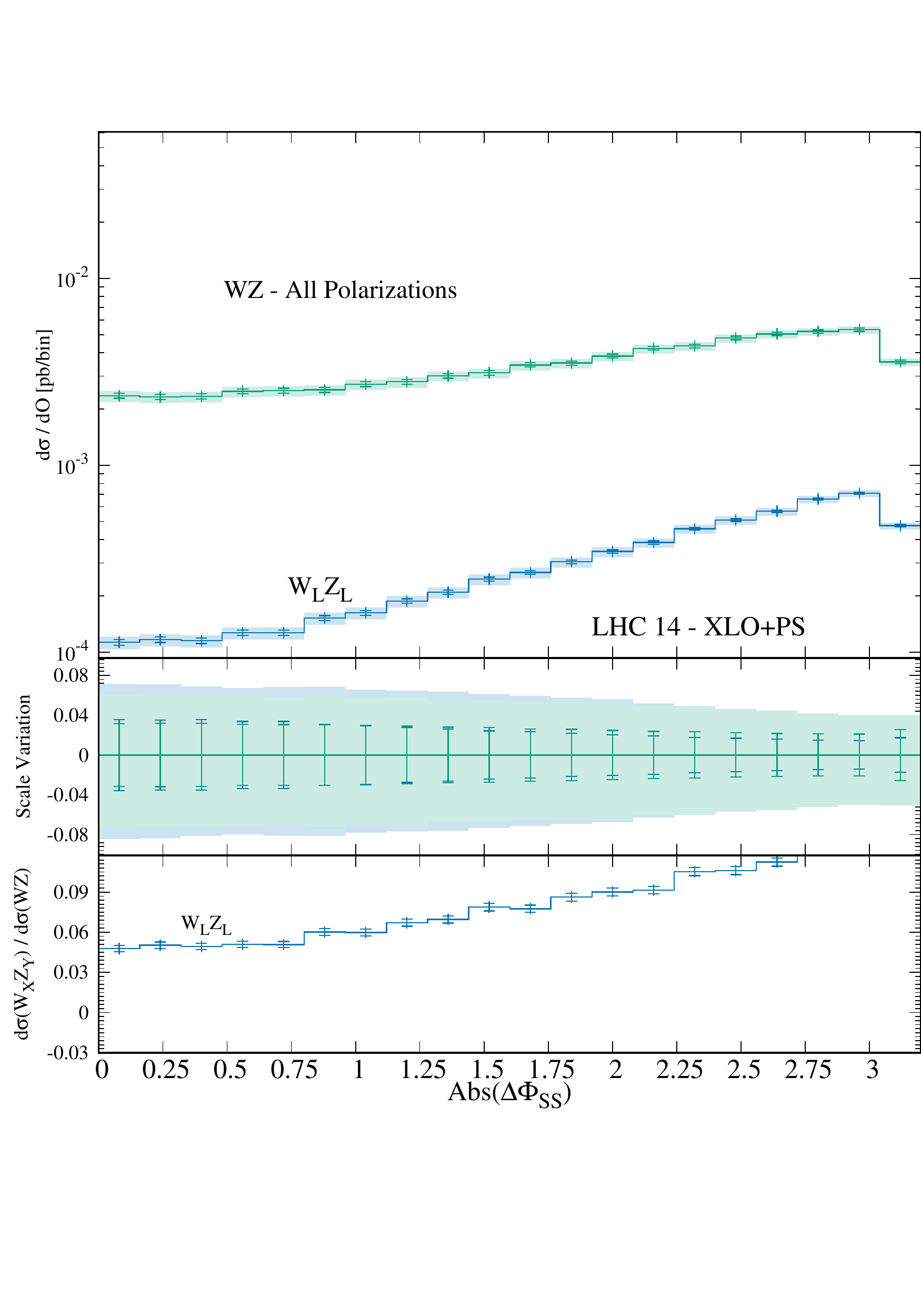}			\label{fig:mgPolar_pp_WZ_3lX_absDPhiSS_XLOPS}
\hfill	
\includegraphics[width=.45\textwidth]{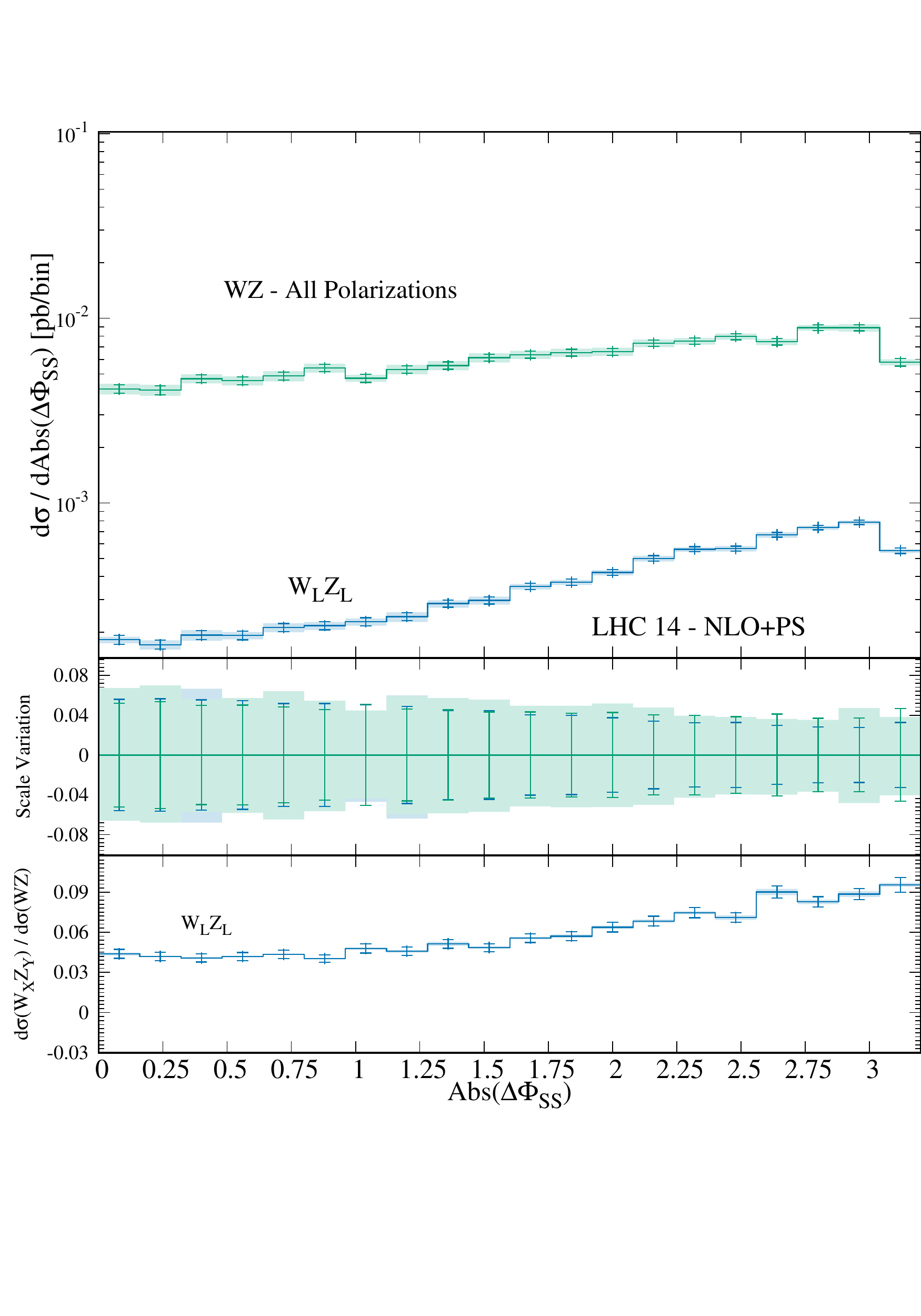}			\label{fig:mgPolar_pp_WZ_3lX_absDPhiSS_NLOPS}	
\end{center}
\caption{
Preliminary differential distribution of $\vert \Delta\Phi_{\rm SS}\vert$ as (left) LO+PS and (right) NLO+PS.
}
\label{fig:mgPolar_pp_WZ_3lX_absDPhiSS}
\end{figure*}

In this framework, it is possible to access the decay products to perform analyses and plot differential distributions.
In \reffi{fig:mgPolar_pp_WZ_3lX_absDPhiSS} we show the distributions in the absolute azimuthal separation of the same-sign dilepton system,
\begin{equation}
\vert \Delta\Phi_{\rm SS}\vert = \vert \phi(\tau^\pm) - \phi(l^\pm) \vert,
\end{equation}
where $l\in\{e,\mu\}$ originates from the $W$ boson decay.
The following kinematical cuts have been required
\begin{eqnarray}
\vert \eta^\ell \vert < 2.4, \quad p_T^{\ell} > 20\GeV,\quad \vert m(\tau\tau) - M_Z \vert < 10\GeV.
\label{eq:dibson_fidkinCuts}
\end{eqnarray}

\subsection*{Conclusion}
We present a framework within the \mgFull~program to deal with polarized states. 
Here we give a brief description of the implementation, reporting the syntax and examples.
In particular, we show applications: high energy VBS in composite Higgs models and polarized diboson production at NLO+PS.
There are many other applications of great interest for both the experimental and theoretical communities.
During the workshop, several discussions took place for the application of the \mgamc~framework to the study of dimension-6 operators sensitive to polarization observables.
This is a path to be pursued in the near future.

\section{Polarization studies in $WZ$ production at the LHC\footnote{speaker: Le Duc Ninh; authors: Julien Baglio, Emmanuel Sauvan}}

Polarization observables can be useful to get 
a deeper understanding of the gauge structure of the Standard Model and 
to look for new physics effects. In this work we discuss, from a practical point
of view, how polarization observables of a massive gauge boson can be
defined. Two observables are presented and calculated for the case of 
$pp \to WZ \to 3\ell\nu + X$ up to the next-to-leading order (NLO) QCD + electroweak (EW) level, 
called template-fitted and 
fiducial polarization fractions. A comparison 
with ATLAS simulation and experimental results at a 
center-of-mass energy of 13 TeV for the
case of the template-fitted fractions is also presented. 
The EW corrections are calculated using a double-pole
approximation and found to be large.

\subsection*{Introduction}

Understanding the polarizations of the gauge bosons in 
the process $pp \to W^\pm Z \to 3\ell \nu + X$ is an important step towards 
the understanding of similar effects in vector boson scattering processes 
$VV \to VV$ at the LHC, $V=W,Z$.
This is because diboson production is the simplest process where the
correlations between the polarizations of the two gauge bosons can be investigated.
These correlation observables can be more sensitive to possible new physics
effects in $VV \to VV$ scatterings. 
For example, the double longitudinal fraction $f_{LL} = d\sigma(V_L V_L)/d\sigma(VV)$ is 
interesting, because it is the leading contribution in the high energy limit ($E_V \to \infty$). 

In the first step, when statistics is still limited, more inclusive
observables should be considered. Therefore, polarizations of one gauge
boson, where the polarizations of the other gauge boson are summed
over, have been recently calculated with NLO 
QCD and EW correction effects taken into account
\cite{Baglio:2018rcu} and have also been measured at ATLAS
\cite{Aaboud:2019gxl}. However, comparisons between these two works
are not possible as different definitions of polarization observables
are used.

In this work we discuss the issue of defining polarization
observables of a massive gauge boson taking into account NLO QCD and EW corrections. 
The two observables used in those two works are both considered, called 
template-fitted \cite{Aaboud:2019gxl} and fiducial \cite{Baglio:2018rcu} polarization fractions. 
For the numerical results, differently from \cite{Baglio:2018rcu}, new results calculated in the 
modified helicity coordinate system are presented. 
For the first time, a comparison to the ATLAS measurement \cite{Aaboud:2019gxl} taking into account 
NLO EW corrections is also provided. 
The correlations between the polarizations of the two gauge bosons are
not further addressed and are left for future work.

\subsection*{Defining polarization observables}
\label{sec:define_polarization}
Polarizations of a massive gauge boson cannot be directly measured in
experiments. In practice, we have to infer them from the angular
distribution of its decay product, typically a charged lepton ($e$ or
$\mu$). In the following discussion, we will first discuss this
distribution in the context of the process $pp \to W^\pm Z \to e^\pm
\mu^+ \mu^- \nu + X$ in the double-pole approximation (DPA)
and at leading order (LO), where the origin of a final-state lepton
can be traced back to a single intermediate gauge boson. Based on
this, polarization observables for the full process $pp \to e^\pm
\mu^+ \mu^- \nu + X$ with off-shell and higher-order corrections
effects taken into account will be defined.

The master equation widely used in the literature reads (see
e.g. \cite{Aad:2016izn})
\begin{align}
\fr{d\sigma}{dp_T^V dy^V dm^V d\!\cos\theta d\phi} &= \fr{3}{16\pi} \fr{d\sigma}{dp_T^V dy^V dm^V}\crn
&\times \Big[ 
(1+\cos^2\theta) + \hat{A}_0 \fr{1}{2}(1-3\cos^2\theta)
+ \hat{A}_1 \sin(2\theta)\cos\phi  \crn
& + \hat{A}_2 \fr{1}{2} \sin^2\theta \cos(2\phi)
+ \hat{A}_3 \sin\theta\cos\phi + \hat{A}_4 \cos\theta \crn
& + \hat{A}_5 \sin^2\theta \sin(2\phi) 
+ \hat{A}_6 \sin(2\theta) \sin\phi + \hat{A}_7 \sin\theta \sin\phi
\Big],
\label{eq:definition_Ai_tot}
\end{align}
describing the $\cos\theta$-$\phi$ angular distribution for every bin
in the ($p_T^V$, $y^V$, $m^V$) space. Here, $p_T^V$, $y^V$, $m^V$ are
the transverse momentum, rapidity, and invariant mass of a gauge boson
($W$ or $Z$), respectively. $\theta$ and $\phi$ are the polar and
azimuthal angles of a charged lepton, determined in a given coordinate
system applied in the rest frame of the gauge boson under consideration. The
angular coefficients 
$\hat{A}_i$, 
$i=0,\ldots ,7$,
are functions of $p_T^V$, $y^V$ and
$m^V$, but independent of $\theta$ and $\phi$. 

Integrating over $p_T^V$, $y^V$, $m^V$, we get
\begin{align}
\fr{d\sigma}{\sigma d\cos\theta d\phi} &= \fr{3}{16\pi} 
\Big[ 
(1+\cos^2\theta) + A_0 \fr{1}{2}(1-3\cos^2\theta)
+ A_1 \sin(2\theta)\cos\phi  \crn
& + A_2 \fr{1}{2} \sin^2\theta \cos(2\phi)
+ A_3 \sin\theta\cos\phi + A_4 \cos\theta \crn
& + A_5 \sin^2\theta \sin(2\phi) 
+ A_6 \sin(2\theta) \sin\phi + A_7 \sin\theta \sin\phi
\Big],\label{eq:definition_Aix}
\end{align}
where the integrated angular coefficients $A_i$ depend on the
integration range of $p_T^V$, $y^V$, $m^V$. We can of course choose to
integrate over two variables (e.g. $y^V$ and $m^V$) to study the
dependence on the remaining variable ($p_T^V$). 

Further integrating over $\phi$ we obtain, using the index $3$ to
indicate the electron and $6$ the muon,
\begin{align}
\fr{d\sigma}{\sigma d\cos\theta_3} &\equiv \fr{3}{8} 
\Big[ 
(1\mp \cos\theta_3)^2 f^{W^{\pm}}_L + (1\pm \cos\theta_3)^2 f^{W^{\pm}}_R 
+ 2 \sin^2\theta_3 f^{W^{\pm}}_0 \Big],\label{eq:f_L0R_W}\\
\fr{d\sigma}{\sigma d\cos\theta_6} &\equiv \fr{3}{8} 
\Big[ 
(1 + \cos^2\theta_6 + 2c\cos\theta_6) f^{Z}_L + (1 + \cos^2\theta_6 - 2c\cos\theta_6) f^{Z}_R\crn
&+ 2 \sin^2\theta_6 f^{Z}_0 \Big],\label{eq:f_L0R_Z}\\
f^{V}_L 
  &= \fr{1}{4}(2- A^{V}_0 + d_V A^{V}_4),\;\; f^{V}_R
    = \fr{1}{4}(2- A^{V}_0 - d_V A^{V}_4),\;\;
f^{V}_0 = \fr{1}{2}A^{V}_0,\crn
f^{V}_L - f^{V}_R &= \fr{d_V}{2} A^V_4, \quad d_Z = \fr{1}{c},\;\; d_{W^\pm}=\mp 1,\crn
c &= \fr{g_L^2 - g_R^2}{g_L^2 + g_R^2} = \fr{1-4s^2_W}{1-4s^2_W+8s^4_W} \approx 0.21,\quad s^2_W = 1 - \fr{M_W^2}{M_Z^2}.
\end{align}
These fractions, satisfying $f^V_L + f^V_R + f^V_0 =1$, depend on the
selection cuts on $p_T^V$, $y^V$, $m^V$ and also on other cuts on the
kinematics of the decay products of the other gauge boson. For
example, the values of $f^Z_{L,0,R}$ depend not only on the cuts on
$p_T^Z$, $y^Z$, $m^Z$ but also on other cuts on the kinematics of the
electron from the $W$ decay (e.g. $p_{T,e}$ or $\eta_e$).

An important condition for Eqs.~(\ref{eq:definition_Ai_tot}),
(\ref{eq:definition_Aix}), (\ref{eq:f_L0R_W}), and (\ref{eq:f_L0R_Z})
to hold is that the phase space of the decay lepton is not
restricted. For example, when calculating the polarization fractions
of the $Z$ boson, kinematic cuts on the individual muons such as cuts
on $p_{T,\mu^{\pm}}$ or $\eta_{\mu^{\pm}}$ are not allowed.

If such a non-restricted (or inclusive) angular distribution
${d\sigma}/{d\cos\theta}$ is known, the fractions can be
easily calculated using the following projection method,
\begin{align}
\langle f(\theta) \rangle &= \int_{-1}^{1} d\cos\theta f(\theta) \fr{1}{\sigma} \fr{d\sigma}{d\cos\theta},\label{eq:def_projection} \\
f^{W^{\pm}}_L 
  &= -\fr{1}{2} \mp \langle\cos\theta_3\rangle +
    \fr{5}{2}\langle\cos^2\theta_3\rangle, \; 
f^{W^{\pm}}_R = -\fr{1}{2} \pm \langle\cos\theta_3\rangle +
    \fr{5}{2}\langle\cos^2\theta_3\rangle, \crn 
f^{W^{\pm}}_0 &= 2 - 5 \langle\cos^2\theta_3\rangle,\label{eq:cal_fLR0_W}\\
f^Z_L 
  &= -\fr{1}{2} + \fr{1}{c}\langle\cos\theta_6\rangle +
    \fr{5}{2}\langle\cos^2\theta_6\rangle, \;
f^Z_R 
  = -\fr{1}{2} - \fr{1}{c}\langle\cos\theta_6\rangle +
    \fr{5}{2}\langle\cos^2\theta_6\rangle, \crn
f^Z_0 
  &= 2 - 5 \langle\cos^2\theta_6\rangle.
\label{eq:cal_fLR0_Z}
\end{align}

In practice, such a non-restricted angular distribution ${d\sigma}/{d\cos\theta}$ cannot be measured. 
Moreover, the measured distribution ${d\sigma^\text{fid}}/{d\cos\theta}$ includes also off-shell, interference, and radiative correction effects. 
Can the polarization information be extracted from this distribution? 

To this end, we define two polarization observables: template-fitted and fiducial polarization fractions. 
The template-fitted fractions are calculated by fitting the distribution ${d\sigma^\text{fid}}/{d\cos\theta}$ using three templates, 
as specified in \cite{Aaboud:2019gxl}. The results therefore depend on 
the kinematical cuts as pointed out in the Table 8.1 of
\cite{Burger:2645514} and also on the templates that are used. 

Alternatively, one can replace ${d\sigma}/(\sigma {d\cos\theta})$ by ${d\sigma^\text{fid}}/(\sigma^\text{fid} {d\cos\theta})$ 
in the equations (\ref{eq:def_projection}), (\ref{eq:cal_fLR0_W}) and (\ref{eq:cal_fLR0_Z}) to define fiducial
polarization fractions \cite{Baglio:2018rcu}\footnote{These are called projection results in \cite{Stirling:2012zt}.}. 
The results depend obviously on the kinematical cuts.

\begin{figure}[ht!]
  \centering
  \includegraphics[height=5cm]{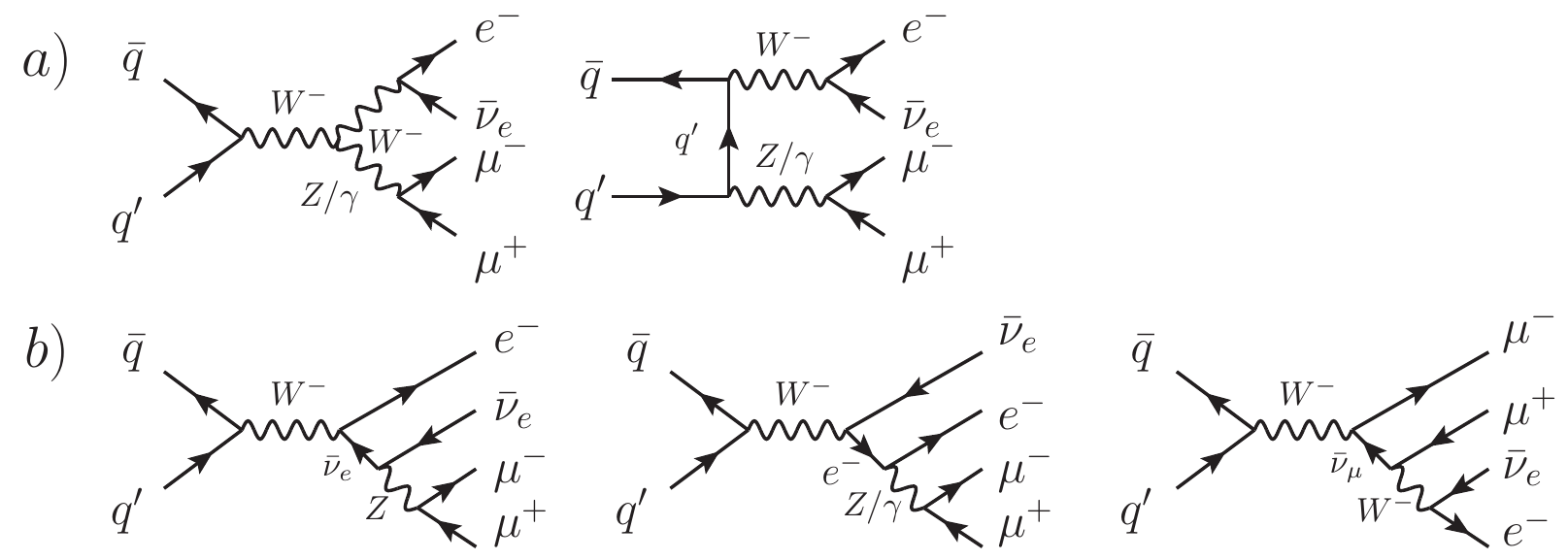}
  \caption{Leading-order Feynman diagrams showing different origins of a final-state charged lepton.}
  \label{fig:LO_diags}
\end{figure}
A word of caution is added here concerning the terminologies, 
in particular for the ``polarization fractions of a particular massive gauge boson''. 
The above definitions have made it clear that those polarization 
fractions are calculated from the angular distribution of a charged
lepton (by using either the template-fit method or projections). 
At LO, the lepton, say the muon, can originate from an intermediate
$Z$, a virtual photon, or even a $W$ as can be seen from
\reffi{fig:LO_diags}. Note that there are also interference effects
between these different mechanisms. Calling the results $Z$
polarization fractions is therefore misleading. However, it can be
acceptable as long as kinematic cuts are applied to enhance the
dominant $Z \to \mu^+ \mu^-$ mechanism. For example, the cut
$\left|m_{\mu^+\mu^-} - M_Z\right| < 10\gev$ used in the
numerical-result section reduces most of the $\gamma^* \to \mu^+
\mu^-$ contribution and part of the $W^- \to \mu^- \mu^+ e^-
\bar{\nu}_e$ effect. Similarly, the electron can be created by
different mechanisms $W^- \to e^- \bar{\nu}_e$ and $W^- \to \mu^-
\mu^+ e^- \bar{\nu}_e$, see \reffi{fig:LO_diags}.

Moreover, the polarization fractions depend on the coordinate system
where the polar angle $\theta$ is determined. It becomes therefore
clear that these polarization observables are process,
kinematical-cuts, and coordinate-system dependent. Their values do not
represent any universal property of the polarization of a massive
gauge boson as the names may suggest, but can be nonetheless a 
useful piece of information to unravel the polarization structure 
of the given gauge boson and/or to discover new physics.

\subsection*{Numerical results}
\label{sect:numset}
The input parameters are the same as in \cite{Baglio:2018rcu}, in
particular $\sqrt{s} = 13\tev$ and the scales $\mu_R = \mu_F = (M_W+M_Z)/2$. We will use the ATLAS fiducial cuts
defined in \cite{Aaboud:2019gxl}:
\begin{align}
\begin{matrix}
        p_{T,\mu} > 15\gev, \quad p_{T,e} > 20\gev, \quad 
        |\eta_\ell|<2.5,\\
        \Delta R\left(e,\mu^\pm\right) > 0.3, \quad \Delta
        R\left(\mu^+,\mu^-\right) > 0.2,\\
        \left|m_{\mu^+\mu^-} - M_Z\right| < 10\gev, \quad m_{T,W} >
        30\gev .
\end{matrix}
\label{eq:kin_cuts}
\end{align}
We remark that the $\Delta R(e, \mu^\pm)$ cut restricts both the phase
space of the electron and the muons at the same time, thereby
affecting the polarization fractions of both $W$ and $Z$ bosons.
For those cuts and also for binning the lepton angular distributions, dressed leptons are 
used as in \cite{Baglio:2018rcu}. A dressed lepton momentum is defined as $p'_\ell = p_\ell +
p_\gamma$ for $\Delta R(\ell,\gamma) \equiv \sqrt{(\Delta\eta)^2+(\Delta\phi)^2}< 0.1$ with 
$p_\ell$ being the momentum of the charged lepton after QED final-state radiation. 

The key element in the calculation of either the template-fitted or
fiducial polarization fractions is the fiducial distribution
$d\sigma^\text{fid}/d\!\cos\theta$. This has been calculated up to full
NLO QCD accuracy using the {\tt VBFNLO} program
\cite{Arnold:2008rz,Baglio:2014uba}. NLO EW corrections using 
the DPA presented in \cite{Baglio:2018rcu}
are also included. The important features of this EW calculation are the following:
\begin{itemize}
\item The virtual and real corrections to the production
  part are included;
\item The virtual and real corrections to the decays
  are included;
\item The quark-photon induced processes, e.g. $q \gamma
  \to WZ q' \to 4l q'$, are included;
\item The non-factorizable contributions are not
  included;
\item Off-shell effects are not included.
\end{itemize}
The exact definition of the various corrections and of the
non-factorizable contribution are given in
\cite{Baglio:2018rcu}. Comparisons between this approximation and the
full NLO EW corrections \cite{Biedermann:2017oae} have been presented in
\cite{Baglio:2018rcu,Baglio:2019ivi}, showing very good
agreements. For the $\cos\theta$ distribution needed here, very good
agreement is expected as it has been known that off-shell effects are
usually very small for angular distributions and can be significant
only for some transverse momentum or invariant mass distributions
\cite{Baglio:2018rcu} (see also \cite{Biedermann:2016guo} for the case of $pp \to W^+ W^- \to 4\,\text{leptons}$).

In the following, results for all polarization fractions are obtained 
at NLO QCD (full amplitudes including off-shell effects) + EW (DPA) 
using the modified helicity coordinate system, which is similar to the
helicity coordinate system defined in \cite{Bern:2011ie}. The only
difference is the direction of the $z$ axis: instead of being the
gauge boson flight direction in the laboratory frame as chosen in
\cite{Bern:2011ie}, it is now the gauge-boson flight direction in the
$WZ$ centre-of-mass frame. This modified helicity coordinate system is
also used by ATLAS in \cite{Aaboud:2019gxl}. In addition, LO results obtained with 
the full amplitudes are provided.\bigskip

\noindent{\bf Template-fitted polarization fractions\\}
In Tables~\ref{tab:coeff_fL0R_WpZ_fits},
\ref{tab:coeff_fL0R_WmZ_fits} results for the polarization fractions
obtained by fitting the fiducial distribution
${d\sigma^\text{fid}}/{d\cos\theta}$ are presented. \footnote{In our talk given at
  the VBSCan Mid-Term Scientific Meeting \cite{Ninh:2019vbsscan} they
  were called inclusive fractions. The word ``inclusive'' is actually
  not appropriate as the results depend on the fiducial cuts.}

\begin{table}[ht!]
  \renewcommand{\arraystretch}{1.3}
\begin{center}
    \fontsize{8}{8}
\begin{tabular}{|c|c|c||c|c|}\hline
$\text{Method}$  & $f^{W^+}_0$ & $f^{W^+}_L-f^{W^+}_R$ & $f^Z_0$ & $f^Z_L-f^Z_R$\\
\hline
$\text{ATLAS data}$ & $0.26(8)$ & $-0.02(4)$ & $0.27(5)$ & $-0.32(21)$\\
$\text{ATLAS POWHEG+PYTHIA}$ & $0.233(4)$ & $0.091(4)$ & $0.225(4)$ & $-0.297(21)$\\
$\text{ATLAS MATRIX}$ & $0.2448(10)$ & $0.0868(14)$ & $0.2401(14)$ & $-0.262(9)$\\
\hline
$\text{NLO QCD EW}$ & $0.244$ & $0.078$ & $0.237$ & $-0.244$\\
$\text{NLO QCD}$ & $0.241$ & $0.082$ & $0.232$ & $-0.307$\\
$\text{LO}$ & $0.247$ & $0.126$ & $0.214$ & $-0.472$\\
\hline
\end{tabular}
\caption{\small Template-fitted $W^+$ and $Z$ polarization fractions
  at LO, NLO QCD and  NLO QCD+EW in comparison with ATLAS measurements
  (data) and simulations (POWHEG+PYTHIA and MATRIX). ATLAS results are
  taken from \cite{Aaboud:2019gxl}.
}
\label{tab:coeff_fL0R_WpZ_fits}
\end{center}
\end{table}
\begin{table}[ht!]
  \renewcommand{\arraystretch}{1.3}
\begin{center}
    \fontsize{8}{8}
\begin{tabular}{|c|c|c||c|c|}\hline
$\text{Method}$  & $f^{W^-}_0$ & $f^{W^-}_L-f^{W^-}_R$ & $f^Z_0$ & $f^Z_L-f^Z_R$\\
\hline
$\text{ATLAS data}$ & $0.32(9)$ & $-0.05(5)$ & $0.21(6)$ & $-0.46(25)$\\
$\text{ATLAS POWHEG+PYTHIA}$ & $0.245(5)$ & $-0.063(6)$ & $0.235(5)$ & $0.052(23)$\\
$\text{ATLAS MATRIX}$ & $0.2651(15)$ & $-0.034(4)$ & $0.2389(15)$ & $0.0468(34)$\\
\hline
$\text{NLO QCD EW}$ & $0.259$ & $-0.045$ & $0.236$ & $0.050$\\
$\text{NLO QCD}$ & $0.257$ & $-0.049$ & $0.232$ & $0.079$\\ 
$\text{LO}$ & $0.252$ & $-0.163$ & $0.209$ & $0.063$\\
\hline
\end{tabular}
    \caption{\small Same as \refta{tab:coeff_fL0R_WpZ_fits} but for the $W^- Z$ process.}
    \label{tab:coeff_fL0R_WmZ_fits}
\end{center}
\end{table}

Our results are obtained from fitting the LO, NLO QCD and NLO QCD + EW distributions using templates of the $d\sigma^\text{fid}/d\cos{\theta}$ distribution for the three helicity states. 
The templates are generated using POWHEG+PYTHIA Monte Carlo events at particle level. 
POWHEG-BOX~v2 \cite{Nason:2004rx,Frixione:2007vw,Alioli:2010xd,Melia:2011tj} is used for generating the four-lepton hard processes at NLO QCD, while PYTHIA~8.210 \cite{Sjostrand:2014zea} is used for the simulation of parton showering, hadronization, and underlying events.
Generated events fulfilling kinematic criteria of Eq.~(\ref{eq:kin_cuts}) with dressed leptons are selected and 
used to determine the $d\sigma^\text{fid}/d\cos{\theta}$ template distributions. 
ATLAS measurement and simulation results taken from \cite{Aaboud:2019gxl} are also shown for the sake of comparison.
We note that the POWHEG+PYTHIA results include NLO QCD corrections in the hard matrix elements and parton-shower effects, while 
the MATRIX \cite{Grazzini:2017mhc} results are obtained by fitting the NNLO QCD $\cos\theta$ distributions using the POWHEG+PYTHIA templates. 
It is also important to note that Born-level leptons are used in the ATLAS results, while 
dressed leptons have to be used in our calculation to ensure infrared (IR) safety in the NLO EW corrections.
It was possible for ATLAS to use Born-level leptons 
because NLO EW corrections were not included in their simulation.

\begin{table}[ht!]
  \renewcommand{\arraystretch}{1.3}
\begin{center}
    \fontsize{8}{8}
\begin{tabular}{|c|c|c||c|c|}\hline
$\text{Correction}~[\%]$  & $f^{W}_0$ & $f^{W}_L-f^{W}_R$ & $f^Z_0$ & $f^Z_L-f^Z_R$\\
\hline
$\delta_\text{QCD}\; (W^+Z)$ & $-2.4$ & $-34.9$ & $+8.4$ & $-35.0$\\
$\delta_\text{EW}\; (W^+Z)$ & $+1.2$ & $-4.9$ & $2.2$ & $-20.5$\\
\hline
\hline
$\delta_\text{QCD}\; (W^-Z)$ & $+2.0$ & $-69.9$ & $+11.0$ & $+25.4$\\
$\delta_\text{EW}\; (W^-Z)$ & $+0.8$ & $-8.2$ & $+1.7$ & $-36.7$\\
\hline
\end{tabular}
\caption{\small Relative QCD and EW corrections to the template-fitted polarization fractions. 
QCD corrections are compared to the LO results, while EW corrections to the NLO QCD ones.}
\label{tab:coeff_fL0R_WpmZ_fits_corrections}
\end{center}
\end{table}
QCD corrections defined as $\delta_\text{QCD} = (f_\text{NLO QCD} - f_\text{LO})/f_\text{LO}$
and EW corrections $\delta_\text{EW} = (f_\text{NLO QCD EW} - f_\text{NLO QCD})/f_\text{NLO QCD}$ 
are shown in \refta{tab:coeff_fL0R_WpmZ_fits_corrections}. We observe
that  the EW corrections are most significant on $f^Z_L-f^Z_R$,
reaching $-21\%$ (-37\%) for $W^+Z$ ($W^-Z$) channels. This effect
comes from the EW corrections to the $Z$ decay into charged leptons as
pointed out in \cite{Baglio:2018rcu}.
This effect is much larger than the $W$ case 
probably because of the following two reasons. The $Z$ boson couples to both left and right-handed leptons while the $W$ only to left-handed leptons, and 
there is a cut on the $\mu^+\mu^-$ invariant mass. Final-state QED radiation shifts the peak position, thereby inducing an effect on the $Z$ decay.
\begin{table}[ht!]
  \renewcommand{\arraystretch}{1.3}
\begin{center}
    \fontsize{8}{8}
\begin{tabular}{|c|c|c||c|c|}\hline
$\text{Pull}$  & $f^{W^+}_0$ & $f^{W^+}_L-f^{W^+}_R$ & $f^Z_0$ & $f^Z_L-f^Z_R$\\
\hline
$\text{ATLAS POWHEG+PYTHIA}$ & $-0.3$ & $+2.8$ & $-0.9$ & $+0.1$\\
$\text{ATLAS MATRIX}$ & $-0.2$ & $+2.7$ & $-0.6$ & $+0.3$\\
\hline
$\text{NLO QCD EW}$ & $-0.2$ & $+2.5$ & $-0.7$ & $+0.4$\\
$\text{NLO QCD}$ & $-0.2$ & $+2.6$ & $-0.8$ & $+0.1$\\
$\text{LO}$ & $-0.2$ & $+3.7$ & $-1.1$ & $-0.7$\\
\hline
\end{tabular}
\caption{\small Pulls of template-fitted $W^+$ and $Z$ polarization fractions 
in comparison with ATLAS data, calculated from \refta{tab:coeff_fL0R_WpZ_fits}.}
\label{tab:coeff_fL0R_WpZ_fits_pull}
\end{center}
\end{table}
\begin{table}[ht!]
  \renewcommand{\arraystretch}{1.3}
\begin{center}
    \fontsize{8}{8}
\begin{tabular}{|c|c|c||c|c|}\hline
$\text{Pull}$  & $f^{W^-}_0$ & $f^{W^-}_L-f^{W^-}_R$ & $f^Z_0$ & $f^Z_L-f^Z_R$\\
\hline
$\text{ATLAS POWHEG+PYTHIA}$ & $-0.8$ & $-0.3$ & $+0.4$ & $+2.0$\\
$\text{ATLAS MATRIX}$ & $-0.6$ & $+0.3$ & $+0.5$ & $+2.0$\\
\hline
$\text{NLO QCD EW}$ & $-0.7$ & $+0.1$ & $+0.4$ & $+2.0$\\
$\text{NLO QCD}$ & $-0.7$ & $0.0$ & $+0.4$ & $+2.2$\\ 
$\text{LO}$ & $-0.8$ & $-2.3$ & $0.0$ & $+2.1$\\
\hline
\end{tabular}
    \caption{\small Same as \refta{tab:coeff_fL0R_WpZ_fits_pull} but for the $W^- Z$ process.}
    \label{tab:coeff_fL0R_WmZ_fits_pull}
\end{center}
\end{table}

Pulls between various theoretical predictions and the ATLAS data are
presented in Tables \ref{tab:coeff_fL0R_WpZ_fits_pull} and
\ref{tab:coeff_fL0R_WmZ_fits_pull}  for the $W^+Z$ and $W^-Z$
channels, respectively. Theoretical errors are neglected as they are
typically one order of magnitude smaller compared to the experimental
errors. We observe that there is a significant deviation between the
SM prediction and the data for the case of $f^{W^+}_L-f^{W^+}_R$, at
the level of $2.5\sigma$ at the next-to-next-to-leading order (NNLO)
QCD (ATLAS MATRIX) or NLO QCD + EW accuracy. However, the relative
uncertainty on this measurement is very large, at the level of $200\%$. \\

\noindent{\bf Fiducial polarization fractions}\\
In \refta{tab:coeff_fL0R_WpmZ_fid} results for the fiducial polarization
fractions obtained by using direct projections of the fiducial
distribution ${d\sigma^\text{fid}}/{d\cos\theta}$ are presented. Similar results
in the helicity and Collins-Soper coordinate systems have been
provided in \cite{Baglio:2018rcu}. We again see that EW corrections
are largest for the case of $f^Z_L-f^Z_R$, reaching $-32\%$ (-15\%)
for the $W^+Z$ ($W^-Z$) channels, due to the EW corrections to the
$Z\to \mu^+ \mu^-$ decay. To compare these results with data, unfolded
angular distributions would be needed. This comparison may shed light
on the above $2.5\sigma$ deviation observed in the template-fitted
fractions.
\begin{table}[ht!]
  \renewcommand{\arraystretch}{1.3}
\begin{center}
    \fontsize{8}{8}
\begin{tabular}{|c|c|c||c|c|}\hline
$\text{Method}$  & $f^{W}_0$ & $f^{W}_L-f^{W}_R$ & $f^Z_0$ & $f^Z_L-f^Z_R$\\
\hline
$\text{LO}\; (W^+Z)$ & $0.482$ & $0.14$ & $0.429$ & $-0.217$\\
$\text{NLO QCD}\; (W^+Z)$ & $0.483$ & $0.091$ & $0.441$ & $-0.123$\\
$\text{NLO QCD EW}\; (W^+Z)$ & $0.485$ & $0.089$ & $0.444$ & $-0.084$\\
\hline
\hline
$\text{LO}\; (W^-Z)$ & $0.518$ & $-0.058$ & $0.405$ & $0.108$\\
$\text{NLO QCD}\; (W^-Z)$ & $0.498$ & $0.05$ & $0.422$ & $0.116$\\
$\text{NLO QCD EW}\; (W^-Z)$ & $0.498$ & $0.054$ & $0.425$ & $0.099$\\
\hline
\end{tabular}
\caption{\small Fiducial $W^\pm$ and $Z$ polarization fractions at LO,
  NLO QCD and NLO QCD+EW accuracy.}
\label{tab:coeff_fL0R_WpmZ_fid}
\end{center}
\end{table}

\subsection*{Summary}
\label{sect:summary}
In this work the issue of defining individual gauge boson polarization observables
has been discussed. Two observables have been defined, template-fitted
and fiducial polarization fractions. The template-fitted fractions are
obtained from fitting the fiducial angular distribution of a decay
lepton and the NLO EW corrections for these fractions are presented here for the first time.
The fiducial fractions are constructed using projections of
this distribution. The fiducial observables are much easier for
theorists to calculate and would be also trivial for experimentalists
when particle-level (unfolded) angular distributions are available. 

Results at full NLO QCD accuracy and also at NLO QCD + EW level, where
EW corrections are calculated in the DPA,
for the $pp \to WZ \to 3\ell \nu + X$ process have been provided. It
is expected that the DPA is an excellent approximation for the
polarization observables. We have found that EW corrections are
important for $f^Z_{L,R}$ observables due to radiative decay.

Comparisons with ATLAS data have been provided for the case of
template-fitted fractions. Different theoretical predictions at NNLO
QCD or NLO QCD + EW accuracy are in good agreement. The largest
deviation compared to the data is about $2.5\sigma$ for the
$f^{W^+}_L-f^{W^+}_R$ observable, albeit with a huge error on the
data. We think that similar comparisons for the fiducial fractions can
help to identify possible issues. This is currently not possible as
measurements for these observables are not available.

For the future, it would be interesting to see whether the fiducial
polarization observables are more sensitive to new physics
(e.g. anomalous triple gauge couplings) than the template-fitted ones.

\section{Multi-parton interactions, Colour Reconnection and Hadronization Effects in Central jet veto \footnote{speaker: Simon Pl\"atzer}}

\subsection*{Status of (N)LO+PS predictions}

After the dedicated comparison of fixed-order partonic as well as
parton shower matched predictions in \cite{Ballestrero:2018anz},
several effects have been understood and NLO+PS predictions at parton
level can be seen as well under control, as long as the hardest two jets are concerned.
The third jet is improved by NLO matching, however differences remain in there, which need to be investigated in a comprehensive study addressing QCD activity relevant to the central jet veto.
Variations of the scales in the hard process and the parton showering suggest that these
predictions are also reliable, and agreeing across different
generators at the level of 10\% for observables involving the hard
tagging jets and the vector bosons. Similar findings are currently
reported in ongoing studies of VBF Higgs production
\cite{HXSWGTalk}. In view of a lack of a systematic expansion within
parton shower algorithms, benchmark comparisons between different
matching and parton shower algorithms are vital to establish that the
variations we typically consider indeed have something to do with an
estimate of uncertainty in the predictions.

\subsection*{Soft QCD effects and event generator uncertainties}

A full simulation of QCD effects in VBS and VBF processes does of
course need to comprise the parton-shower evolution, which performs
the resummation of Sudakov logarithms for a large class of
observables, possibly even at the next-to-leading logarithmic level
concerning global properties of the final state such as jet transverse-momentum spectra.
If dipole-type algorithms are employed, a
resummation of the leading non-global logarithms in jet vetoes can be
expected, at least in the large-$N$ limit, though these algorithms
have known issues themselves.

\begin{figure}
  \begin{center}
    \includegraphics[scale=0.6]{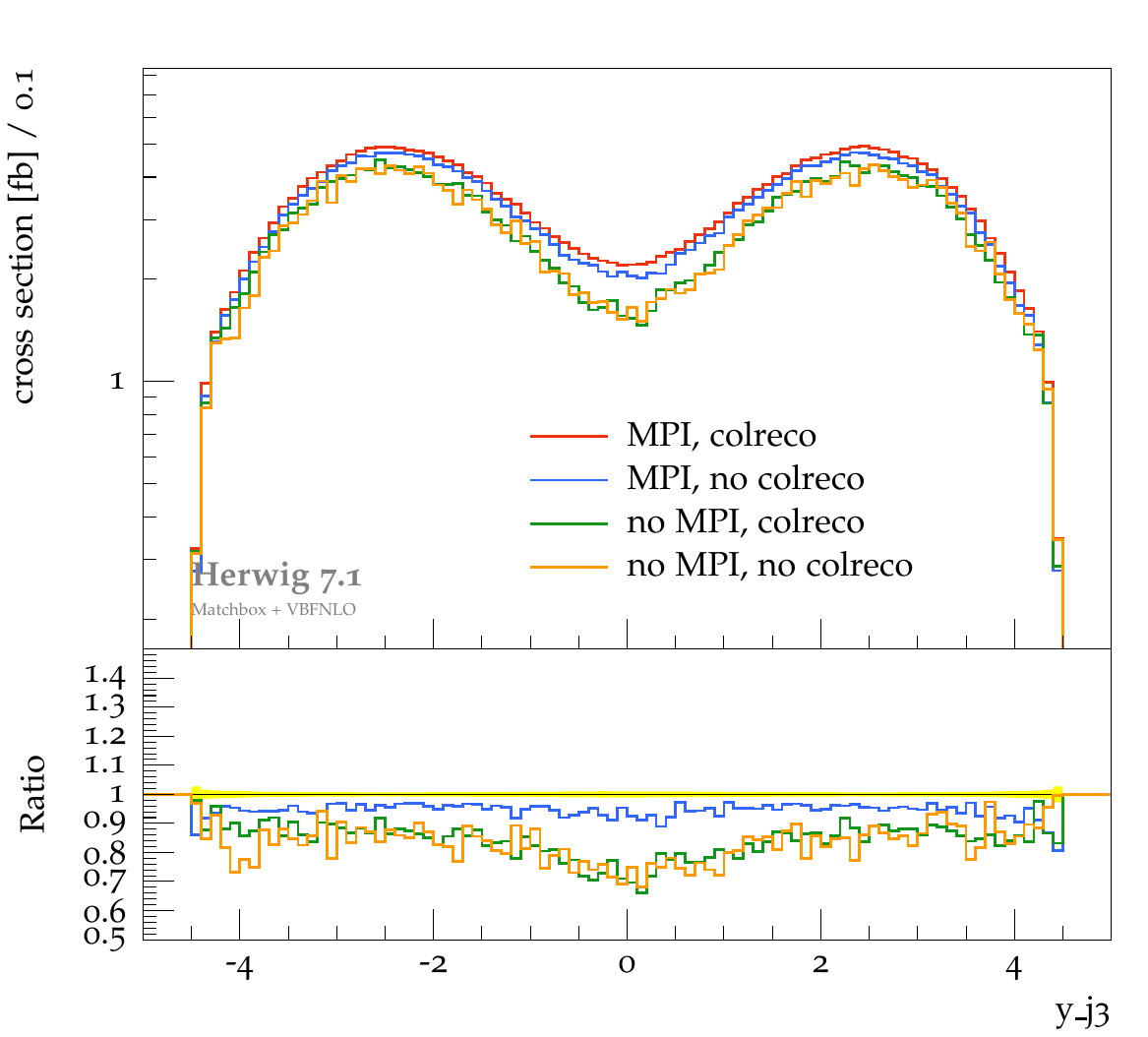}
    \includegraphics[scale=0.6]{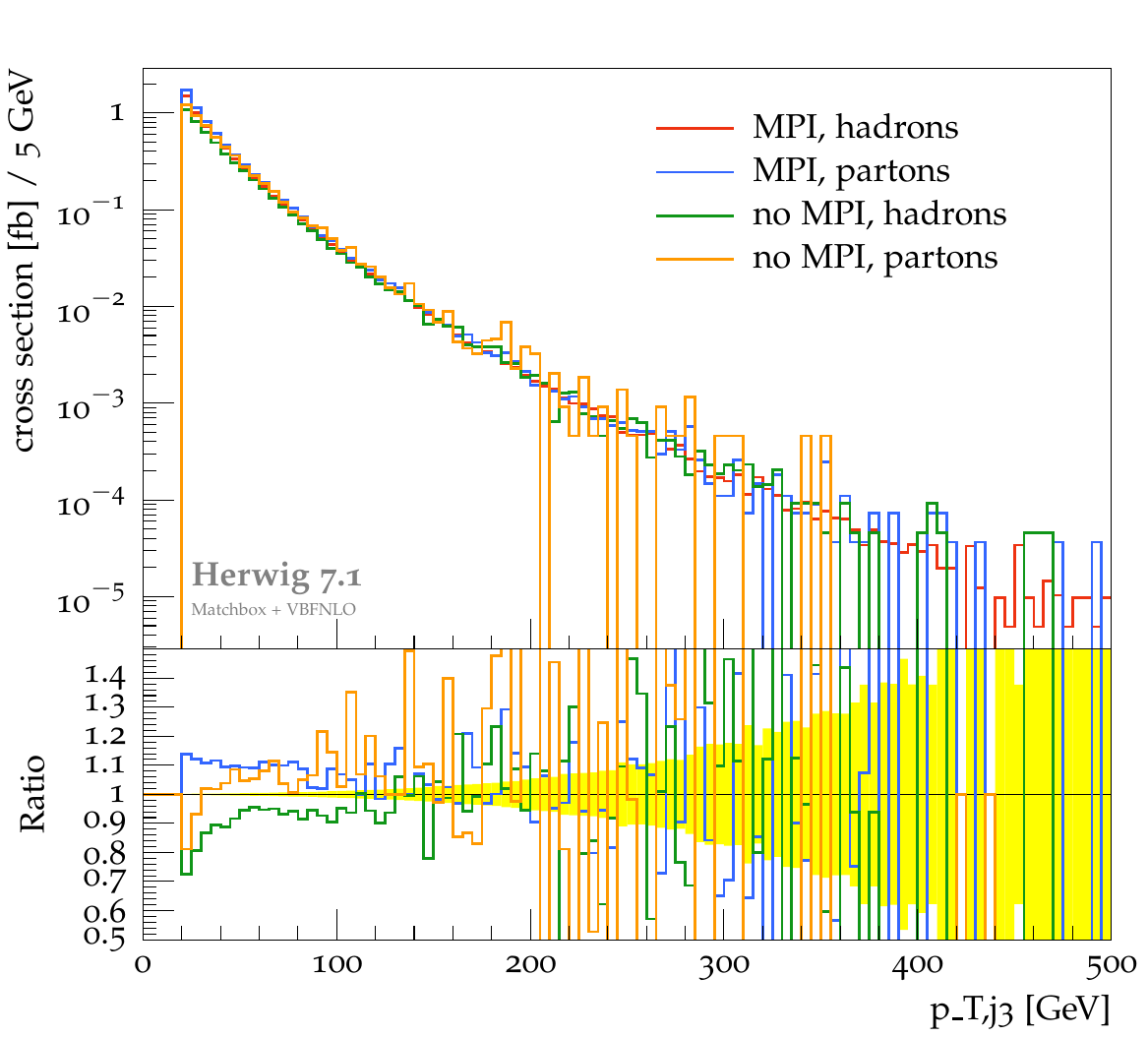}
  \end{center}
  \caption{\label{fig:thirdjet}Rapidity and $p_\perp$ spectrum of the
    third jet in VBF Z production. We compare the impact of
    hadronization, multi-parton interactions and colour reconnection.}
\end{figure}

For a realistic picture, multi-parton interactions (MPI) in between
the colliding protons need to be taken into account, as well as the
effect of colour reconnection and hadronization. While we expect a
significant impact of MPI in contaminating the typical VBF/VBS
signature by additional central activity and a change in jet
properties, an impact of colour reconnection is not immediately
obvious. Its role might become more important in both the presence of
MPI as well as different colour flows in the hard process contributing
to VBS-topology diagrams, or ``s-channel'' contributions. Evaluating
the impact of these contributions by an on-off exercise will set the
typical order of magnitude of variations we can expect, and indicate
effects at the level of 10 to 30 \% in typical observables like the
third jet rapidity and transverse momentum spectra, see
Fig.~\ref{fig:thirdjet} in the case of the Herwig 7 soft QCD models.
In this case we use VBFNLO interfaced to Herwig via the Matchbox module and matched to the angular ordered shower.
The rapidity of the thirs jet (left) and the transverse momentum of the third jet (right) can see significant contributions from soft QCD effects such as multi-parton interaction (MPI) and colour reconnection (colreco).

For a leading-order plus parton-shower (LO+PS) simulation the impact
of the models, as well as variations thereof, will entirely be covered
by parton-shower scale variations, notably the hard shower veto or
starting scale, which probes the shower impact in the phase space
region of hard emissions. It is therefore vital to switch to
next-to-leading order matched (NLO+PS) simulation which will provide
comparable shower uncertainties along for third jet spectra and the
interplay of the shower and model variations can be quantified in
greater detail. First steps in this direction have been undertaken and
have been presented at the MBI Workshop in Thessaloniki \cite{MBI19}.

\subsection*{Interplay with VBF approximation and colour mixing}

Since the VBF approximation is severely limiting the colour flows
which can contribute to the amplitude, the dynamics of showering as
well as colour reconnection subsequently taking place is happening in
a very constrained setting. While this is certainly a valid assumption
for a tight VBF selection and observables mainly concerned with the
hard tagging jets and electroweak bosons, the mixing with other colour
structures might be significant for more inclusive selections even if
a fixed-order comparison would signal that the VBF approximation is
still acceptable. Details of these physics, and their connection to
soft gluon dynamics have been discussed in great detail at the Vienna
workshop \cite{Vienna19} and will be followed up in future work, at least
for processes for which a full calculation is available.

\section{Vector boson  polarizations in the fully leptonic WZjj channel at the LHC\footnote{speaker: Ezio Maina; authors: Alessandro Ballestrero, Giovanni Pelliccioli}}

\subsection{Introduction}
\label{intro}
In Run 2 CMS and ATLAS have finally produced clear evidence that VBS actually takes place at the LHC 
\cite{Sirunyan:2017fvv,Sirunyan:2017ret,Aaboud:2018ddq,Aaboud:2019nmv,Sirunyan:2019ksz}.
Unfortunately, the statistics is still too small to analyze vector-boson polarizations.
The higher rates which will be available after the Long Shutdown in 2019 and 2020 and later in the High-Luminosity phase of the LHC will hopefully allow more detailed studies~\cite{Azzi:2019yne}.
A polarization analysis of VBS nicely complements the large invariant mass VBS study
\cite{Ballestrero:2009vw, Ballestrero:2010vp, Ballestrero:2011pe} and the study of Higgs boson properties
in the effort to fully characterize the EWSB mechanism.

In two recent papers \cite{Ballestrero:2017bxn,Ballestrero:2019qoy}, we have shown that it is possible to 
define, in a simple and natural way, cross sections corresponding to vector bosons with definite polarization.
We have demonstrated that the sum of polarized cross sections, 
even in the presence of cuts on the final state leptons, describes reasonably well the full total 
cross section and most of the differential distributions. As a consequence, it is possible to fit 
the data using single polarized templates and the interference, to extract polarization fractions.

In the Feynman amplitudes which describe VBS, all information about the
vector boson polarization is confined to the polarization sum in the corresponding propagators. 

\begin{equation}
-g^{\mu\nu} + \frac{k^{\mu}k^{\nu}}{M^2} = \sum_{\lambda = 1}^4 \varepsilon^{\mu}_\lambda(k)
\varepsilon^{\nu^*}_{\lambda}(k)\,\,.
\label{eq:polexpansion}
\end{equation}

When squaring the amplitude,
the individual polarizations interfere among themselves. These interference contributions cancel 
exactly only when an integration over the full azimuth of the decay products is performed. 
Because of acceptance cuts, however, cancellations cannot be complete. 
In the following, we call single polarized amplitude with polarization
$\lambda$ an amplitude in which the sum on the left hand side of \eq{eq:polexpansion}
is substituted by one of the terms on the right hand side.

\begin{figure}[bth]
\includegraphics[scale=0.45]{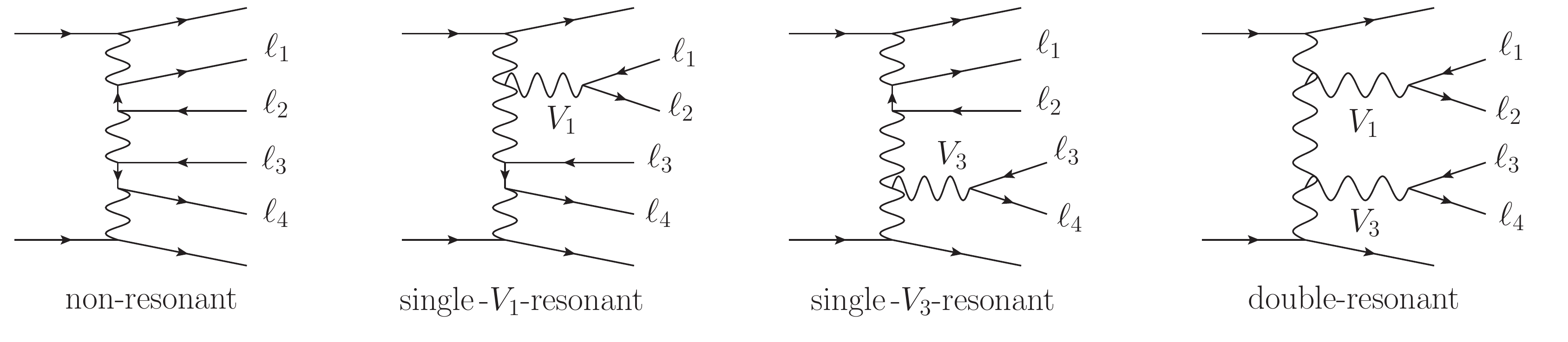}
\caption{Sample tree level diagrams for VBS at the LHC.
Scattering diagrams (like the rightmost one) are only a subset of double resonant diagrams.}
\label{fig:FeynRes}
\end{figure}

In addition, electroweak boson production processes are, in general, described by amplitudes including non 
resonant diagrams, see \reffi{fig:FeynRes},
which cannot be interpreted as production times decay of any vector boson.
These diagrams are essential for gauge invariance and cannot be ignored. For them, separating polarizations
is simply meaningless.
In order to obtain an expression which preserves gauge invariance and is interpretable in terms of vector boson polarization we:
\begin{itemize}
\item Drop all non resonant diagrams
\item Project on the vector boson mass shell the momentum flowing through the resonant propagators in the numerator of the diagrams,
leaving the denominator untouched.
\end{itemize}
We refer to \cite{Ballestrero:2017bxn,Ballestrero:2019qoy} for the details. 

Following  this procedure, the normalized cross section, after integration over the azimuthal angle of the decay products,
can be expressed, in the absence of cuts on decay leptons, as follows:
\begin{eqnarray}\nonumber
 \frac{1}{\frac{d\sigma(X)}{d X}} \,\,\frac{d\sigma(\theta,X)}{d\cos\theta\, d X}
 &=&\frac{3}{8}  f_L(X) \bigg(1+\cos^2{\theta}-
\frac{2(c_L^2-c_R^2)}{(c_L^2+c_R^2)}\cos\theta\bigg)\\
&+&\frac{3}{8} f_R(X) \bigg(1+\cos^2{\theta}
+\frac{2(c_L^2-c_R^2)}{(c_L^2-c_R^2)}\cos\theta\bigg)
+\frac{3}{4} f_0(X) \sin^2\theta,
\label{eq:diffeqV}
\end{eqnarray}
where $X$ stands for all additional phase space variables in addition to the decay angle $\theta$ and
$c_L,\, c_R$ are the weak couplings. 
Hence, each physical polarization is uniquely associated with a specific angular distribution of the charged
lepton, even when the vector boson is off mass shell.

\subsection{$W\!Z$ scattering}
\label{sec:WZ}

The $W\!Z$ channel is strongly sensitive to the EWSB mechanism.
The presence of a new resonance coupling to $W$ and $Z$ bosons or a modified Higgs sector
would interfere with the delicate cancellation of large contributions,
enhancing the longitudinal cross section at high energies.

In this section we investigate
the phenomenology of polarized $W^{+}\!Z$ scattering in the fully leptonic decay channel at the LHC@13TeV.
All simulations have been performed at parton level with 
\texttt{PHANTOM~1.6} \cite{Ballestrero:2007xq,Ballestrero:1994jn}, employing
\texttt{NNPDF30\_lo\_as\_0130} PDFs \cite{Ball:2014uwa}, with factorization scale $\mu=M_{4\ell}/\sqrt{2}$.

We have applied the following kinematic cuts:
$|\eta_j|<5$;
$p_t^j>20$ GeV;
 $M_{jj}>500$ GeV;
$|\Delta\eta_{jj}|> 2.5$;
$|M_{e^+e^-}-M_Z|<15$ GeV;
 $M_{W\!Z}>200$ GeV.
The results presented in Sect.~\ref{subsec:lepcutwz} include three additional cuts:
$|\eta_\ell|<2.5$; $p_t^\ell>20$ GeV; $p_t^{\rm miss}>40\,\GeV$.
In Sect.~\ref{subsec:wznocut} the $M_{W\!Z}$ cut is imposed directly on the generated, not reconstructed,
momenta. In  Sect.~\ref{subsec:lepcutwz} the cut is applied after neutrino reconstruction.

\subsection{Single polarized results and their validation in the absence of lepton cuts}
\label{subsec:wznocut}
In order to verify that polarizations can be separated at the amplitude level  while reproducing
properly the full result, we consider the ideal kinematic setup in which no cut on charged
leptons and neutrinos is applied, apart from $|M_{e^+e^-}-M_Z|<15$ GeV and $M_{W\!Z}>200\,\GeV$.

We select only
the $W$($Z$) resonant diagrams (single and double resonant) out of the
full set of contributions. Then we apply the single On-Shell Projection on the
$W$($Z$) boson (OSP1-W(Z)), to avoid any cut on the $\mu^+\nu_\mu $ system invariant
mass. We have shown that for $Z$ resonant diagrams, OSP1 has no visible
effect, but, nonetheless, we apply it for consistency. In all the following we will refer
to OSP1-W(Z) projected $W$($Z$) resonant calculation simply as resonant
calculation.

\begin{figure}[!htb]
\centering
\subfigure[\label{fig:legcw200}]{\includegraphics[scale=0.37]{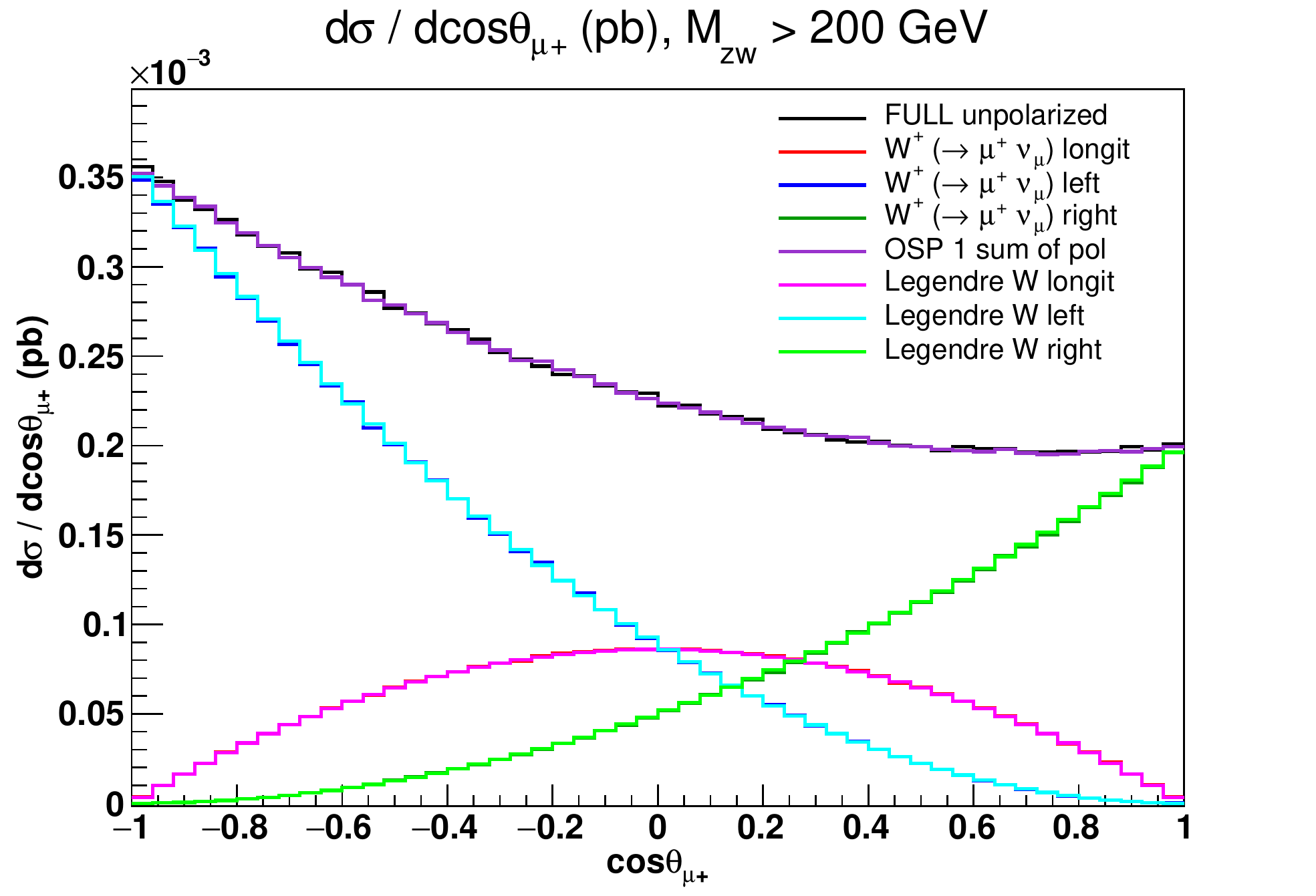}}
\subfigure[\label{fig:wpolfr}]{\includegraphics[scale=0.35]{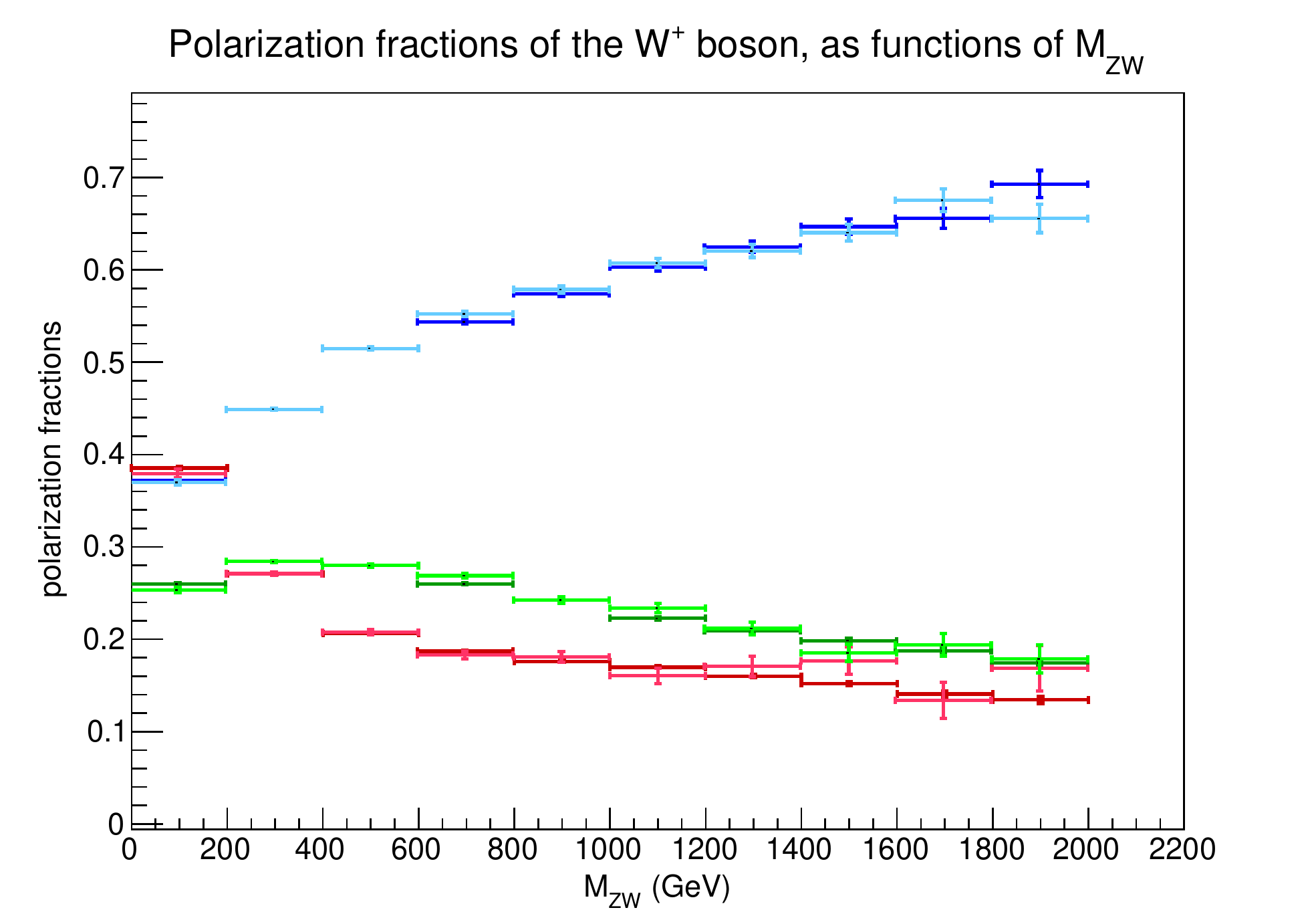}}
\caption{$W^+\!Z$ scattering: $\cos\theta_{\mu^+}$ distributions for a polarized $W^+$, in the region
$M_{W\!Z}>200\,\GeV$ (left), and
polarization fractions as functions of $M_{W\!Z}$ (right). Comparison between Monte Carlo distributions and
results extracted from the full $\cos\theta_{\mu^+}$ distribution by projecting into the first three
Legendre polynomials.
No lepton cuts, no neutrino reconstruction.}\label{fig:legw}
\end{figure}

The total cross section computed with full matrix elements is $486.4(2)\,\ab$.
The unpolarized OSP1-W resonant result is only 0.2\% smaller.
Similarly, the OSP1-Z resonant computation underestimates by
0.7\% the full result.
Differential distributions are also in
good agreement. Discrepancies are smaller than 2\% bin by bin.

We then separate the polarizations of the $W^+$ boson. In \reffi{fig:legcw200} we consider the
$\cos\theta_{\mu^+}$ distributions in the full $M_{W\!Z}>200\,\GeV$ range
in the absence of lepton cuts and without neutrino reconstruction.

The distributions obtained with polarized amplitudes (red, blue and dark green, for longitudinal,
left and right polarization, respectively) are compared with the components extracted from
the full distribution (magenta, azure and light green) by projecting onto the first three Legendre polynomials.
The agreement is very good: both the normalization and the quadratic dependence on $\cos\theta_{\mu^+}$ 
is perfectly reproduced for each polarization state.
Very similar conclusions can be drawn when separating the polarization of the $Z$ boson.

\begin{figure}[!h]
\centering
\subfigure[$M_{W\!Z}$\label{fig:lcmwz}]
{\includegraphics[scale=0.35]{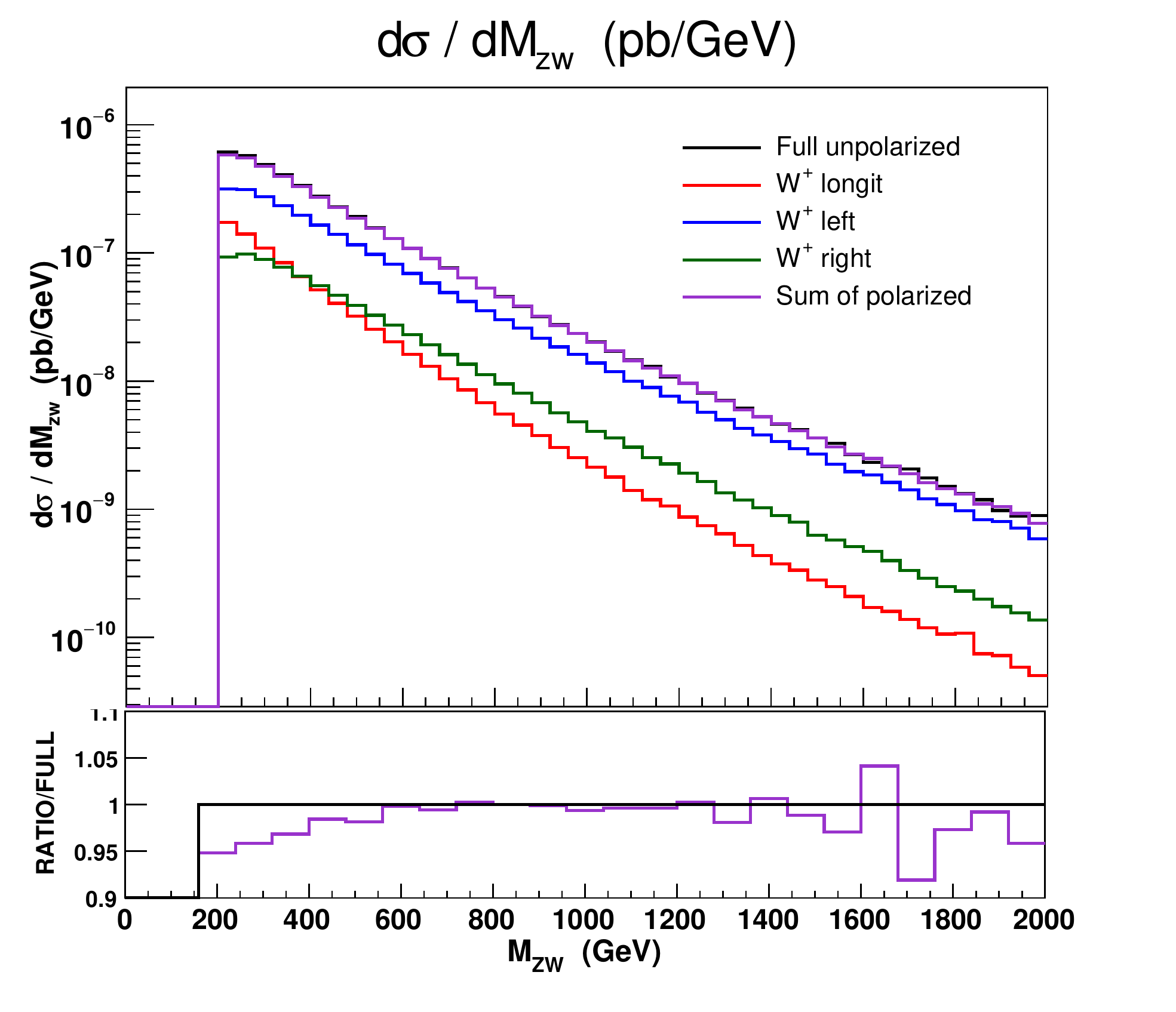}}
\subfigure[$\cos\theta_{\mu^+}$\label{fig:lccth}]
{\includegraphics[scale=0.35]{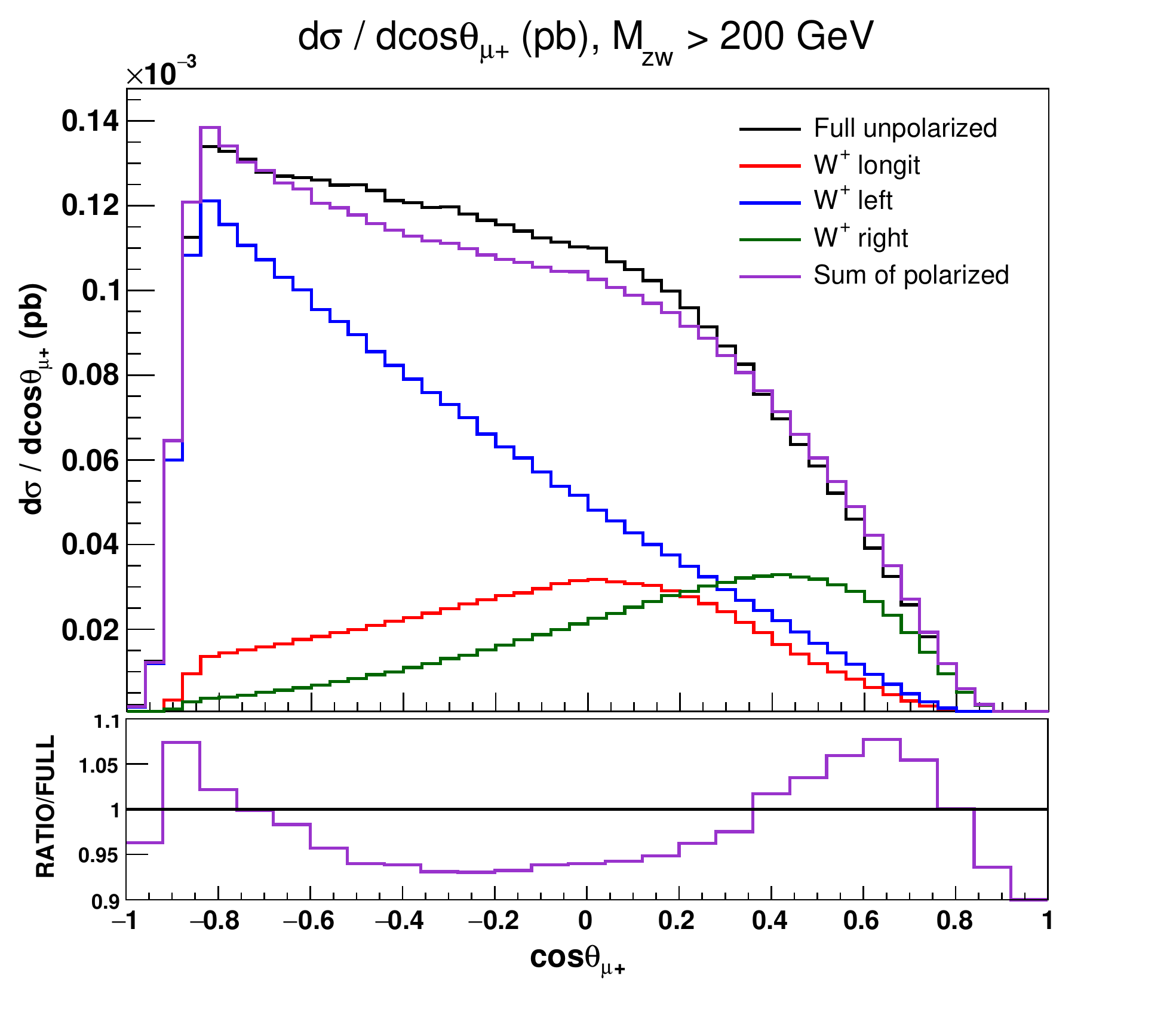}}\\
\subfigure[$\eta_{W}$\label{fig:lcetaw}]
{\includegraphics[scale=0.35]{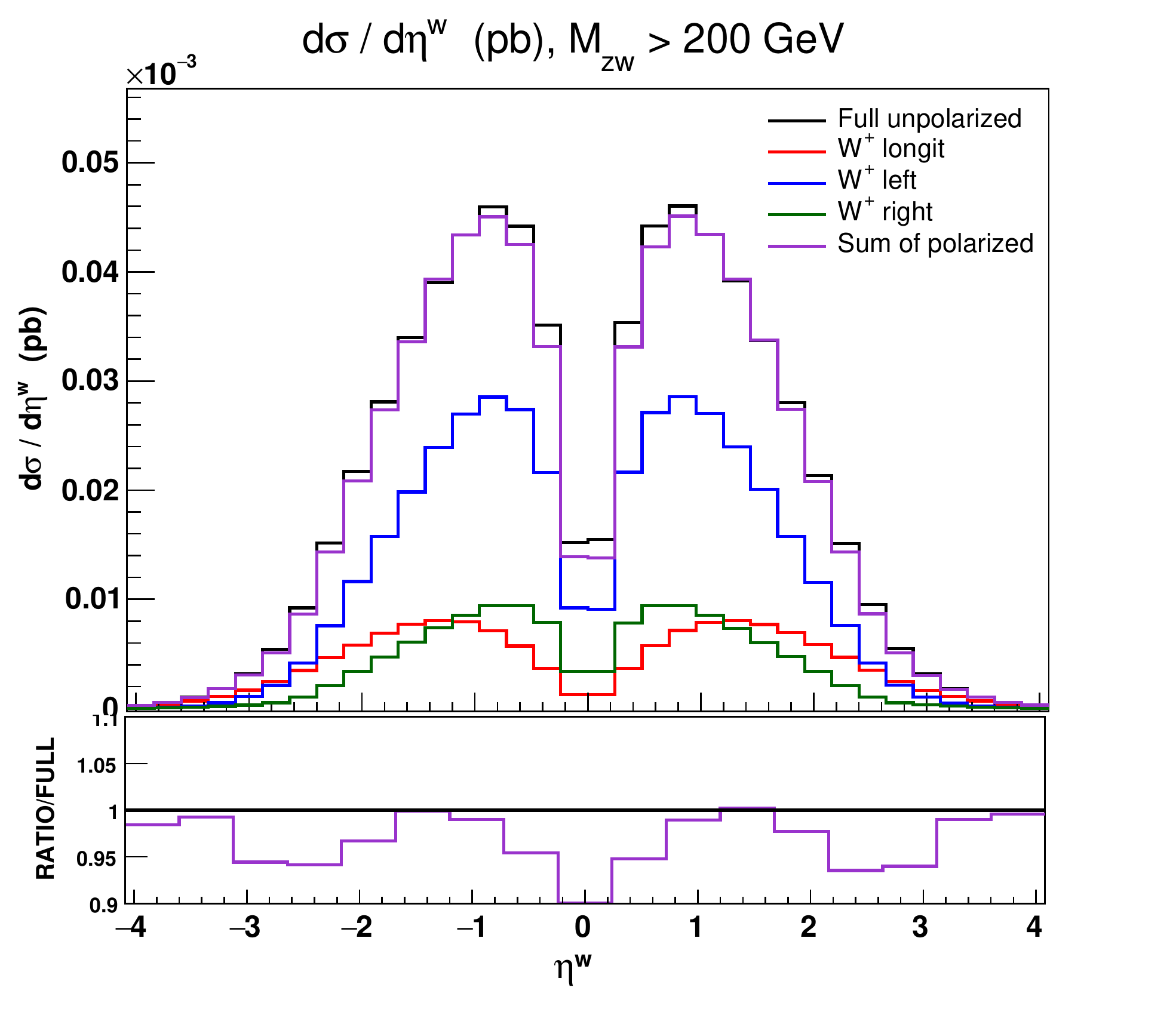}}
\subfigure[$p_t^{\mu^+}$\label{fig:lcptmu}]
{\includegraphics[scale=0.35]{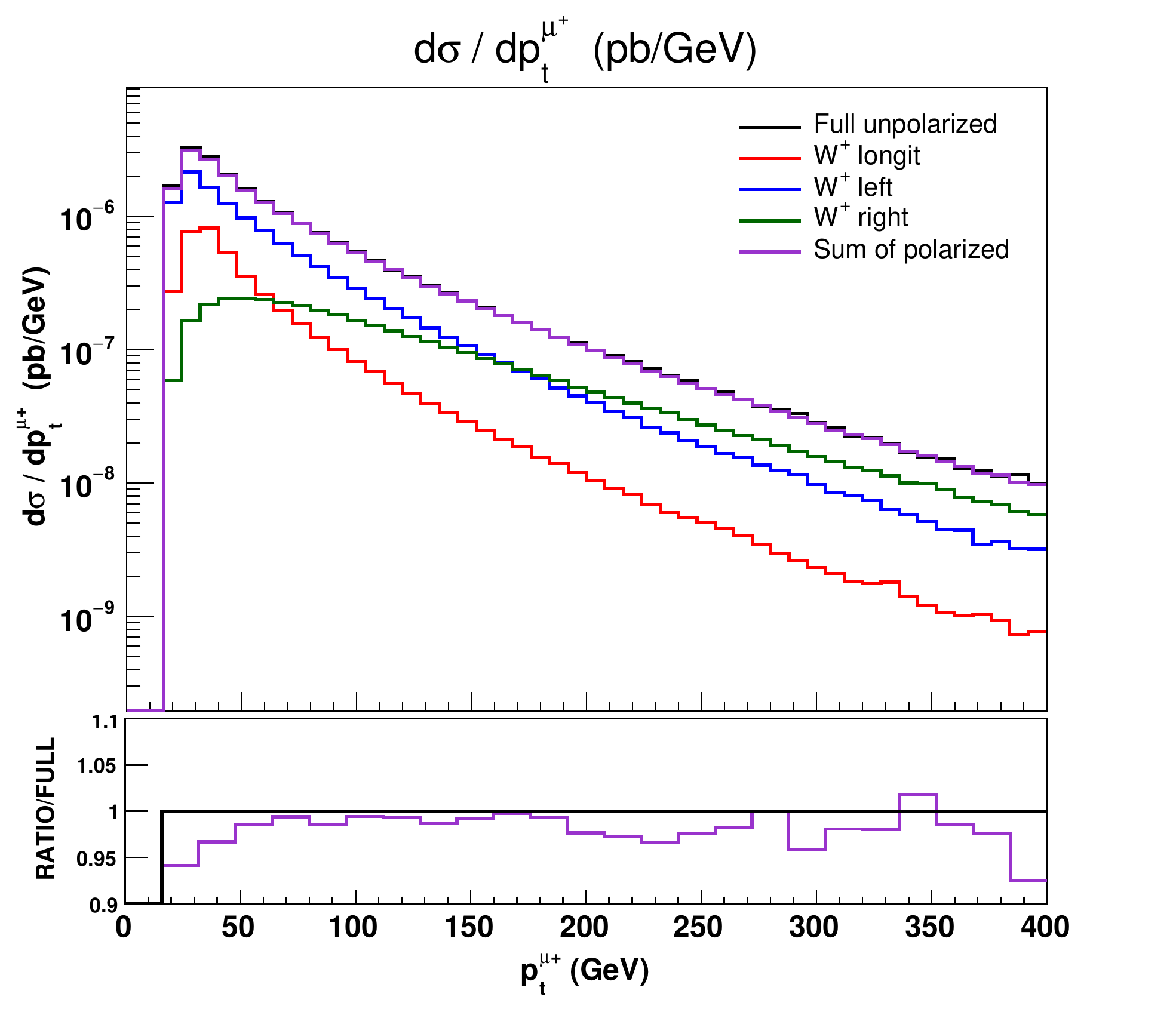}}
\caption{$W^+\!Z$ scattering: differential cross sections for a polarized $W^+$ boson, in the presence of
lepton cuts and neutrino reconstruction.
We show the full result (black), the single polarized distributions (red: longitudinal, blue: left handed, green:
right handed) and the incoherent sum of the polarized results (violet).
The pull plot shows
the ratio of the sum of polarized distributions to the full one.}\label{fig:wzlepcut_w}
\end{figure}

\subsection{Effects of lepton cuts and neutrino reconstruction on polarized distributions}
\label{subsec:lepcutwz}
In this section we present polarized differential distributions in the presence of lepton
cuts and neutrino reconstruction for a number of relevant kinematic variables.
The specific neutrino reconstruction scheme that is applied in the following
(\texttt{CoM} + \texttt{transvMlv}) is described in ref.~\cite{Ballestrero:2019qoy}.

We start from the total cross section. In order to evaluate separately the effect of dropping
the non resonant diagrams and the effect of neglecting interferences among different polarization modes,
we have computed the cross section with the full matrix element and with OSP1-W(Z) projected resonant
diagrams.
The difference between these two results provides an estimate of non resonant effects.
The difference between the resonant unpolarized cross section and the
sum of the single polarized ones (either for a polarized $W^+$ or for a polarized $Z$) provides an
estimate of the interference among polarizations, which is non zero because of the leptonic cuts. Numerical
results are shown in Tab.~\ref{tab:WZSM}.

\begin{table}[hbt]
\begin{center}
\begin{tabular}{|c|c|c|}
\hline
\multicolumn{3}{|c|}{Total cross sections [$\ab $]}\\
\hline
& {polarized $W^+$} & { polarized $Z\,\,$}\\
\hline
longitudinal (res. OSP1) &   33.21(3) & 42.56(3)\\
\hline
left handed  (res. OSP1) &  96.31(8)  & 76.87(6)\\
\hline
right handed (res. OSP1) &  30.93(2)  & 40.54(3)\\
\hline
sum of polarized &  160.45(9) & 159.97(8)\\
\hline
unpolarized (res. OSP1)  &  164.2(2 ) & 164.0(2) \\
\hline
non res. effects     &  0.9(2)   & 1.1(2)\\
\hline
pol. interferences     &  3.8(2)   & 4.0(2)\\
\hline
full         &  165.1(1)  &165.1(1)\\
\hline
\end{tabular}
\end{center}
\caption{Polarized and unpolarized total cross sections ($\ab $) for $W^+\!Z$ scattering in the fiducial
region.} \label{tab:WZSM}
\end{table}

The resonant unpolarized calculation has been performed selecting single $W$($Z$) resonant diagrams, and
then applying the corresponding
single On Shell projection. In both cases non resonant effects are smaller than 1\% , implying that the
resonant approximation works rather well.
Interference among polarization states amounts to
2.5\% .

Concerning polarized total cross sections, the $W^+$ is mainly left handed (58.3\%),
while the longitudinal and right handed contributions are of the same order of magnitude (20.1\% and
18.7\%, respectively). For the $Z$ boson, the left polarization is again the largest (46.6\%) while the
longitudinal and right components account respectively for 25.8\% and 24.6\%.

In \reffi{fig:wzlepcut_w} we present differential distributions for a polarized $W^+$ boson for a variety of
kinematic variables, which provide a more detailed description of the polarized signals.
For each variable, we show single polarized distributions,
their incoherent sum, and the distribution of the full result.
The colour code is as follows: the full result is in black;
the longitudinal, left and right single polarized distributions are in red blue and green respectively; the
incoherent sum of the polarized results is in violet.
Pull plots show the bin by bin ratio of the  incoherent sum of polarized distributions
to the full one.

\reffi{fig:lcmwz} presents the distribution of the invariant mass of the four leptons.
The interference and non-resonant effects account for less than 5\% of the full result (bin by bin) in the
whole $W^+\!Z$
invariant mass spectrum. The longitudinal fraction decreases rapidly with increasing energy. The left handed
component is the largest one over the whole range.

The angular distributions in $\cos\theta_{\mu^+}$ are strongly affected by the neutrino reconstruction
and the lepton cuts, as can be seen comparing \reffi{fig:legcw200} with \reffi{fig:lccth}.
The difference is mainly due to the $p_t$ cuts on the muon and the corresponding neutrino,
which deplete the peaks
at $\theta_{\mu^+}=0,\,\pi$ of the transverse modes and make the longitudinal shape asymmetric.
The sum of polarized
distributions underestimates the full result by about 5\%, except for the regions of rapid change.

In \reffi{fig:lcetaw} we show distributions of the reconstructed $W^+$ pseudorapidity. Neutrino
reconstruction leads to a marked depletion of the central region.
Interferences and non-resonant effects account for less than
6\% of the full
result over all the pseudorapidity range, apart from the central bin where they reach 10\%.

The muon transverse momentum distribution (\reffi{fig:lcptmu}) is minimally affected by
neutrino reconstruction. The longitudinal component is of
the same order of magnitude as the left handed one for $p_t$ values slightly above the cut, while
for large values it decreases faster than the transverse distributions.
For $p_t^{\mu^+}>160\,\GeV$ the right handed component
becomes larger than the left handed one. Interferences are small over the full range.

\subsection{Polarized amplitudes and reweighting approach}
\label{subsec:reweight}

Reweighting is an approximate procedure which has been
widely used by experimental collaborations to obtain polarized
samples, starting from unpolarized Monte Carlo events.
In this section we evaluate how well the reweighting method can
separate polarized samples and describe polarized distributions in the case
of $W^+\!Z$ scattering, by comparing its results with those presented in
Sect.~\ref{sec:WZ}, which have been obtained using polarized
amplitudes computed by the Monte Carlo.

Let us consider a generic process which involves a $W^+$ boson decaying into leptons
(similar considerations apply to the $Z$). The reweighting
procedure is based on the partition of the $W^+$-boson phase space in two dimensional
$\{p_t,\eta\}$ regions. In the absence of lepton cuts and
neutrino reconstruction, polarization fractions $f_0^{(i)},\,f_L^{(i)}$, and $f_R^{(i)}$
are computed in each $\{p_t^W,\eta_W\}$ region $i$, expanding
the full, unpolarized $\cos\theta_{\mu^+}$ distribution in Legendre
polynomials.
For each event in region $i$ with $\cos\theta_{\mu^+}=x$, three weights are computed,
\bea
w_{0,L,R}=\frac{
\frac{1}{\sigma}\frac{d\sigma}{dx}\Big|_{0,L,R}
}{ \frac{3}{4}(1-x^2)f_{0}^{(i)}+ \frac{3}{8}(1-x)^2f_{L}^{(i)} + + \frac{3}{8}(1+x)^2f_{R}^{(i)}}
\label{eq:weightfactor}
\eea
where,
\bea
\frac{1}{\sigma}\frac{d\sigma}{dx}
\Big|_{0}=\frac{3}{4}(1-x^2)f_{0}^{(i)},
\qquad \frac{1}{\sigma}\frac{d\sigma}{dx}
\Big|_{L/R}=\frac{3}{8}(1\mp x)^2f_{L/R}^{(i)}\,. \nnb
\eea
The event is assigned to the longitudinal, left or right polarized sample with probability
$w_0,\,w_L,\,w_R$. The three samples are then analyzed separately, applying
lepton cuts and performing neutrino reconstruction.

\begin{figure}[!htb]
\centering
\subfigure[Differential cross sections \label{fig:rew1}]
{\includegraphics[scale=0.37]{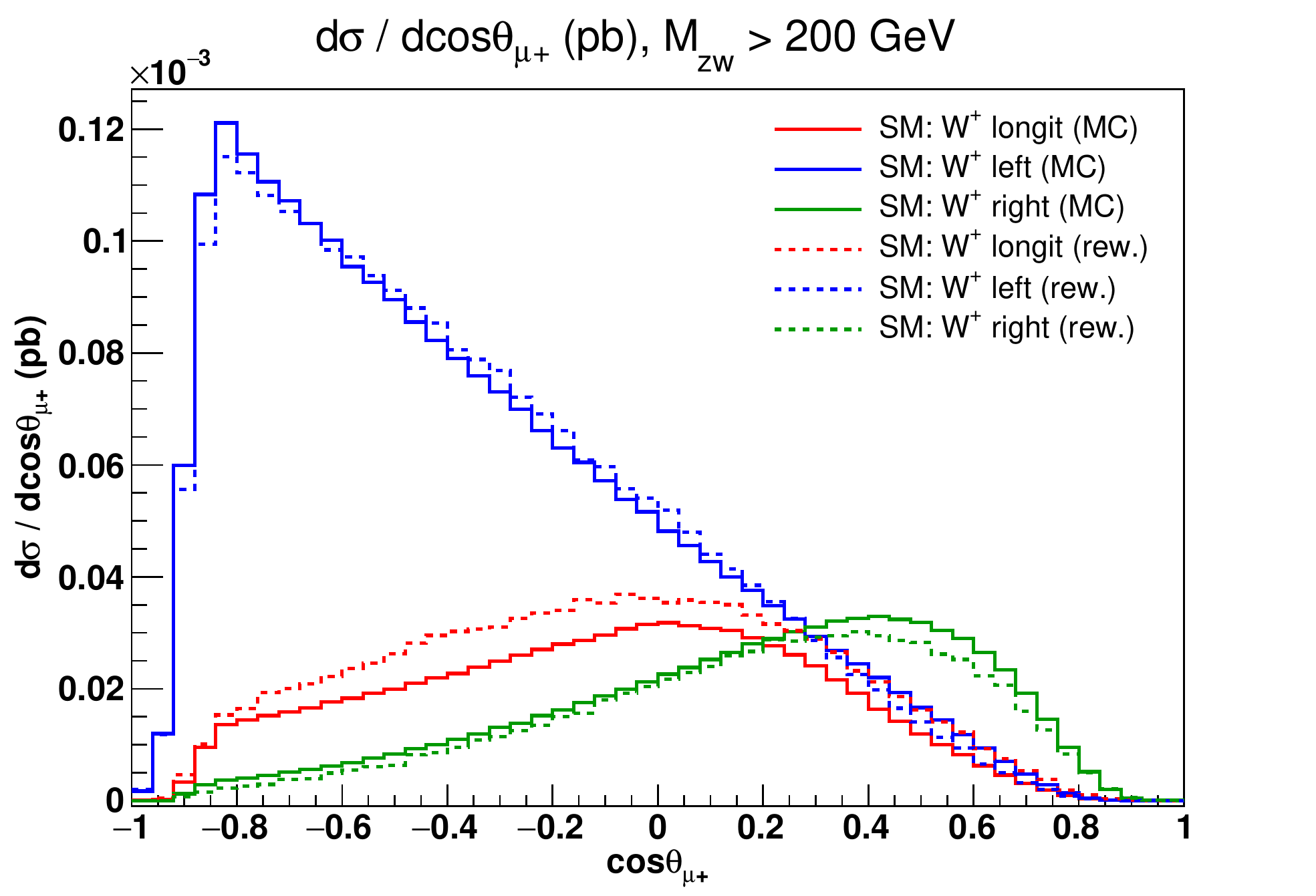}}
\subfigure[Normalized shapes \label{rew2}
]{\includegraphics[scale=0.37]{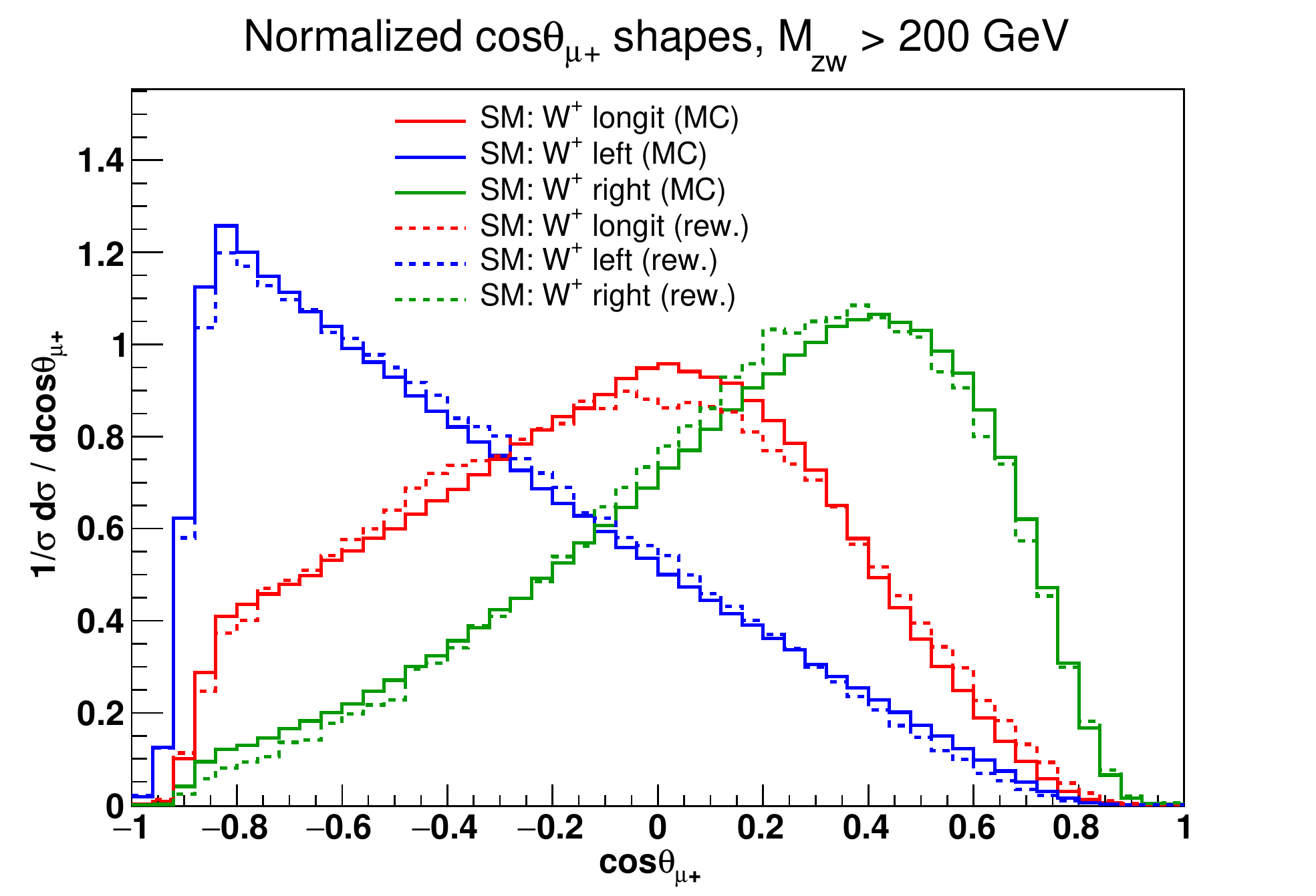}}
\caption{$W^+\!Z$ scattering: polarized $\cos\theta_{\mu^+}$ distributions in the region 
$M_{W\!Z}>200\,\GeV$. Results of the reweighting procedure compared with results of the MC calculation 
with polarized amplitudes. The full set of cuts and neutrino reconstruction are understood.} 
\label{fig:MCvsREW1}
\end{figure}

\begin{figure}[!htb]
\centering
\subfigure[Distributions \label{fig:rew3}]{\includegraphics[scale=0.37]{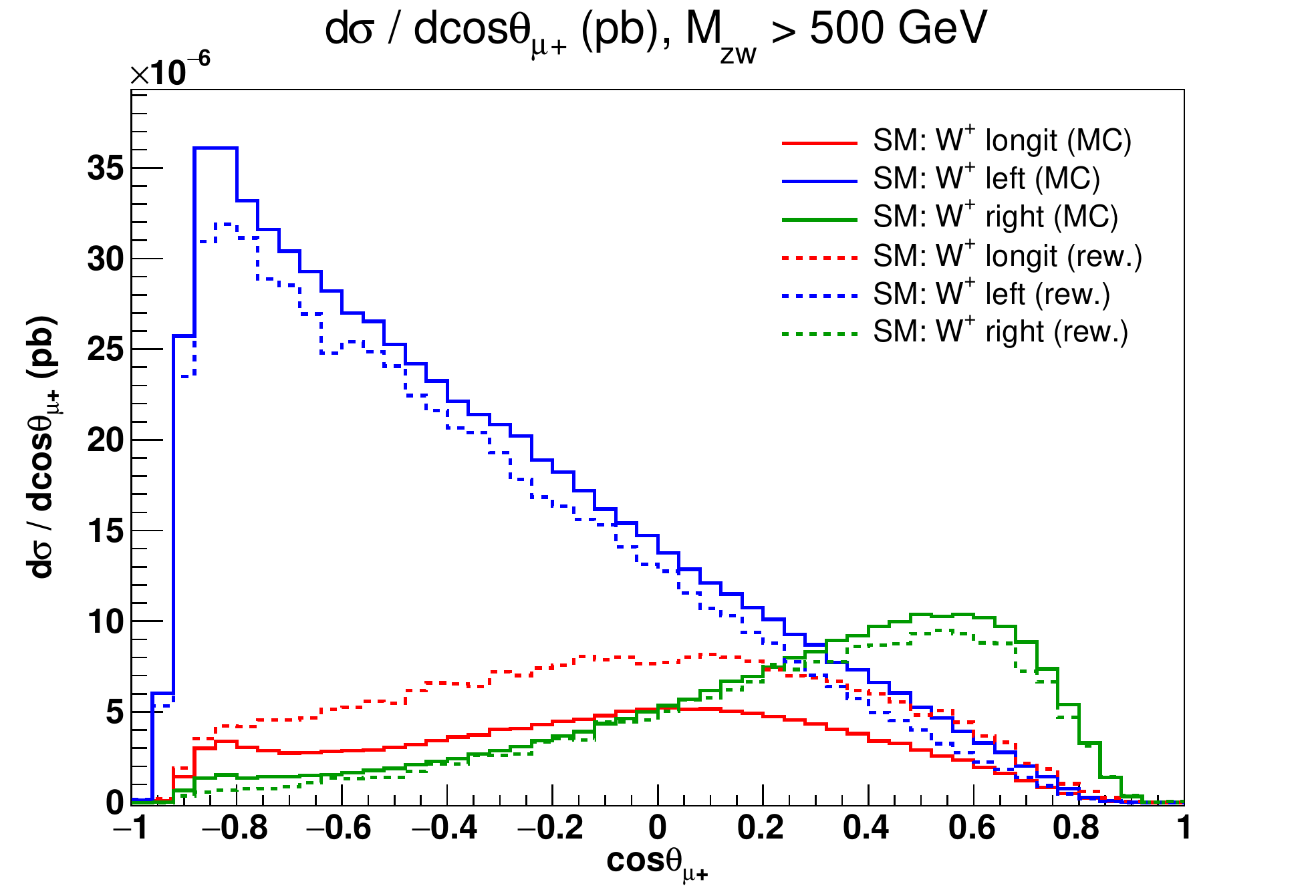}}
\subfigure[Normalized shapes \label{rew4}]
{\includegraphics[scale=0.37]{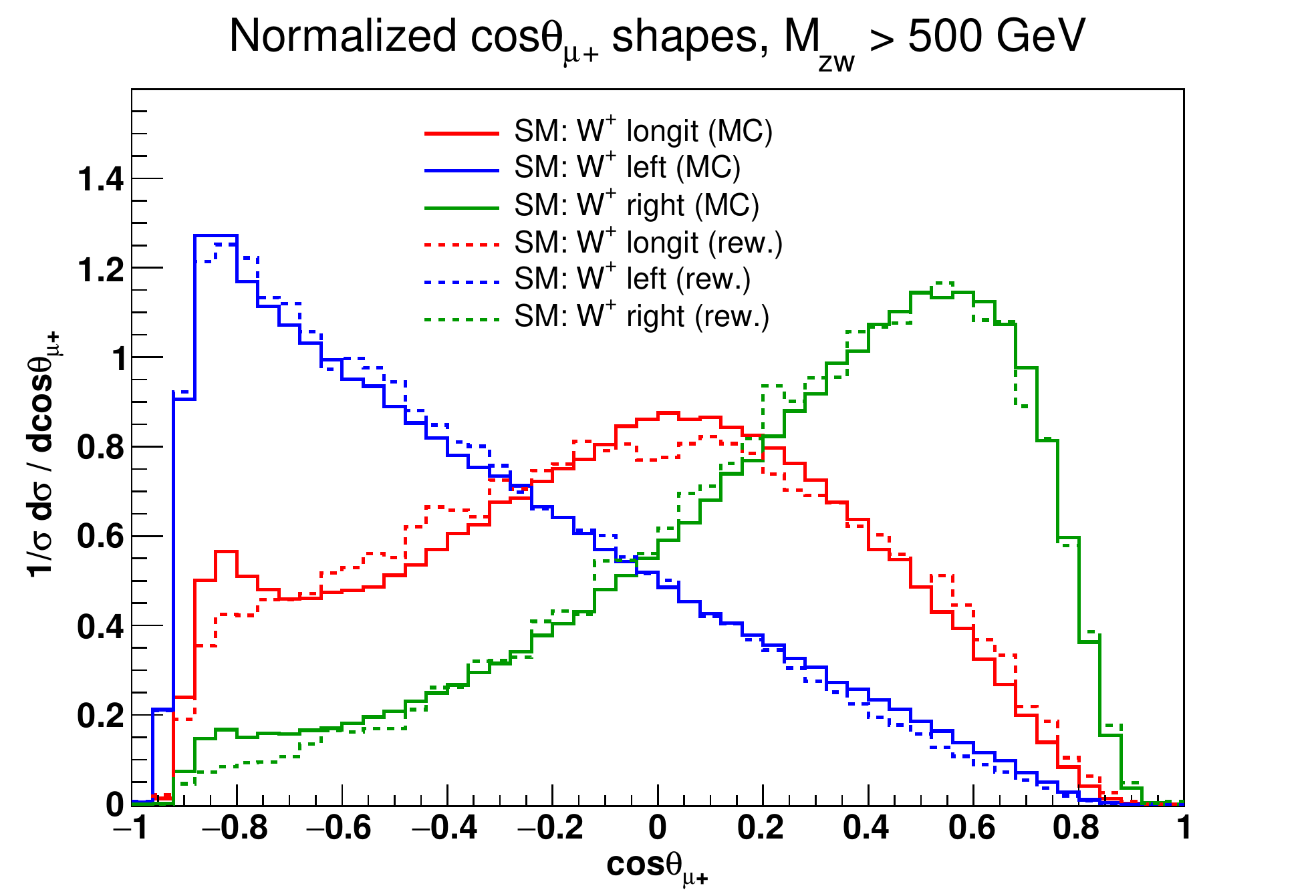}}
\caption{$W^+\!Z$ scattering: polarized $\cos\theta_{\mu^+}$ distributions in the region $M_{W\!Z}>500\,\GeV$. Results of the reweighting procedure compared with results of the MC calculation with polarized amplitudes. The full set of cuts and neutrino reconstruction are understood.} \label{fig:MCvsREW2}
\end{figure}

We have applied the reweighting method to $pp\rightarrow jj  e^+e^- \mu^+\nu_\mu$. In the
absence of lepton cuts and neutrino reconstruction (see Sect.~\ref{subsec:wznocut}), we have
computed polarization fractions for the full process with the following partitioning of the
$\{p_t^W,\eta_W\}$ phase space:
\begin{itemize}
\item[-] $p^{W}_t<30\GeV$, $30\GeV\,<p^W_t<60\GeV$, $60\GeV\,<p^W_t<90\GeV$, $p^W_t>90\GeV$;
\item[-] $|\eta_{W}|<1$, $1<|\eta_{W}|<2$, $2<|\eta_{W}|<3$, $|\eta_{W}|>3$.
\end{itemize}
In each region, we have separated the full unpolarized sample into three polarized samples, using
the algorithm described above. Then we have applied
the full set of leptonic cuts and performed neutrino reconstruction, obtaining
approximate polarized distributions which can be compared with those presented
in Sect.~\ref{subsec:lepcutwz}.
We have compared total cross sections (Tab.~\ref{tab:MCvsREW2})
and reconstructed $\cos\theta_{\mu^+}$ differential distributions (Figs.~\ref{fig:MCvsREW1}, \ref{fig:MCvsREW2}).
\begin{table}[h!]
\begin{center}
\begin{tabular}{|c|c|c|}
\hline
polarization & MC polarized & Reweighting \\
\hline
\multicolumn{3}{|c|}{$M_{W\!Z}> 200$ GeV}\\
\hline
longit. & 33.21(3)  &     41.02(3) \\
\hline
left &  96.31(8)   &     95.97(2)  \\
\hline
right & 30.93(2)  &     27.87(3)  \\
\hline
\multicolumn{3}{|c|}{$M_{W\!Z}> 500$ GeV}\\
\hline
longit. &  5.96(2)  &     9.94(4) \\
\hline
left    &  28.38(3)  &     25.49(3)  \\
\hline
right   &  9.06(3)  &     8.13(3)  \\
\hline
\end{tabular}
\caption{Polarized total cross sections ($\ab $) for $W^+\!Z$ scattering in the region
$M_{W\!Z}>200\,\GeV$ and $M_{W\!Z}>500\,\GeV$: results of the reweighting procedure compared with 
results of the MC calculation
with polarized amplitudes. The full set of cuts and neutrino reconstruction are understood.} 
\label{tab:MCvsREW2}
\end{center}
\end{table}

In the whole fiducial region ($M_{W\!Z}>200\,\GeV$), the left polarized $\cos\theta_{\mu^+}$ distribution
obtained with the reweighting procedure describes fairly well the analogous
distribution obtained with polarized amplitudes, both in total cross section
and in shape (${\sigma}^{-1}\,{d\sigma(X)}/{dX}$).
On the contrary, the longitudinal total cross section is
overestimated by 23\% and the right polarized cross section
is underestimated by 10\%, as shown in Tab.~\ref{tab:MCvsREW2}. Even larger
discrepancies show up when analyzing the $\cos\theta_{\mu^+}$ differential
cross section and shape (Fig.~\ref{fig:MCvsREW1}).

It is important to observe that the sum of the three cross sections
obtained with polarized amplitudes (central column of Tab.~\ref{tab:MCvsREW2}),
is not equal to the full unpolarized cross section, since the interferences
among polarizations account for 5\% of the full result.
Interferences are completely neglected in the reweighting method
(rightmost column of Tab.~\ref{tab:MCvsREW2}). As a consequence,
the sum of the three polarized cross sections is, by construction,
equal to the full unpolarized one.

The inaccuracy of the reweighting procedure becomes even more evident at high
energies, as Fig.~\ref{fig:MCvsREW2} and Tab.~\ref{tab:MCvsREW2} show. For
$M_{W\!Z}>500\,\GeV$, the  reweighting predictions are absolutely unreliable. In
particular, the longitudinal cross section is overestimated by 70\%, and
the corresponding $\cos\theta_{\mu^+}$ shape is other than the Monte Carlo polarized
prediction. At large diboson masses, the polarization interferences are smaller than at lower masses, however
neglecting them contributes to the low precision of the reweighting method.

\begin{figure}[!htb]
\centering
\subfigure[Monte Carlo \label{fig:rew1dejj}]{\includegraphics[scale=0.37]{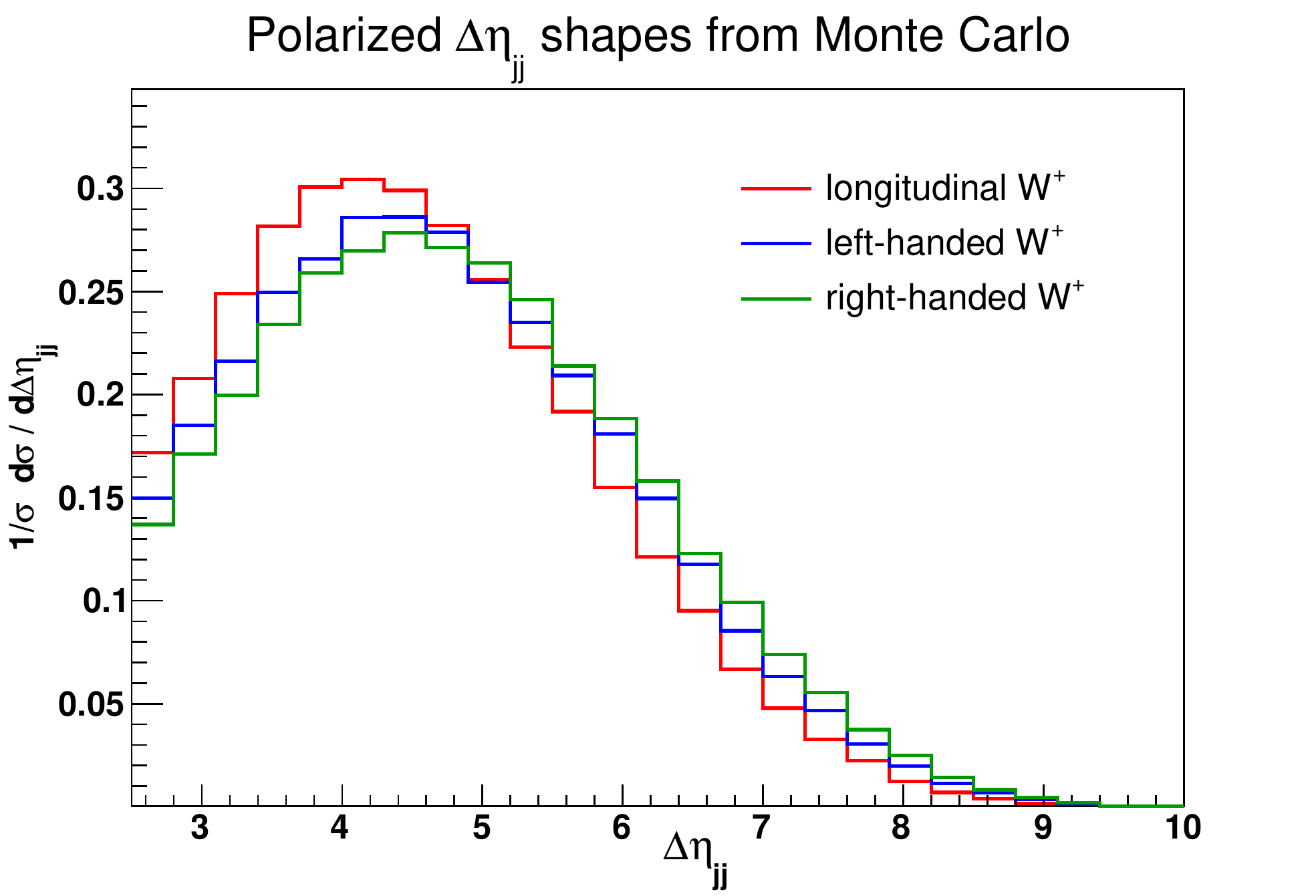}}
\subfigure[Reweighting \label{fig:rew2dejj}]{\includegraphics[scale=0.37]{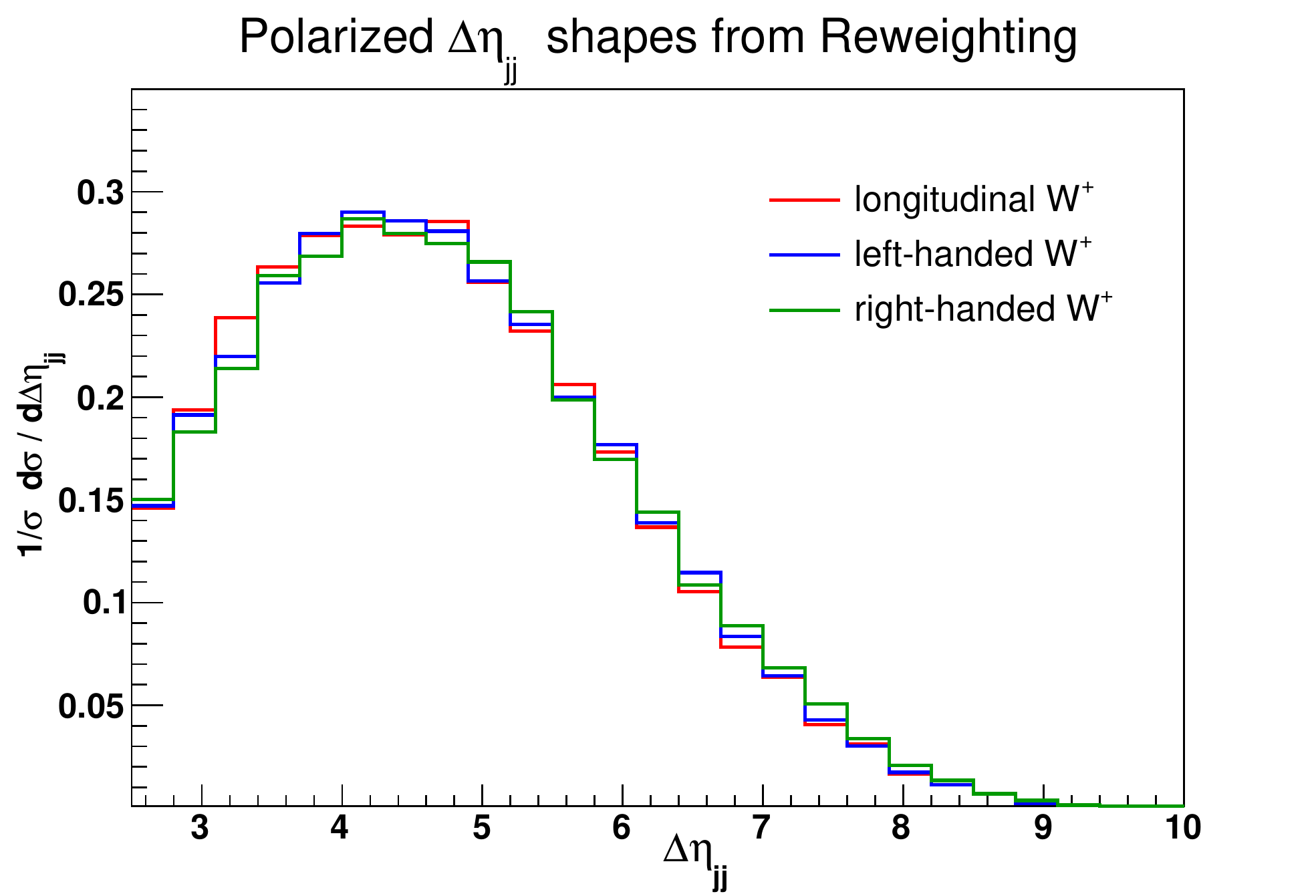}}
\\
\subfigure[Monte Carlo \label{fig:rew1dejj}]{\includegraphics[scale=0.37]{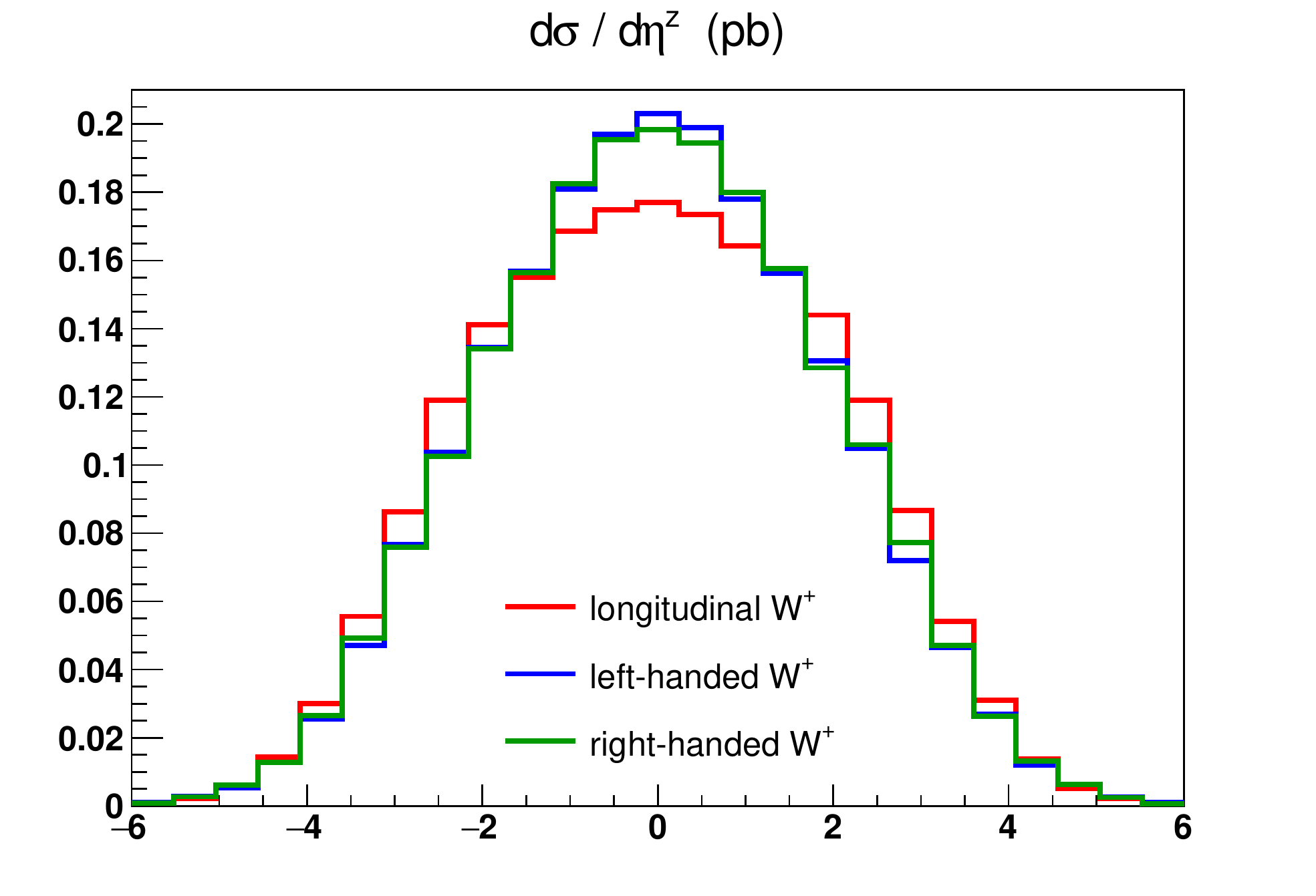}}
\subfigure[Reweighting \label{fig:rew2dejj}]{\includegraphics[scale=0.37]{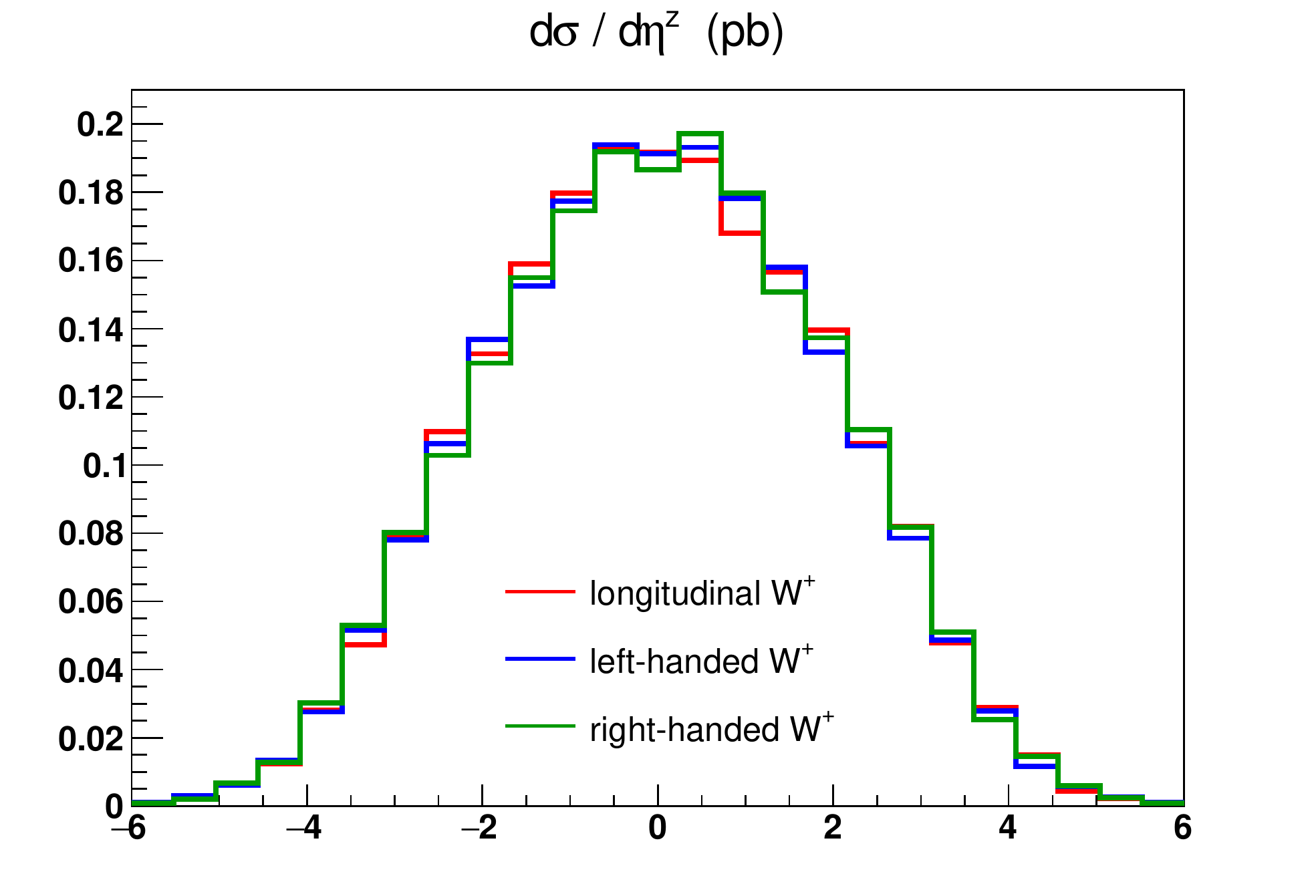}}
\\
\subfigure[Monte Carlo \label{fig:rew1dejj}]{\includegraphics[scale=0.37]{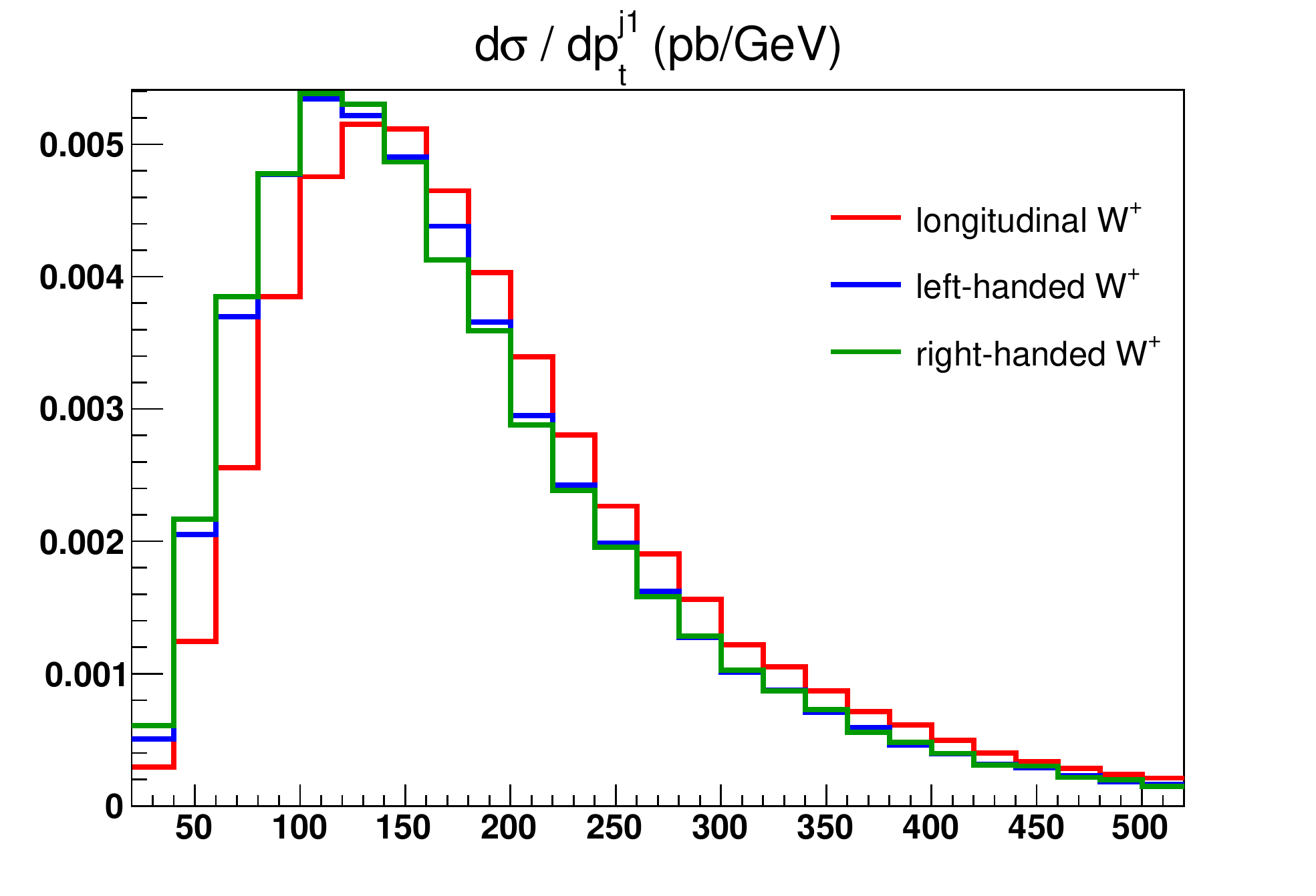}}
\subfigure[Reweighting \label{fig:rew2dejj}]{\includegraphics[scale=0.37]{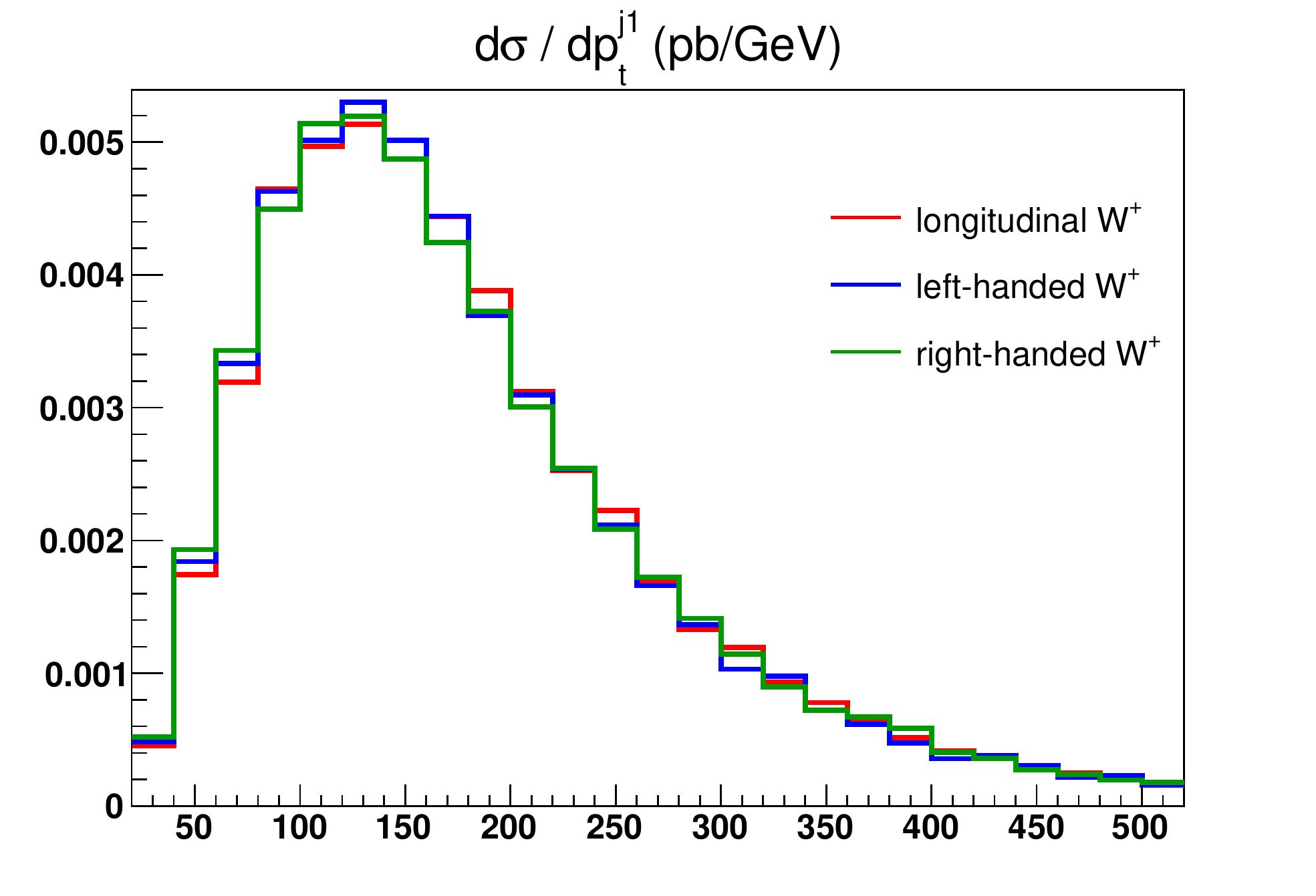}}
\caption{$W^+\!Z$ scattering: $\lvert\Delta\eta_{jj}\rvert$, $\eta^Z$, $p_t^{j1}$ normalized distributions for a polarized $W^+$,
obtained with polarized amplitudes (left side) and with
the reweighting procedure (right side), in the region $60\,\GeV\! <\! p_t^W \!<\!90\,\GeV,1\!<\!\lvert\eta_W\rvert\!<\!2$, in the absence of lepton cuts and
without neutrino reconstruction.} \label{fig:deltaetajj_sec6}
\end{figure}
The main bottleneck of the reweighting procedure is represented by
the phase-space dependence of the polarization fractions.
In the absence of lepton cuts, each polarization gives the same lepton
angular distribution in the $W$ rest frame in any phase space point.
However, the relative weight of the three polarizations varies from point to point.
When assigning a polarization state to a single event, the reweighting
procedure assigns to each event belonging to a $\{p_t^{W},\,\eta_W\}$ cell
the average weight over the whole region. As a consequence, the reweighting method
is not capable of reproducing the correct dependence on kinematic variables different
from $\cos\theta_{\ell^+}$.

To show that this is the case even in the absence of lepton cuts and neutrino reconstruction,
we have compared, in the region $60\,\GeV\! <\! p_t^W \!<\!90\,\GeV,1\!<\!|\eta_W|\!<\!2$,
the longitudinal, left, and right distributions obtained from reweighting
with those computed directly with polarized amplitudes, for a number of variables
that do not depend on the decay products of the polarized $W^+$.
In Fig.~\ref{fig:deltaetajj_sec6}, we show the normalized distributions of
the rapidity difference between the two tagging jets.
The polarized shapes on the left, obtained with polarized amplitudes,
are clearly different from each other: the longitudinal one is peaked at a
smaller value of $\lvert\Delta\eta_{jj}\rvert$ than the two transverse components, and
decreases faster in the distribution tail.
The analogous polarized shapes from reweighting,  on
the right of Fig.~\ref{fig:deltaetajj_sec6}, are similar to each other,
confirming that, even when considering a small  $\{p_t^W,\,\eta_W\}$
region, reweighting corresponds to averaging on the dependence on other variables, washing out
the differences, even in the absence of leptonic cuts.

This becomes even more problematic when lepton cuts are imposed on
the polarized samples, since selection cuts have different effects on different
polarizations.
The conceptual issue is that the polarized samples are obtained without lepton cuts, and
then are analyzed in the presence of cuts. The computation of polarization fractions
and the application of lepton cuts are non commuting procedures.
Notice that the correct description of all kinematic variables is
mandatory for a Multi Variate Analysis.

We have shown that the reweighting method to separate an unpolarized
event sample into three polarized samples provides only approximate predictions,
which can be quite far from being accurate, particularly at high energies.
Therefore it would be better, both for phenomenological and for experimental analyses,
to produce polarized event samples employing directly polarized amplitudes.

\subsection{Extracting polarization fractions}
\label{subsec:extracting_ZW_ZZ}

In this section we investigate the possibility of extracting polarization fractions from VBS
events without prior knowledge of the underlying dynamics.
We consider a Standard Model
with no Higgs boson, \emph{i.e.} $M_h\rightarrow \infty$, as instance of  BSM theories.

Polarization fractions are determined with two different methods.
The first one relies on the expectation that the shapes of the
decay angular distributions are not too sensitive to the underlying
dynamics. If this is the case, one can  fit the unpolarized
distribution of a BSM model with a superposition of SM templates, as done in Ref.~\cite{Ballestrero:2017bxn}.
The second exploits the similarity, in shape and normalization, of the transverse distributions across different
models, which allows to extract the longitudinal component by subtracting the SM transverse contribution.
Both methods give acceptable results within a few percent. The difference between the two determinations
provides a rough estimate of the uncertainty in the extraction procedure.
All the results have been obtained applying the complete set of cuts.
Neutrino reconstruction is always implied.

In \refse{sec:WZ}  we have considered left and right contributions
separately. If we consider the coherent sum of left and right polarizations (which we refer to as
transverse), we include the left-right interference term. Therefore, separating only the longitudinal from the
transverse mode is expected to minimize the total interferences among different polarizations.

\begin{figure}[!htb]
\centering
\subfigure[$M_{W\!Z}$\label{fig:extr_compmwz_Z}]{\includegraphics[scale=0.37]{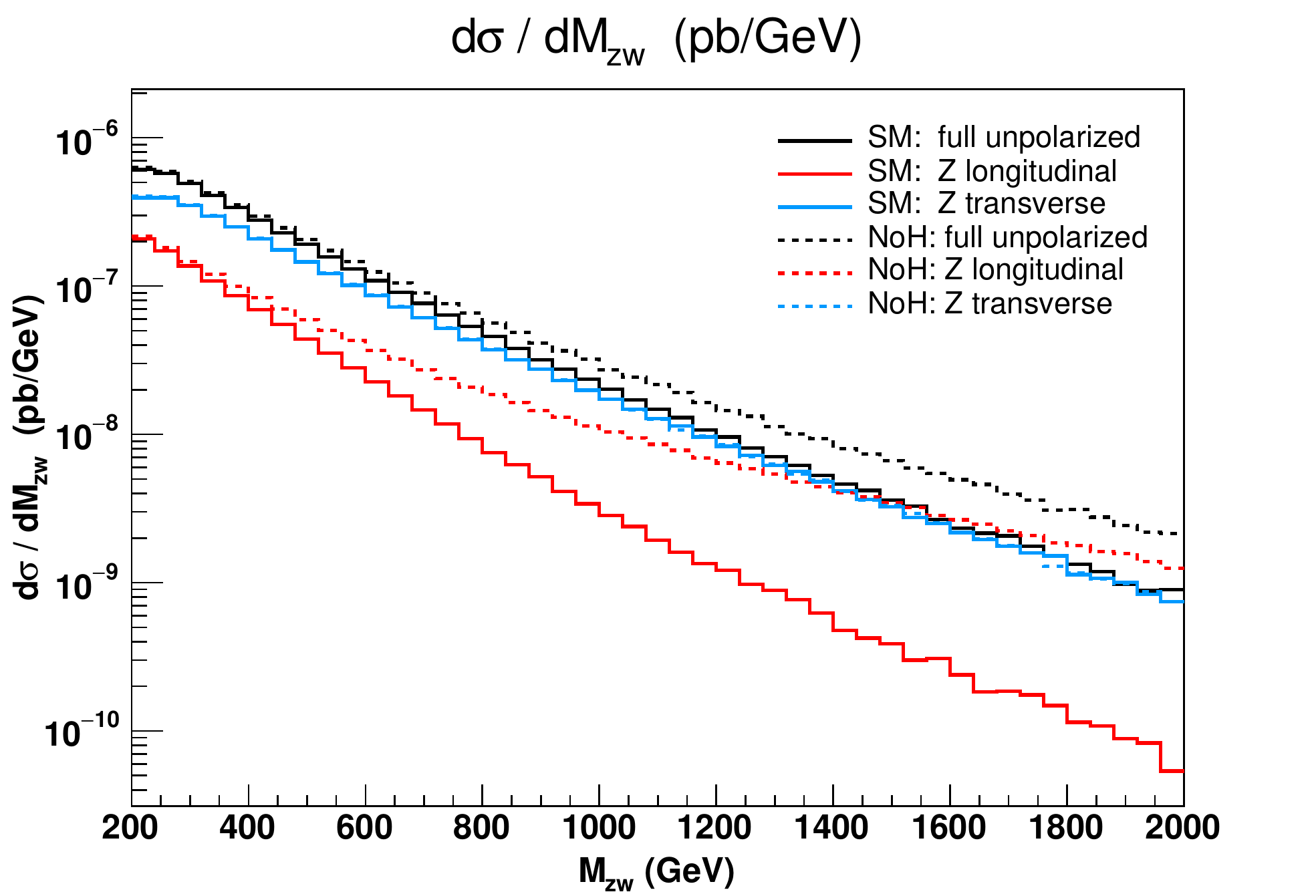}}
\subfigure[$\cos\theta_{e^-}$\label{fig:extr_compcth_Z}]{\includegraphics[scale=0.37]{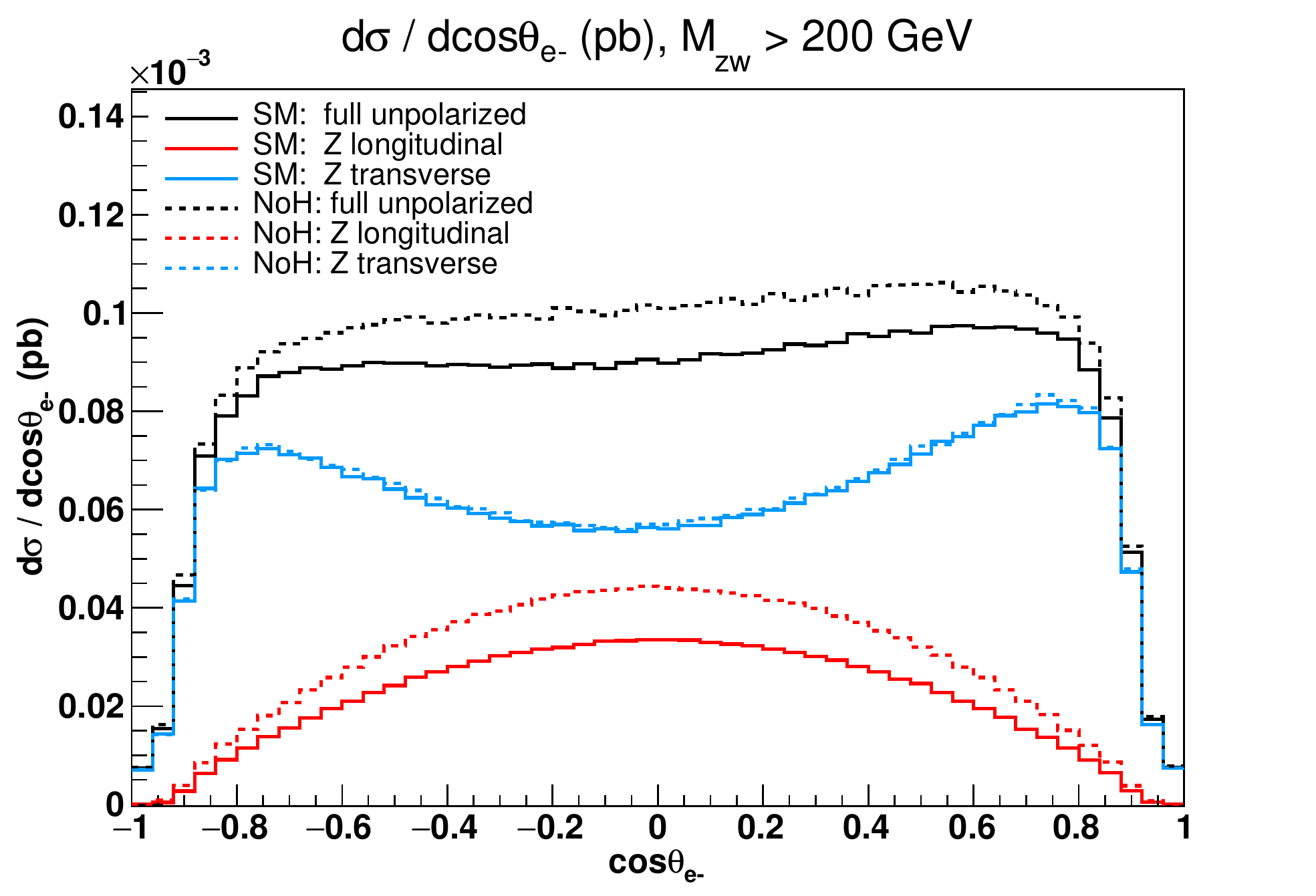}}
\caption{$W^+\!Z$ scattering: comparison of Standard Model (solid) and Higgsless model (dashed) distributions in $M_{W\!Z}$ and
$\cos\theta_{e^-}$. Polarized distributions concern the $Z$ boson. The full set of kinematic cuts
is understood, including lepton and missing transverse momentum cuts, as
well as neutrino reconstruction.}\label{fig:extr_comp_Z}
\end{figure}

In Fig.~\ref{fig:extr_compcth_Z}, we present the $\cos\theta_{e^-}$ distributions for a polarized $Z$ boson
in association  with an unpolarized $W^+$. At large boson invariant mass, the longitudinal
component in the Higgsless model dominates.
The transverse components are almost identical, even at very large four lepton
invariant masses, both in shape and cross section.

\begin{figure}[!htb]
\centering
\subfigure[$M_{W\!Z}>500\,\GeV$\label{fig:extr_fit1_Z}]
{\includegraphics[scale=0.37]{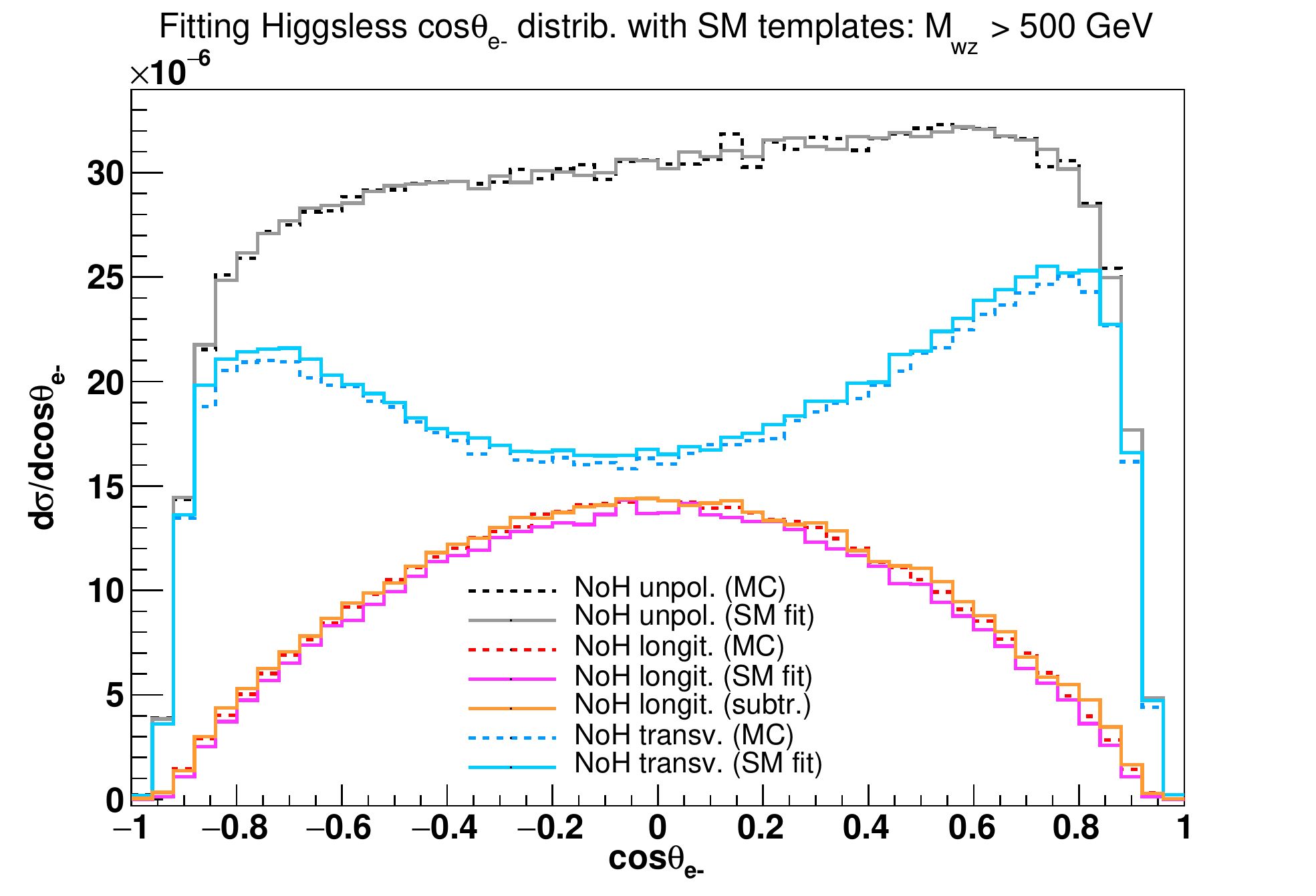}}
\subfigure[$M_{W\!Z}> 200\,\GeV,\,p_t^{Z}> 300\,\GeV$\label{fig:extr_fit2_Z}]
{\includegraphics[scale=0.37]{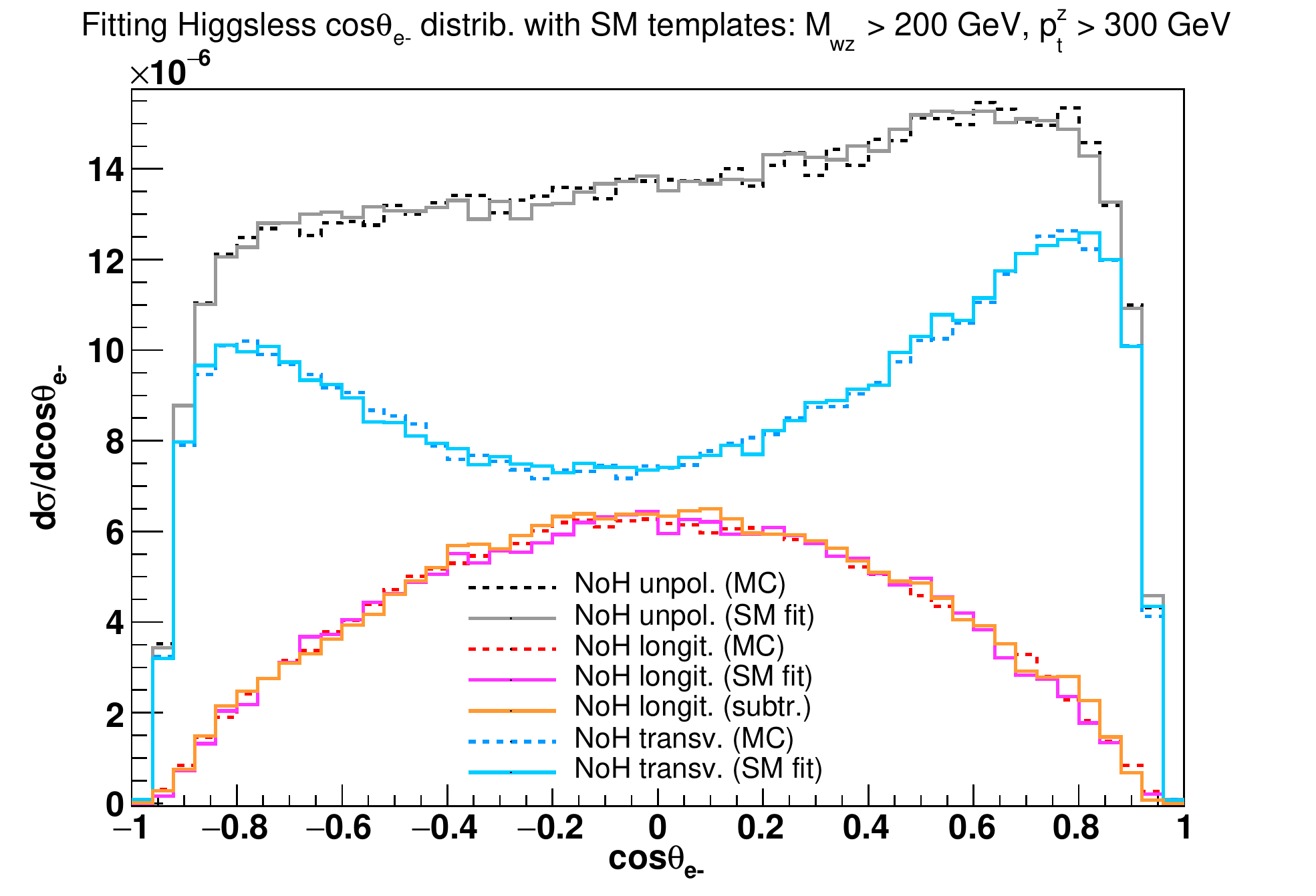}}
\caption{$W^+\!Z$ scattering: fit of Higgsless unpolarized $\cos\theta_{e^-}$ distributions with SM
templates, in two different
kinematic regions (large mass and large $p_t$). For the longitudinal component the result of the fit (magenta)
and the one of the subtraction technique (orange) are compared with the Monte Carlo expectation (dashed
red).}\label{fig:fitsubtrwz_z}
\end{figure}

Both the fit and the subtraction procedure provide longitudinal cross sections for the Higgsless model
which differ from the Monte Carlo expectations by less than 5\%, in all studied kinematic regions.
Numerical results for extracted longitudinal and transverse cross sections are shown in
Tab.~\ref{tab:extr_fitsubtra_Z}. Fitted and expected distributions are
shown in Fig.~\ref{fig:fitsubtrwz_z}, in two specific kinematic regions.

\begin{table}[hbt]
\begin{tabular}{|c|c|c|c||c|c|c|}
\hline
& \multicolumn{6}{|c|}{Polarized cross sections [$\ab $]} \\
\hline
& \multicolumn{3}{|c||}{ Longitudinal} & \multicolumn{3}{|c|}{ Transverse  }\\
\hline
kinematic region  & MC & Fit & Subtr. & MC & Fit & Subtr. \\
\hline	
$M_{W\!Z}>200 \,\GeV$ & 56.27 & 54.88&  57.75 & 122.24 & 124.46 & 120.96\\
\hline
$M_{W\!Z}>500 \,\GeV$ & 18.35 & 17.59& 18.63 & 35.46 & 36.30 & 35.26\\
\hline
$M_{W\!Z}>1000 \,\GeV$ & 4.90 & 4.73& 4.91 & 5.37 & 5.54 & 5.39\\
\hline
\end{tabular}
\caption{Cross sections ($\ab $) for a longitudinal and transverse $Z$ in $W^+\!Z$ scattering, in the Higgsless model,
in several kinematic regions: comparison of MC predictions for the Higgsless
model with results obtained via fit and subtraction procedure. The subtraction procedure
results for a transverse $Z$ coincide with the SM cross sections.} \label{tab:extr_fitsubtra_Z}
\end{table}

As a general trend, the subtraction procedure overestimates by a few percent the expected values.
On the contrary, the fit procedure underestimates by few percent the expected longitudinal cross section in
the various kinematic regions. This results in a very mild enhancement of the transverse component (see azure
and cyan curve in Fig.~\ref{fig:fitsubtrwz_z}).

In the large invariant mass
($M_{W\!Z}>1000\,\GeV$), large $p_t$ ($p_t^Z>400\,\GeV$), and forward rapidity ($|\eta_Z|>2$) region the
subtraction procedure reproduces very well the Monte Carlo expected longitudinal cross sections.

\subsection{Conclusions}
\label{sec:conclusions}

In this note we have presented a procedure to separate polarization states
of massive weak bosons in $WZjj$ VBS processes.
We have shown that with a sufficiently tight cut on the invariant mass of charged lepton pairs around the $Z$ pole
and a single on-shell projection (OSP1) on $W$ resonant diagrams, the signal for a polarized $W$ and $Z$ reproduces accurately the results that can be extracted from full $\cos\theta_{\ell}$ distributions by means of projections onto the first three Legendre polynomials, in the absence of lepton cuts.
After applying a realistic set of leptonic cuts, the sum of polarized signals reproduces the full unpolarized results within a few percent.
The proposed method provides reliable results in the Standard Model.

The reweighting method, which has been widely used to determine approximate polarized signals
in presence of lepton cuts~\cite{Chatrchyan:2011ig,ATLAS:2012au,Aad:2013ksa,Khachatryan:2015paa,Bittrich:2015aia}, 
provides inaccurate predictions particularly at high diboson invariant mass.

Our results suggest that it will be
possible to estimate, with reasonable accuracy, polarization fractions in VBS at the LHC  by using
Standard-Model angular distributions, even in the presence of new physics.


\section{Simplified Models for New Physics Coupled to Transverse Vector Bosons\footnote{speaker: J\"urgen Reuter; authors: Simon Brass, Wolfgang Kilian}}

As the first two runs of the LHC have not been revealing any signal of
new physics beyond the Standard Model (BSM), many searches have been
focusing on setting limits on coefficients in a setup as
model-independent as possible. Any deviation from the SM induced by
heavy new physics (i.e. without any new particle at the electroweak
scale or slightly above) can be formulated in the SM
effective field theory (SMEFT). The SMEFT consists of the SM with its
dim-2 and dim-4 operators together with a whole tower of
higher-dimensional operators built from the SM 
fields obeying the gauge symmetries of the SM (there are variations
regarding the assumptions on the flavour structure). Consistency of
measurements at the LHC (or future colliders) with the SM within their
statistical and systematic uncertainties then translates into
bounds on the operator coefficients. The most sensitive processes are
diboson production and the Higgs measurements where deviations by
dim-6 operators dominate. They also play a role in vector-boson
scattering (VBS) and triple boson production, however, there
dimension-8 operators give the dominating effects for parameters that
cannot be measured in simpler processes like dibosons. The size of the
effects also depends on the type of new physics, where loop
corrections from weakly coupled models yield dominating dim-6
operators while new (strongly coupled) resonances lead to dim-8
operators when integrated out.

Energy-frontier measurements at the LHC like VBS are not necessarily
\begin{figure}[!ht]
  \includegraphics[width=.48\textwidth]{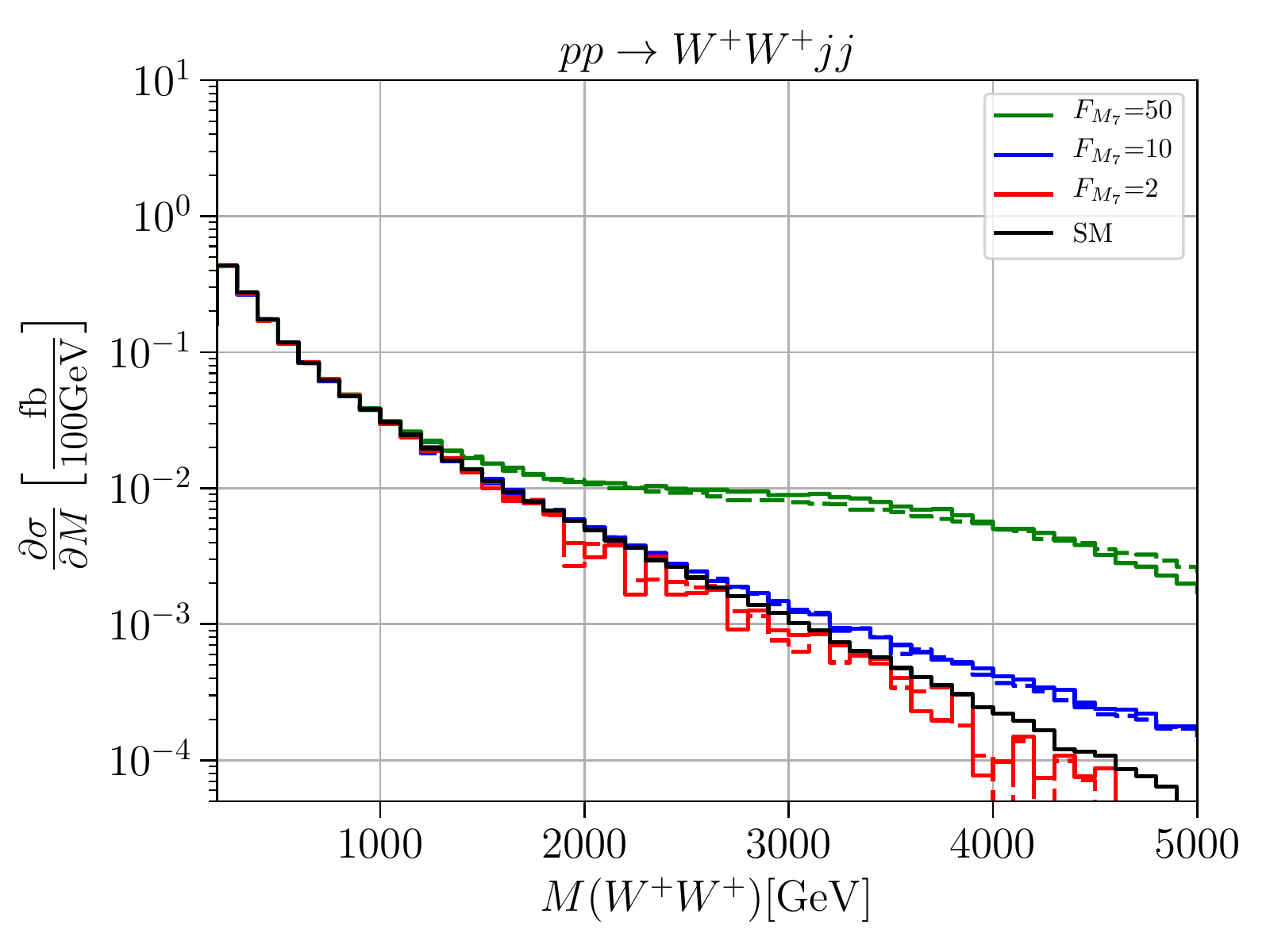}
  \includegraphics[width=.48\textwidth]{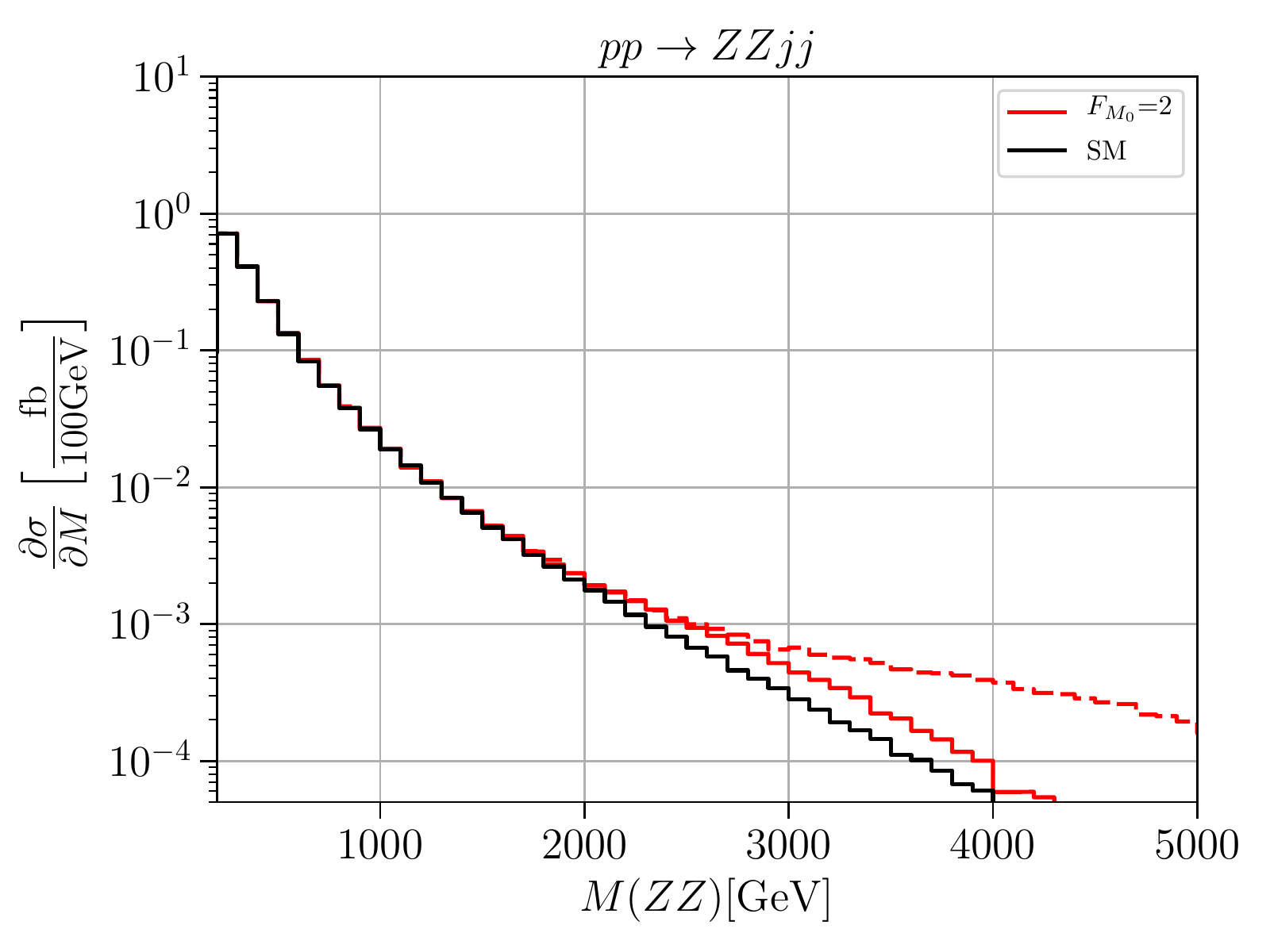}
  \includegraphics[width=.48\textwidth]{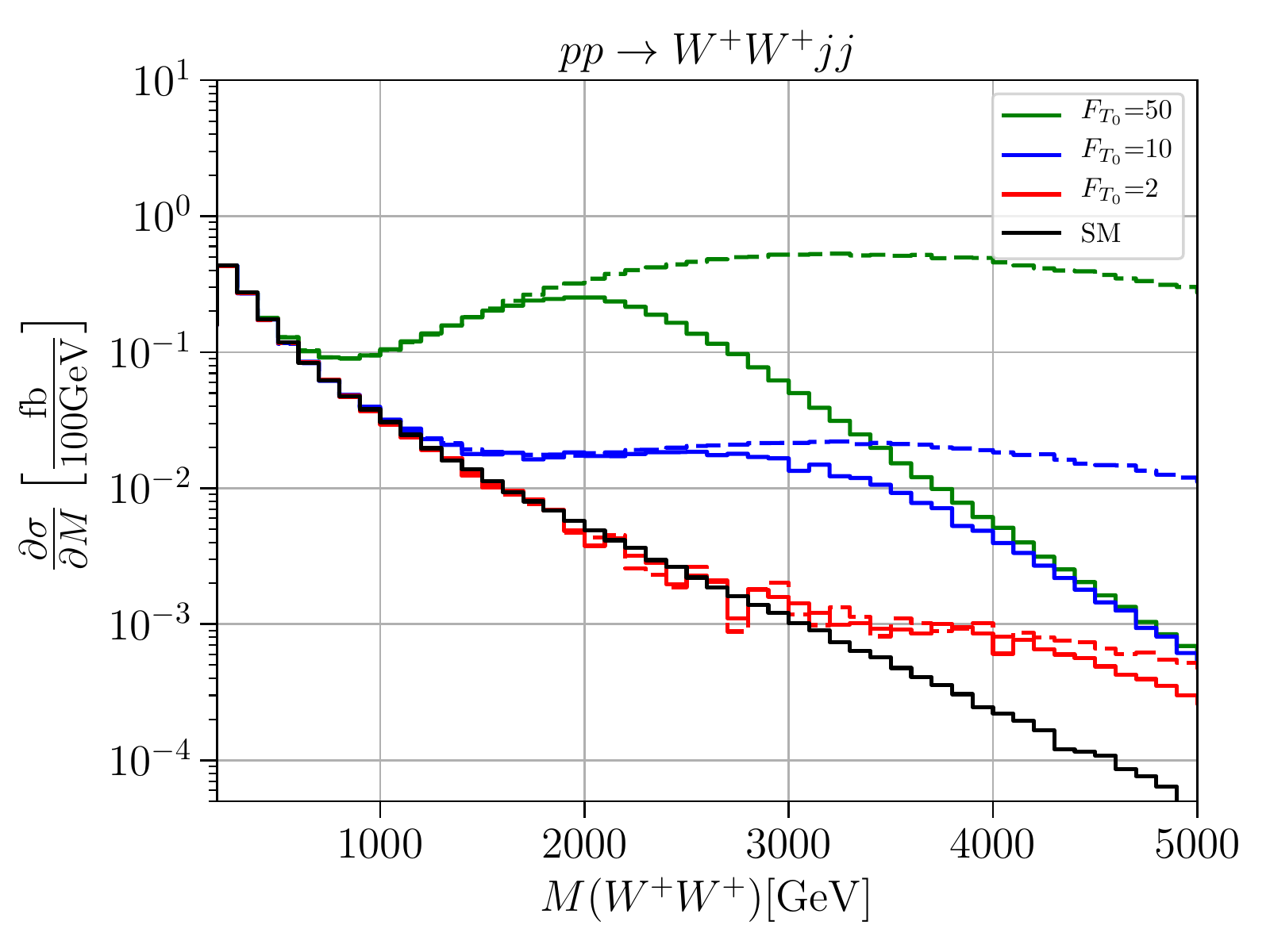}
  \includegraphics[width=.48\textwidth]{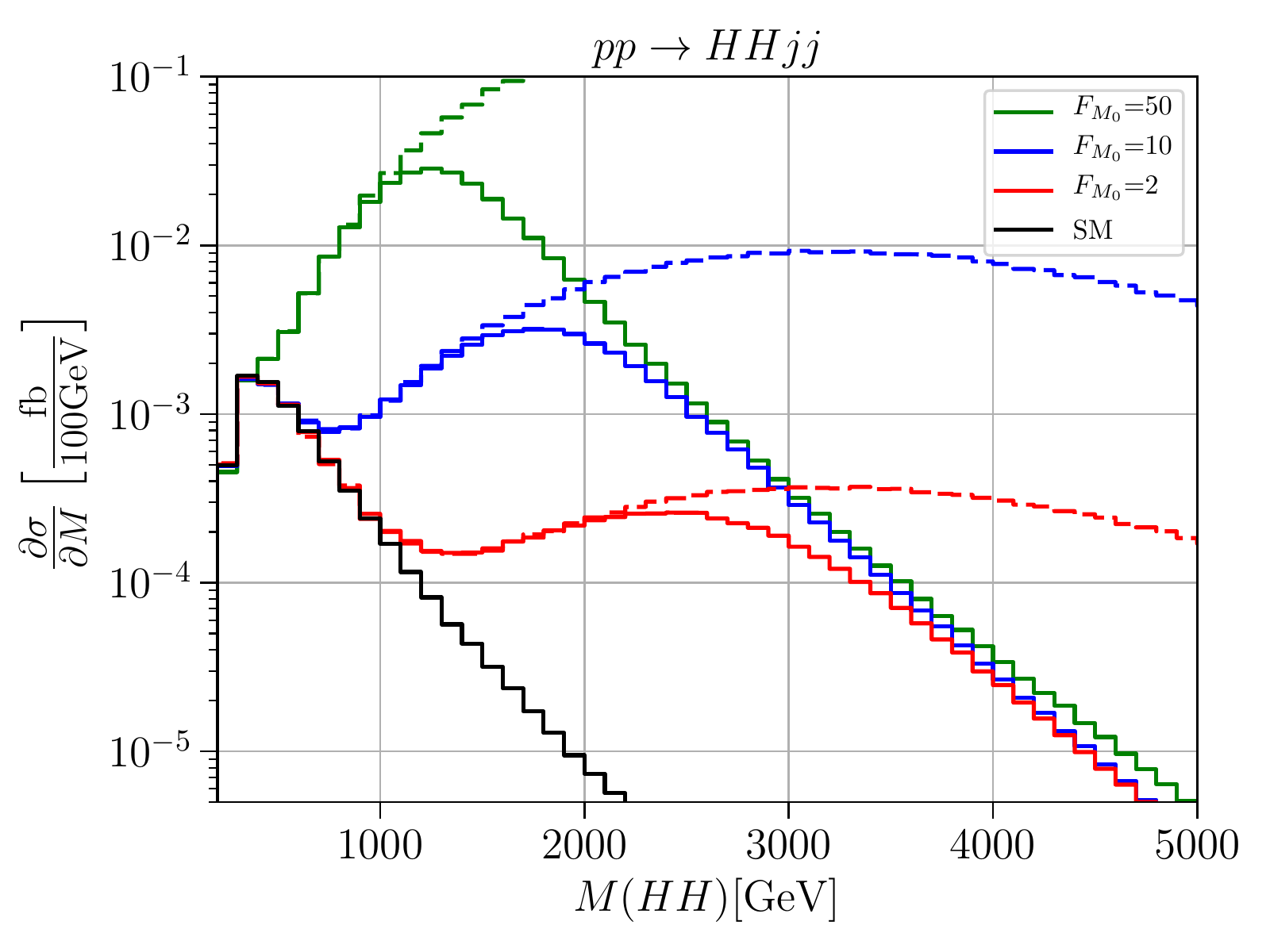}
  \caption{\label{transverse_op}
    New physics effects from different transverse operators in
    VBS: left column: $pp \to jjW^+W^+$, right column $pp \to ZZjj$
    (top), $pp \to HHjj$ (bottom). In the upper row, deviations are
    from mixed operators $\LL_{M,7}$ and $\LL_{M,0}$, in the lower
    row, from $\LL_{T,0}$ and $\LL_{M,0}$, respectively. Black lines
    are the SM distributions, dashed lines are the naive SMEFT
    results, and full colored lines are unitarized results. The value
    for the operator coefficients are in TeV${}^{-4}$.}
\end{figure}
tailor-made applications of EFT expansions, as the parton density
functions (PDFs) effectively scan over a wide range of effective
collision energies. This effect gets enhanced by the effect that the
diboson invariant mass systems can only be fully reconstructed for the
leptonic (or semi-leptonic) $ZZ$ channel or the fully hadronic $WW$
channel. VBS processes are among the rarest SM processes to be probed
at the LHC and hence severely statistics limited (originally, at the
SSC collider 40 TeV had been foreseen for this measurement). This
leads to the scylla and charybdis between operator coefficients that
are too small to be experimentally visible or too large to lie within
the validity range of the EFT expansion. Amplitudes with dim-8
deviations from the SM grow quadratically with energy and lead easily
to violation of perturbative unitarity using such a naive signal
model. This has been e.g. studied in~\cite{Alboteanu:2008my} for the
electroweak chiral Lagrangian and after the Higgs discovery~\cite{Kilian:2014zja}
within the linearized EFT in the Warsaw basis~\cite{Grzadkowski:2010es}. 
In Fig.~\ref{transverse_op}, we show the
effects on the invariant mass spectra of the diboson system in VBS
from the mixed operators 
\begin{align*}
  \mathcal{L}_{M,0}&=-g^2 F_{M_0}\text{tr}\left[(\textbf{D}_\mu
    \textbf{H})^\dagger (\textbf{D}^\mu \textbf{H})\right]
  \text{tr}\left[\textbf{W}_{\nu \rho} \textbf{W}^{\nu \rho}\right]
  \\
  \mathcal{L}_{M,7}&=-g^2 F_{M_7}\text{tr}\left[(\textbf{D}_\mu
    \textbf{H})^\dagger \textbf{W}_{\nu \rho} \textbf{W}^{\nu \mu}
    (\textbf{D}^\rho \textbf{H}) \right]  
\end{align*}
and the transversal operator
\begin{equation*}
  \mathcal{L}_{T_0}=g^4 F_{T_0}\text{tr}\left[\textbf{W}_{\mu \nu}
    \textbf{W}^{\mu \nu}\right]
  \text{tr}\left[\textbf{W}_{\alpha \beta} \textbf{W}^{\alpha \beta}
    \right]   \qquad.
\end{equation*}
For the complete list of longitudinal, mixed and transverse operators
for VBS processes, we refer to~\cite{Brass:2018hfw}. First of all, as in the
SMEFT setup the Higgs is an electroweak doublet, it has to be considered
together with the longitudinal states to construct complete $SU(2)$
tensor product representations for the diboson system. This can be
seen in the lower right. SM predictions are in black, dashed lines are
naive SMEFT distributions, and full coloured lines are using $T$ matrix
unitarization~\cite{Kilian:2014zja,Kilian:2015opv}. This unitarization
method saturates the bound from perturbative unitarity and hence gives
bin per bin the maximal number of events that are possible in any
quantum field theory described by a unitarity $S$ matrix. Note that
also resonance peaks are not allowed to exceed these lines. Similar
studies have also been performed for future high-energy $e^+e^-$
colliders, for the EW chiral Lagrangian~\cite{Beyer:2006hx}
and for SMEFT~\cite{Fleper:2016frz}, respectively. Simulations have
been performed using the \texttt{WHIZARD} event
generator~\cite{Kilian:2007gr}. From the plots
in~\cite{Kilian:2014zja} and here it is obvious that the SM curves for
\begin{figure}
  \includegraphics[width=.48\textwidth]{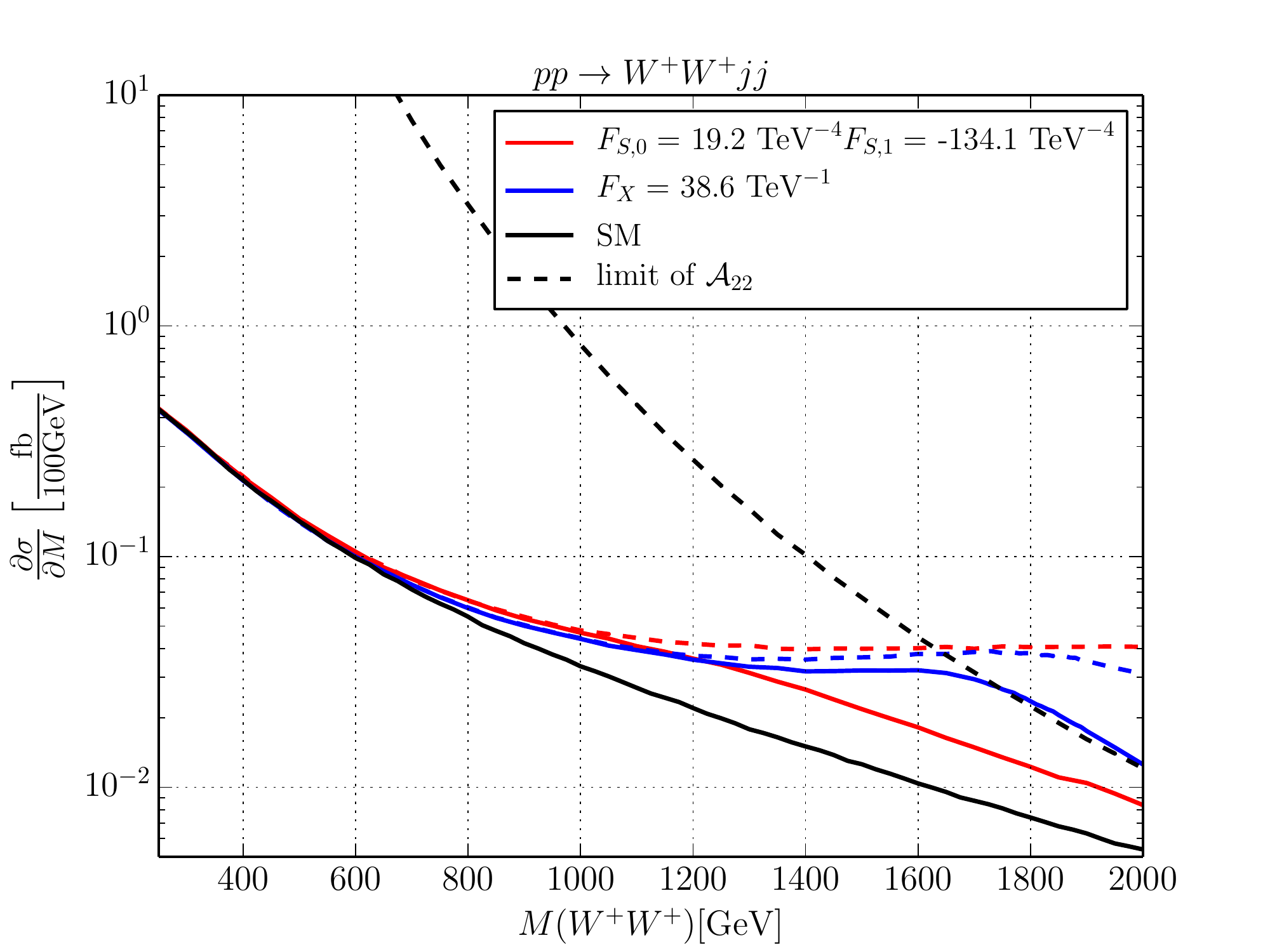}
  \includegraphics[width=.48\textwidth]{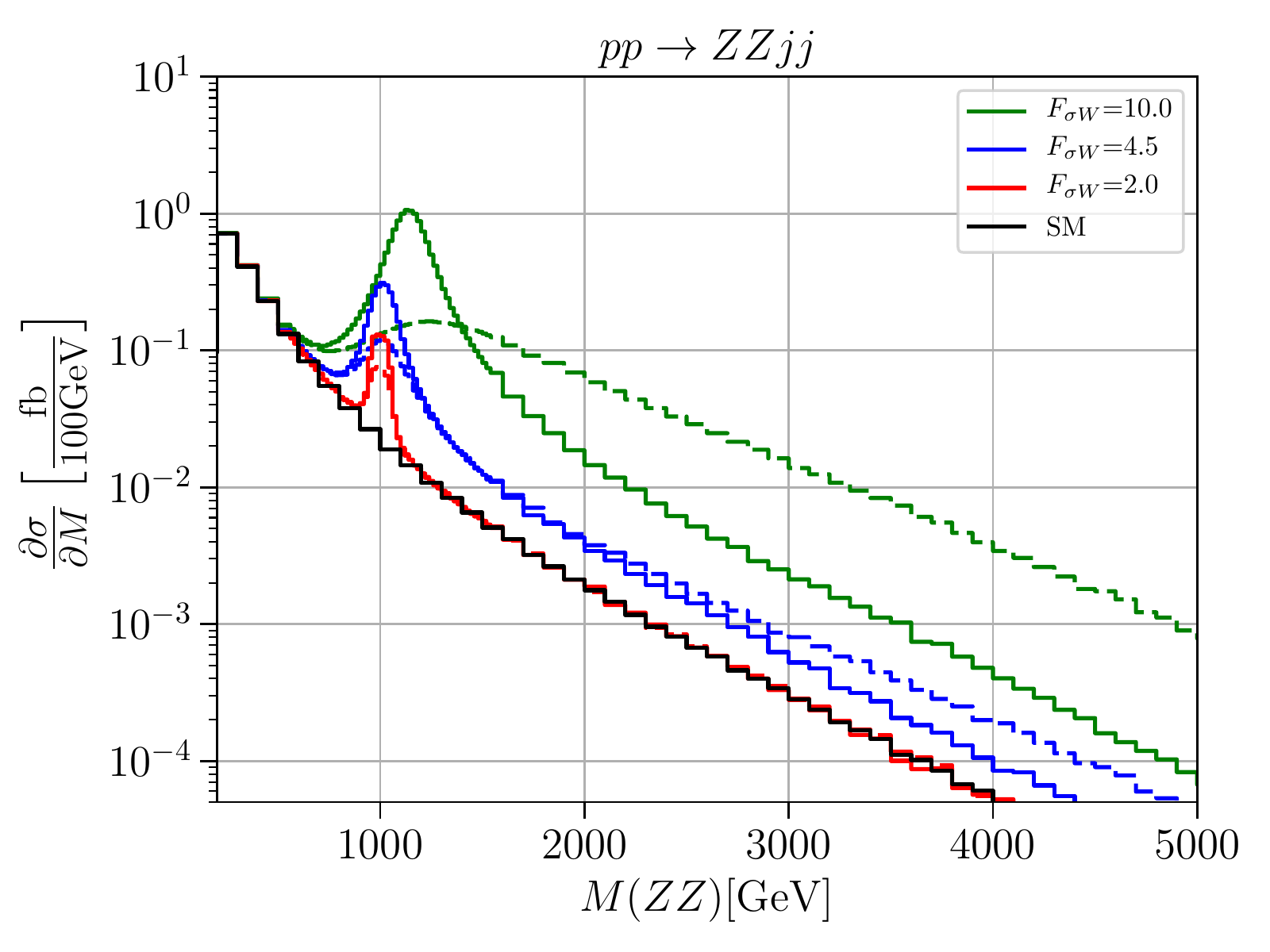}  
  \caption{\label{diboson_res}
    Simplified models with a broad tensor resonance coupled to
    longitudinal EW bosons (left) and a relatively narrow scalar
    resonance coupled to transversal EW bosons (right).}  
\end{figure}
longitudinal final states are much closer to the unitarity bound than
for mixed and transversal operators. Turning the argument around, this
means that there is a lot more space for new physics excesses than in
the longitudinal modes, which are anyway much harder to discriminate
from the transversal SM background. Of course, to achieve sensitivity
in the tails of the distributions (in the overflow bins), a precise
knowledge of the SM distributions including higher order QCD and
electroweak corrections is necessary. These predictions have been
studied in~\cite{Biedermann:2016yds,Biedermann:2017bss,Ballestrero:2018anz}. 

As mentioned above, setting meaningful limits on these operator
coefficients from the overflow bins, is a hard task given the limited
amount of data and the constraints from perturbative unitarity. A less
model-independent approach is to consider simplified models that
describe possible resonances in the diboson sector. From spin and
isospin selection rules, only scalars, vectors and tensors can couple
to the diboson system, which can occur as isoscalars, isovectors and
isotensors. Vector resonances are theoretically more involved as they
can mix with the electroweak gauge bosons (at least after electroweak
symmetry breaking). Fig.~\ref{diboson_res} compares resonances coupled
to longitudinal EW bosons which may become rather broad (left) and
resonances coupled exclusively to transversal bosons (right). The
resonance on the left is a tensor resonance, the one on the right a
scalar resonance. While it is obvious that the resonance coupled to
transversal EW can be properly described in any finite truncation of
the EFT description, the broad tensor resonance resembles at least in
the low-energy tails the EFT description with the resonance being
integrated out and looks like a mismatch in normalization in the
overflow bin at the "peak" of the resonance. Note that also
the simplified models need to be scrutinized regarding their unitarity
constraints as the SM together with single EW resonances is not
necessarily a UV-complete renormalizable model which could violate
perturbative unitarity. Simplified models have only two independent
parameters, either the mass and width of the resonance, or the mass
and its coupling to EW bosons. They allow for a more sophisticated
signal model in searches for new physics effects in VBS and
multi-boson production.

\chapter{Analysis Techniques}
\label{WG2}

\section{Recent ATLAS results in Vector-Boson Scattering\footnote{speaker: Philip Sommer on behalf of the ATLAS Collaboration\\%
Copyright 2020 CERN for the benefit of the ATLAS Collaboration. Reproduction of this section or parts of it is allowed as specified in the CC-BY-4.0 license.}}

\let\bf\textbf
\let\it\textit

\subsection{Introduction}
Vector boson scattering (VBS) is amongst the rarest processes currently accessible experimentally at the Large Hadron Collider (LHC). The ATLAS experiment~\cite{Aad:2008zzm} at the LHC has studied VBS in $pp$ collision data corresponding to $36.1~\text{fb}^{-1}$ at $\sqrt{s}=13$~TeV by measuring the electroweak production of $W^\pm W^\pm$ and $WZ$ bosons, with the $W$ and $Z$ bosons decaying to leptons, as well as the electroweak production of $VV$ ($V=W,Z$) bosons, with one gauge boson decaying to leptons and the other decaying hadronically. At Born level, the electroweak production does not involve the exchange of colour between partons leading to the experimental signature of two high energy jets with large rapidity separation produced in association with the gauge-boson pair.

At the LHC, two gauge bosons in association with two jets can also be produced in mixed strong and electroweak interactions. Interference effects between electroweak and strong production are typically assigned as a systematic uncertainty in the electroweak production, and results of electroweak production, hence, depend on the assumptions made in the theoretical predictions. Such dependencies are reduced when combined electroweak and strong production cross sections are reported.

\subsection{Observation of electroweak $W^\pm W^\pm jj$ production }

\begin{figure}
  \centering
  \subfigure[]{\includegraphics[width=0.5\textwidth]{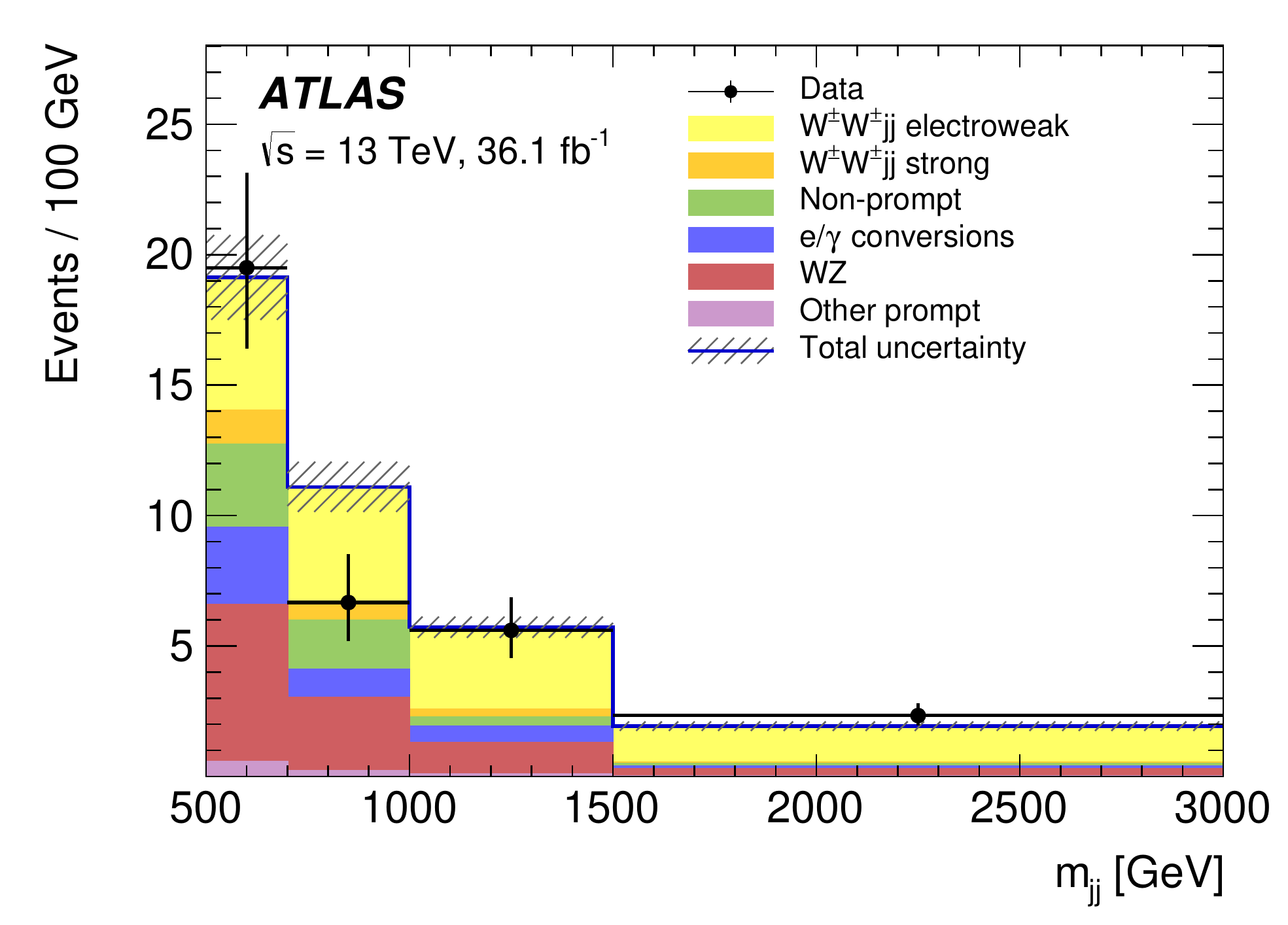}%
    \label{fig:atlas:sswwmeas}} %
  \subfigure[]{\includegraphics[width=0.4\textwidth]{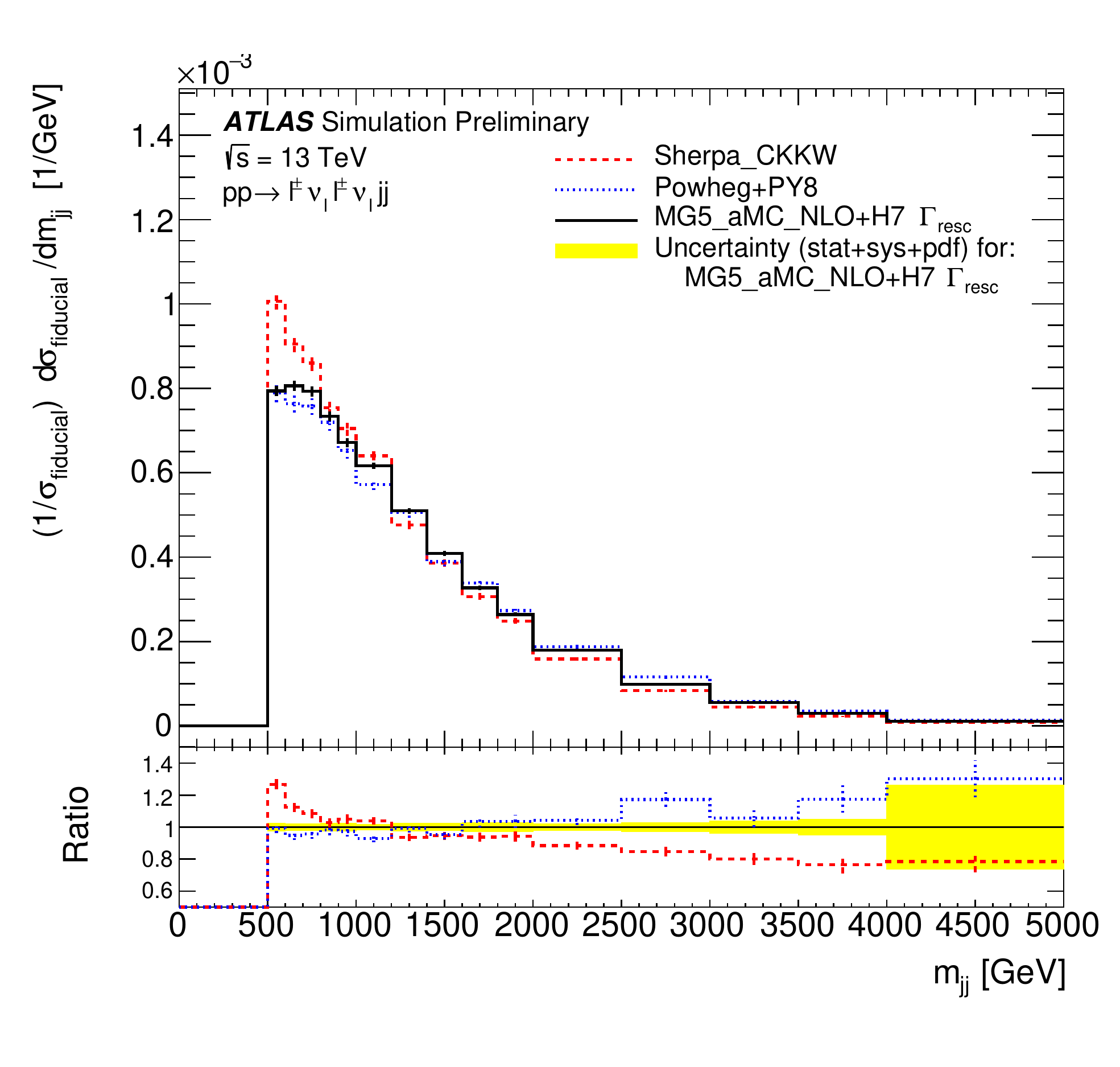}%
    \label{fig:atlas:sswwmc}} \par
  \caption{
    Dijet invariant mass distributions for electroweak $W^\pm W^\pm jj$ production in the signal region. (a) Events selected in data are compared to the estimated signal and background contributions~\cite{Aaboud:2019nmv}. Systematic uncertainties are shown as a hatched band. The signal has been simulated with \texttt{Sherpa~2.2.2}. (b)~Differential fiducial cross sections normalised to unity are shown for various event generators~\cite{ATL-PHYS-PUB-2019-004}. Due to a non-optimal colour flow setting in \texttt{Sherpa}, the total cross section significantly differs from that of \texttt{Powheg} or \texttt{MG5\_aMC@NLO}.
  }
\end{figure}

The $W^\pm W^\pm jj$ process has the largest ratio of electroweak to strong production processes since strong production diagrams are heavily suppressed compared to other diboson processes, e.g.\ gluon initiated diagrams are absent at leading order QCD. Also, purely electroweak diagrams which are not sensitive to gauge-boson self interactions are suppressed. The $W^\pm W^\pm jj$ process is therefore ideally suited for VBS studies.
Events with exactly two leptons, $\ell=e$~or~$\mu$, with the same electric charge, missing transverse momentum induced by the neutrinos, and at least two jets with a large rapidity separation are selected as candidate events~\cite{Aaboud:2019nmv}. 
The electroweak signal is extracted in a fit of simulated signal events and estimated background yields to five bins of the dijet invariant mass distribution, separately for $ee$, $e\mu$ and $\mu\mu$ final states and with positive and negative electric charge. The dijet invariant mass distribution of the selected candidate events in the signal region is shown  in Figure~\ref{fig:atlas:sswwmeas}. The background from $WZjj$ production is constrained in a control sample selected with exactly three leptons and a normalisation factor of $0.86\pm0.07~\text{(stat.)}^{+0.18}_{-0.08}~\text{(exp.\ syst.)}^{+0.31}_{-0.23}~\text{(mod.\ syst.)}$ is obtained.

The selected $W^\pm W^\pm jj$ candidate events show an excess of events with respect to the estimated background. The background-only hypothesis is rejected with a significance of 6.5$\sigma$. The excess is consistent with the electroweak $W^\pm W^\pm jj$ signal for which a fiducial cross section of:

\begin{equation*}
  \sigma^{\text{fid.}}_{W^\pm W^\pm jj\text{-EW}} = 2.89^{+0.51}_{-0.48}~\text{(stat.)}^{+0.24}_{-0.22}~\text{(exp.\ syst.)}^{+0.14}_{-0.16}~\text{(mod.\ syst.)}^{+0.08}_{-0.06}~\text{(lumi.)~fb}
\end{equation*}
is measured. The corresponding cross sections predicted by the \texttt{Sherpa~2.2.2}~\cite{Gleisberg:2008ta}
and \texttt{PowhegBox} + \texttt{Pythia8}~\cite{Melia:2011gk} event generators are
$2.01^{+0.33}_{-0.23}$~fb and $3.08^{+0.45}_{-0.46}$~fb, respectively~\cite{Aaboud:2019nmv}. 

Whilst the prediction from \texttt{PowhegBox+Pythia8} agrees well with the measurement, the prediction from \texttt{Sherpa~2.2.2} is approximately 30\% lower. This can be explained by a non-optimal colour-flow setting for the parton shower in \texttt{Sherpa~2.2.2} in VBS diagrams which leads to an excess of central jet emissions. Since for $W^\pm W^\pm jj$ production a multi-leg configuration is used, these effects are partially mitigated. At the same time, this multileg configuration causes a significant underestimation of the total cross section. Comparisons of theoretical predictions with various configurations of event generators and parton-shower programs~\cite{ATL-PHYS-PUB-2019-004} show that \texttt{Sherpa~2.2.2} differs by up to 40\% in the tails of the dijet invariant mass distribution from predictions using \texttt{PowhegBox+Pythia8} and \texttt{MG5\_aMC@NLO}~\cite{Alwall:2014hca}, as shown in Figure~\ref{fig:atlas:sswwmc}. The comparison of  \texttt{Sherpa~2.2.2} with \texttt{PowhegBox+Pythia8} shows the largest difference. Using the bin boundaries of the analysis, the differences are of the same size or smaller than the statistical uncertainties in the data. Within the modelling uncertainties derived using the dependence of the \texttt{Sherpa~2.2.2} prediction on renormalisation and factorisation scales, different PDF sets, and variations of the matching and resummation scales in the combination of the matrix element with the parton shower, no significant effect was found in the measured cross section.
\subsection{Observation of electroweak $WZjj$ production}
\label{sec:wzjj}

\begin{figure}
  \centering
  \includegraphics[width=.44\textwidth]{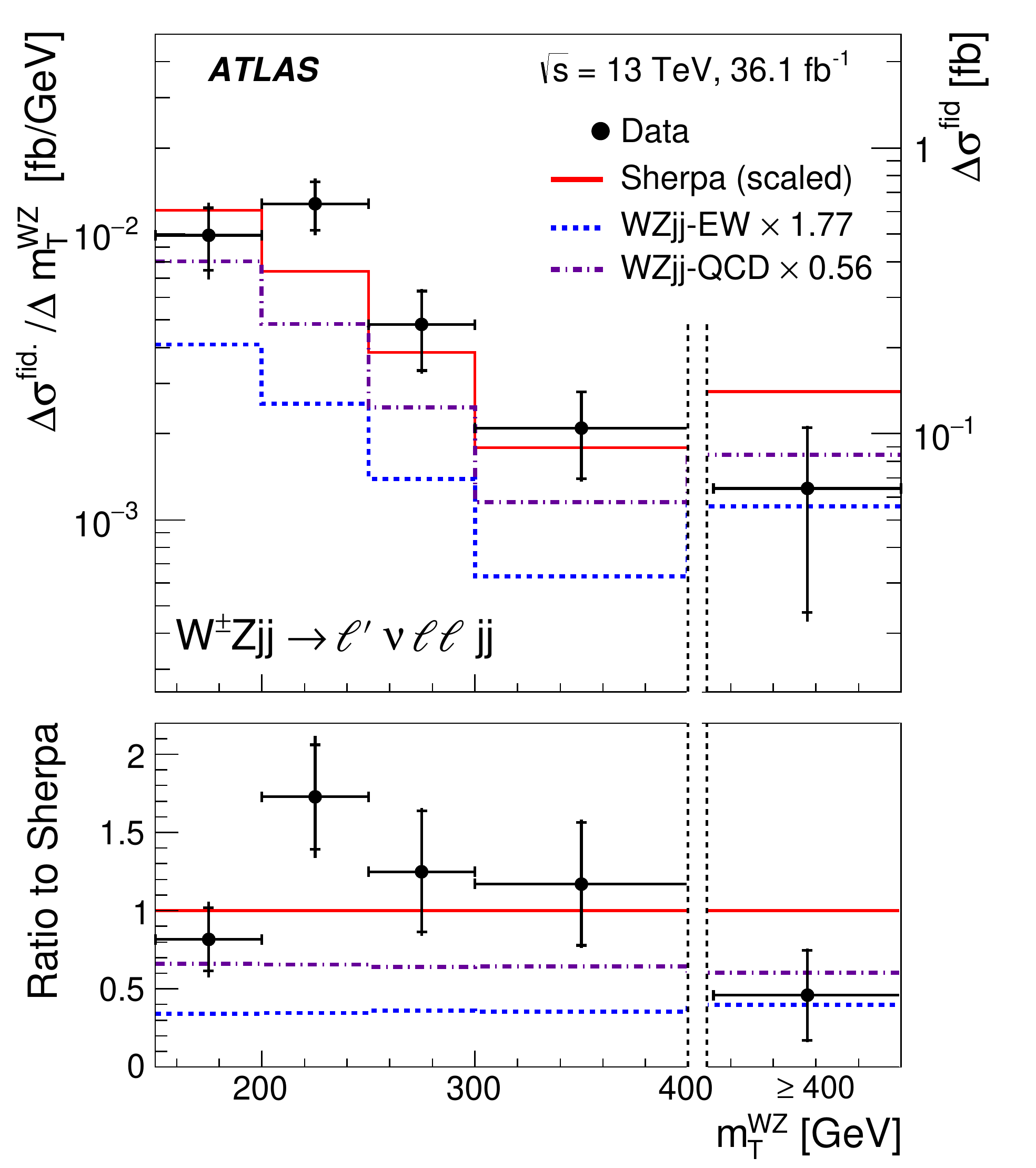}
  \includegraphics[width=.44\textwidth]{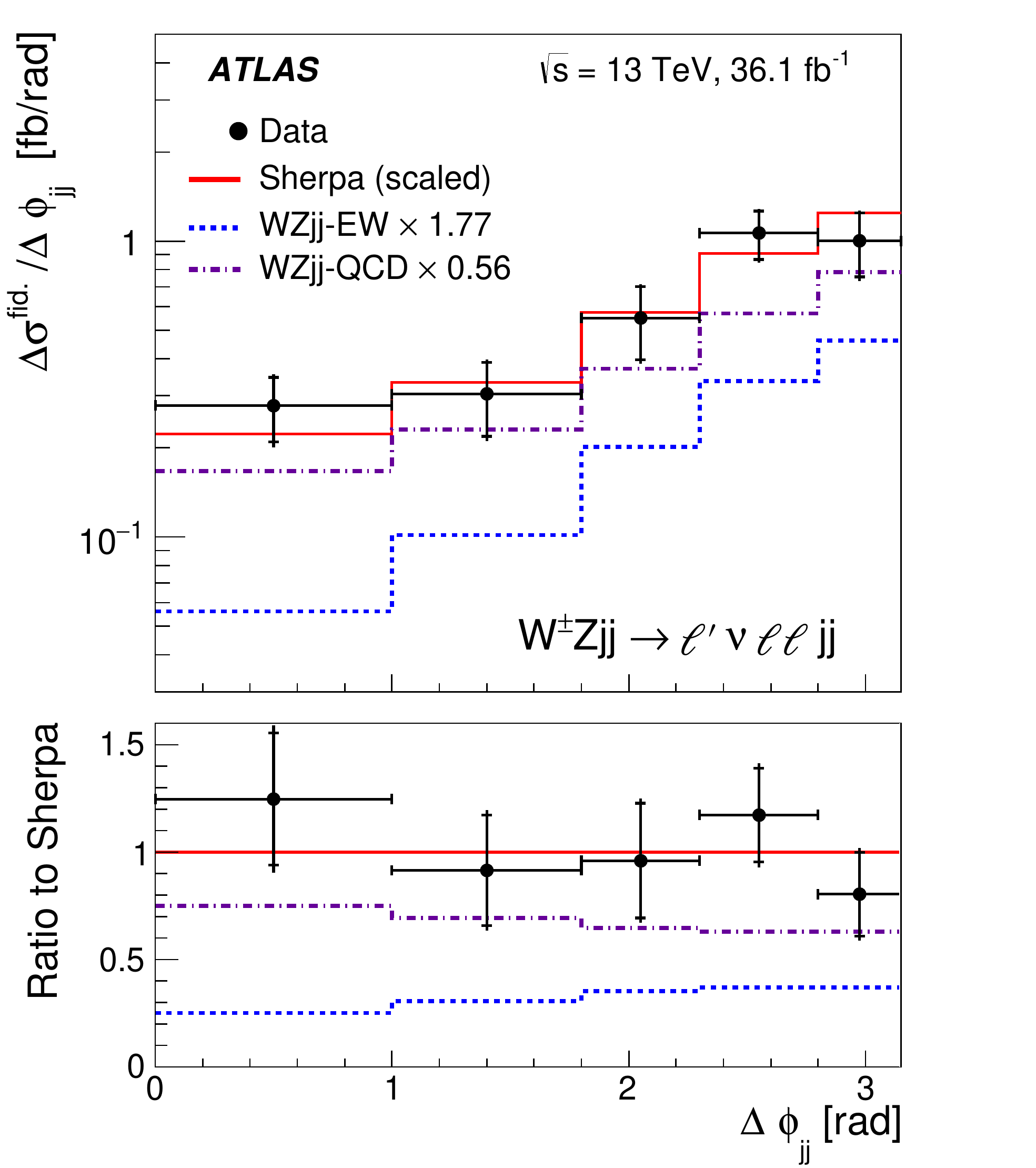}
  \caption{
    Differential $pp\rightarrow WZjj$ cross-sections as a function of the transverse mass of the $WZ$ system, $m_{\mathrm{T}}(WZ)$ (left), and the azimuthal separation of the two jets, $\Delta\phi_{jj}$ (right) measured in data. The measurements are compared to theoretical predictions from \texttt{Sherpa~2.2.2}, separated into strong (QCD) and electroweak production (EWK) as well as the sum of both, and have been scaled by normalisation factors obtained in a fit to measure the electroweak production~\cite{Aaboud:2018ddq}.
  }
  \label{fig:atlas:wzjjunfold}
\end{figure}

The combined strong and electroweak production of $WZjj$, with the $W$ and $Z$ boson decaying to leptons, is measured in a fiducial phase space enriched in events from electroweak production~\cite{Aaboud:2018ddq}. Three leptons, missing transverse momentum induced by a neutrino, and two jets are required. Two of the leptons are required to have the same flavour, opposite electric charge, and an invariant mass consistent with a $Z$ boson, and the other, together with the neutrino, is required to form a transverse mass consistent with a $W$ boson.
Additional kinematic regions are used to constrain the background from $ZZjj$ and $t\bar tV$ production, and the strong production process.
As in the measurement of $W^\pm W^\pm jj$ production, a significant overestimation of the strong $WZjj$ production in simulation is observed and a normalisation factor of $0.56\pm0.16$ is obtained in the fit.
The combined electroweak and strong $WZjj$ production cross section is measured to be:

\begin{equation*}
  \sigma^{\text{fid.}}_{WZjj\text{-QCD+EW}} = 1.68
  \pm 0.16~\text{(stat.)}\pm 0.12~\text{(exp. syst.)}\pm 0.13~\text{(mod. syst.)}\pm 0.044~\text{(lumi.) fb}.
\end{equation*}
In addition, differential cross sections of the combined strong and electroweak production are measured. They allow comparisons of generator level studies of perturbative QCD, electroweak corrections or the presence of anomalous gauge couplings. The result depend only minimally on assumptions due to the classification into strong and electroweak production or their interference. As examples, the measured differential cross sections as a function of the $WZ$ transverse mass and the azimuthal separation of the two jets are shown in Figure~\ref{fig:atlas:wzjjunfold}. Differential cross sections for five other dijet or diboson quantities have also been measured.

\begin{figure}
  \centering
  \includegraphics[width=.44\textwidth]{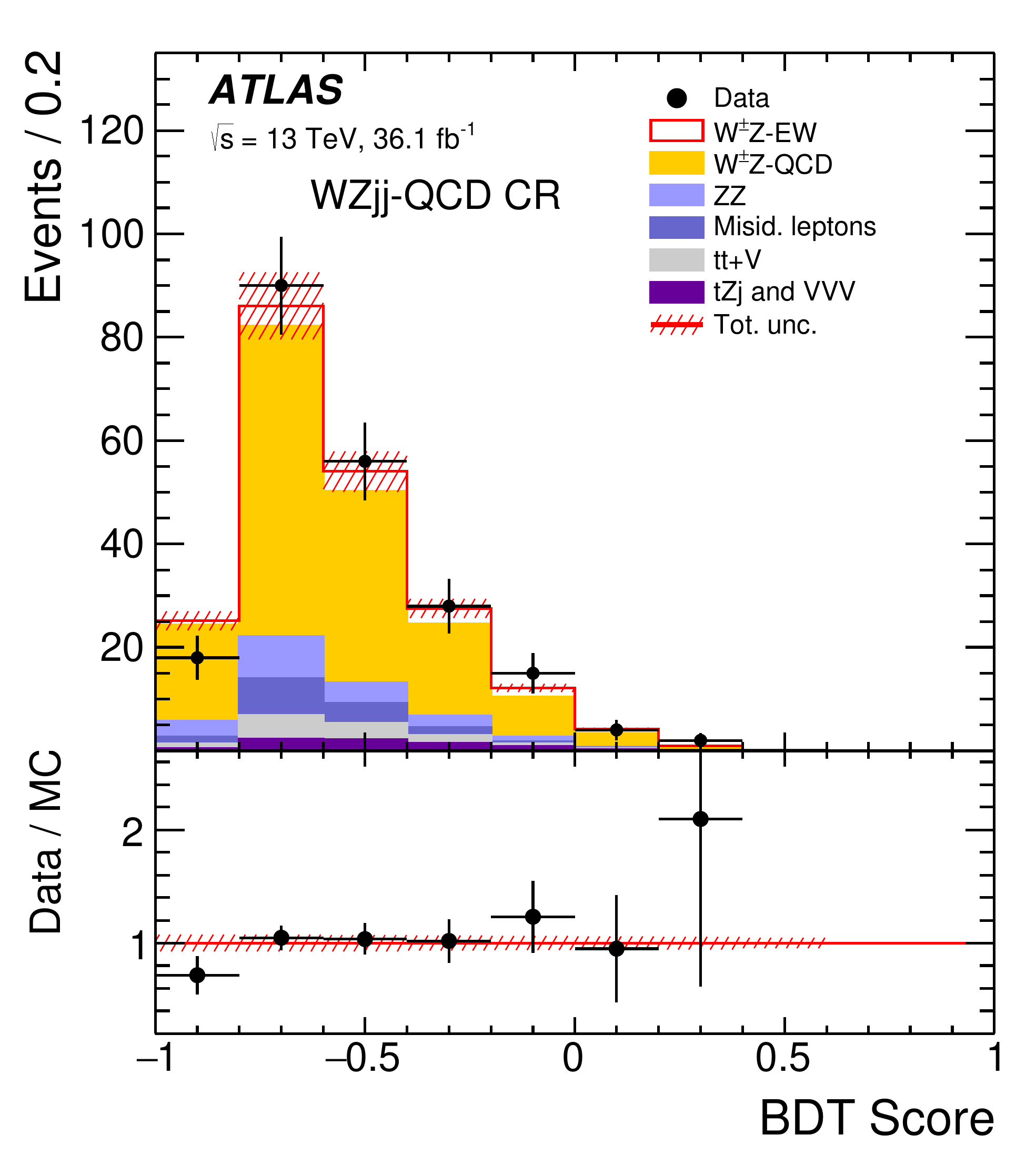}
  \includegraphics[width=.44\textwidth]{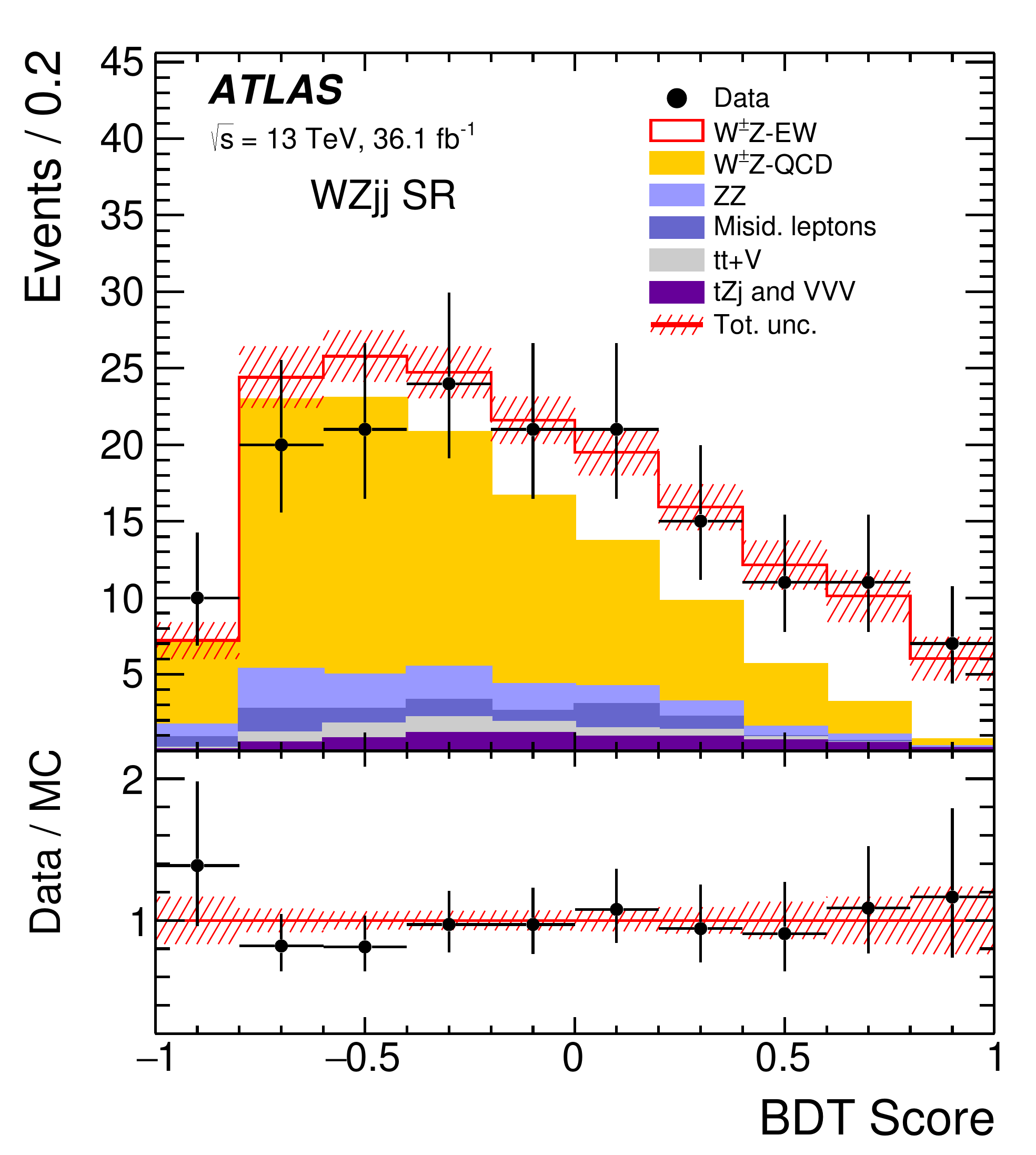}
  \caption{
    Distributions of the BDT discriminant used to extract the signal in the measurement of electroweak $WZjj$ production. The BDT is optimised using fifteen kinematic variables. The distribution is shown in a phase space dominated by strong $WZjj$ production (left) and in the fiducial phase space region (right). The events selected in data are shown with the estimated signal and background contributions. Systematic uncertainties are shown as a hatched band~\cite{Aaboud:2018ddq}.
  }
  \label{fig:atlas:wzjjbdt}
\end{figure}

The ratio of strong and electroweak production is predicted to be much larger in $WZjj$ than in $W^\pm W^\pm jj$ production. To separate the strong and electroweak production, a BDT discriminant is optimised using fifteen variables related to the kinematic properties of the jets, the diboson system and combined jet--diboson variables. 
It is shown in Figure~\ref{fig:atlas:wzjjbdt} in the region enriched in strong production and in the fiducial phase space. The simulation of the strong production is found to model the BDT discriminant well after the normalisation has been corrected.
The electroweak signal is extracted in a fit to the BDT discriminant and the background-only hypothesis is rejected with a significance of 5.3$\sigma$ where a significance of 3.2$\sigma$ are expected. The excess corresponds to a fiducial electroweak $WZjj$ cross section of:

\begin{equation*}
  \sigma^{\text{fid.}}_{WZjj\text{-EW}} = 0.57^{+0.14}_{-0.13}~\text{(stat.)}^{+0.05}_{-0.04}~\text{(exp.\ syst.)}^{+0.05}_{-0.04}~\text{(mod.\ syst.)}\pm 0.01~\text{(lumi.)~fb}
\end{equation*}
compared to  $0.32\pm0.03$~fb predicted by \texttt{Sherpa~2.2.2}.

Since the simulation of the electroweak signal process does not include additional partons in the matrix element, the analysis is more sensitive to the effects of the colour flow configuration in \texttt{Sherpa~2.2.2}. The difference of the BDT distribution simulated with \texttt{Sherpa~2.2.2} and \texttt{MG5\_aMC@NLO}~\cite{Alwall:2014hca} is therefore included as a systematic uncertainty in the modelling of the simulated signal events in the measurement.

\subsection{Search for electroweak diboson production in semileptonic decays}

Analyses of the semileptonic decay of gauge boson pairs reach higher event yields and a higher energy compared to the fully leptonic decay. The study of such decays for a diboson system produced in association with two additional jets have been conducted as well, where the leptonically decaying gauge boson is reconstructed from two neutrinos ($Z\rightarrow \nu\nu$), one charged lepton and one neutrino ($W\rightarrow \ell\nu$), or two charged leptons ($Z\rightarrow \ell\ell$)~\cite{Aad:2019xxo}. The hadronically decaying gauge boson is reconstructed from either one jet reconstructed with the anti-$k_t$ algorithm and radius parameter $R=1.0$, or two jets reconstructed with radius parameter $R=0.4$. Jets reconstructed with $R=1.0$ are further classified into high- and low-purity candidates using jet substructure variables. If no such jet can be found, two $R=0.4$ jets are selected instead. In addition, two high-\pt\ jets reconstructed with $R=0.4$ produced in association with the diboson system are required in all categories.

\begin{figure}
  \centering
  \subfigure[]{
    \includegraphics[width=.35\textwidth]{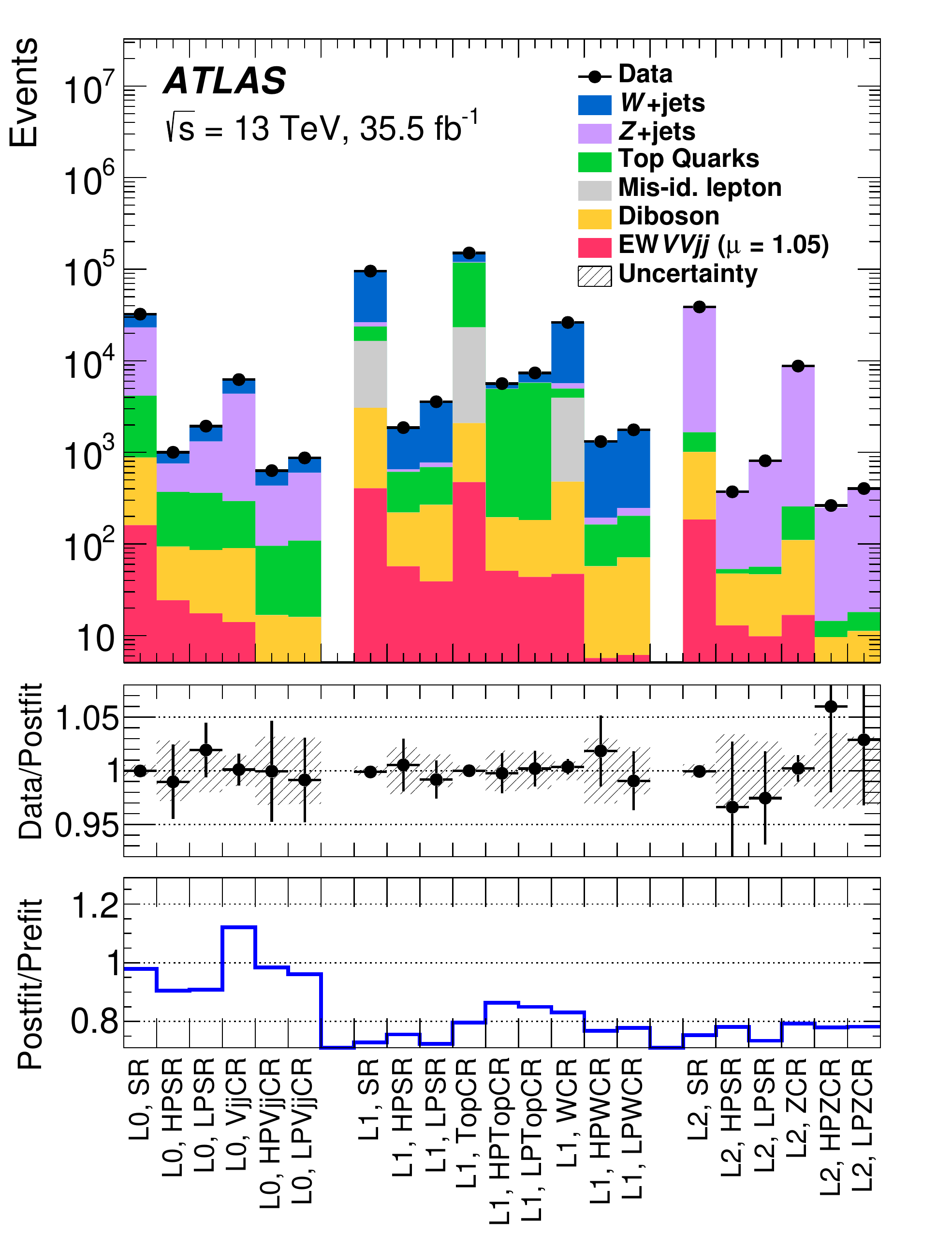}
    \label{fig:atlas:lnjjjjregions}
  }
  \raisebox{0.1\height}{
    \subfigure[]{
      \includegraphics[width=.55\textwidth]{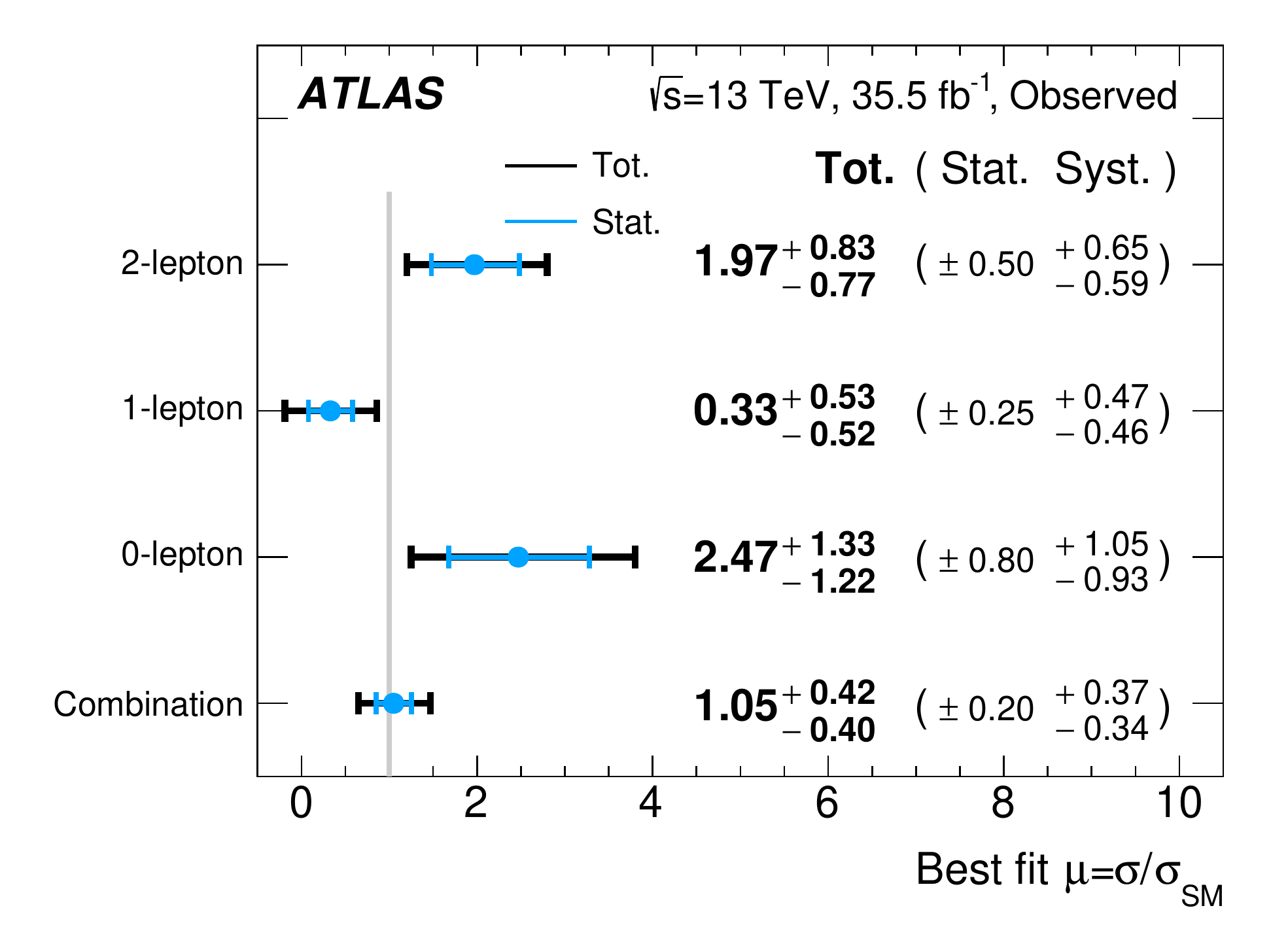}
      \label{fig:atlas:lnjjjjmu}
    }
  }
  \caption{
    (a) Event yields  in 21 kinematic phase space regions used in the analyses of semileptonic $VVjj$ production. The selected data are shown with the estimated  contributions from background and the electroweak $VVjj$ signal. The ratio of the yield in data and the sum of the estimated signal and background contributions, and the sum of the signal and background yields estimated in the fit and predicted from simulation  are shown in the lower pads. The labels L0, L1 and L2 on the  $x$-axis indicate the number of leptons selected and whether the region is primarily defined to extract the signal (SR) or constrain the background (CR)~\cite{Aad:2019xxo}. (b) Ratio of the measured and theoretical cross sections in 0-lepton, 1-lepton and 2-lepton final states, and for the combination of the three final state~\cite{Aad:2019xxo}. The theoretical prediction is obtained from \texttt{MG5\_aMC@NLO}.
  }
\end{figure}

The primary source of background in these events is $V$+jets production where a single gauge-boson is produced in association with jets from initial state radiation. Another large source of background is top quark production. 
The $V$+jets background contributions are constrained in dedicated kinematic regions, one for each of the nine signal regions. Background from top quark production can be significant for 1-lepton final states since top quark pair production involves two $W$ bosons. A dedicated control sample is defined for this background, as well. The data events observed in every signal category and every control sample are shown in Figure~\ref{fig:atlas:lnjjjjregions} with the estimated background contributions.
Respectively 0.5\%, 1\% and 3\% of the selected data are expected to originate from electroweak $VVjj$ production if two $R=0.4$ jets, a low-purity $R=1.0$ jet or a high-purity $R=1.0$ jet are selected.

The signal is extracted in a fit to BDT discriminants where a separate BDT is optimised for every combination of lepton final state, and for jets reconstructed with $R=1.0$ and $R=0.4$.
If the vector boson is reconstructed as a single jet with $R=1.0$, the BDT discriminant is optimised using nine variables, four variables and eight variables in 0-lepton, 1-lepton, and 2-lepton final states, respectively. In case the vector boson is reconstructed as two jets with $R=0.4$, a larger number of variables is used to isolate the electroweak signal, with 13, 16 and 16 variables  used in 0-lepton, 1-lepton, and 2-lepton final states, respectively. Differently from the BDT used for the $WZjj$  study discussed in Section~\ref{sec:wzjj}, many variables related to the kinematics of individual leptons and jets are used, as well as variables related to the properties of the $R=1.0$ and $R=0.4$ jets. In particular,  variables discriminating quark and gluon jets are used to identify $V\rightarrow qq$ decays.

Distributions from all 21 different kinematic regions, shown in Figure~\ref{fig:atlas:lnjjjjregions}, are combined in the fit.
The background-only hypothesis is rejected with a significance of 2.7$\sigma$ where a significance of 2.5$\sigma$ is expected according to the theoretical prediction from \texttt{MG5\_aMC@NLO}. A cross section of:

\begin{equation*}
  \sigma^{\text{fid.}}_{VV jj\text{-EW}} = 45.1\pm8.6~\text{(stat.)}^{+15.9}_{-14.6}~\text{(syst.)~fb}
\end{equation*}
is measured in the combined fiducial phase space region, in agreement with  $43.0\pm 2.4$~fb predicted by theory. Fiducial  cross sections are also reported for each final state individually. The ratio of the measured and predicted cross sections for 0-lepton, 1-lepton and 2-lepton final states is shown in Figure~\ref{fig:atlas:lnjjjjmu}. The largest individual sources of systematic uncertainties are related to the estimation of the $V$+jets backgrounds. There is the possibility to improve some of these uncertainties in future measurements on a larger dataset.

\subsection{Conclusions}

Recent studies of vector-boson scattering in $pp$ collision data, recorded in the years 2015 and 2016 with the ATLAS experiment at the LHC and corresponding to an integrated luminosity of $36.1~\text{fb}^{-1}$, have been presented. The electroweak production of two electroweak gauge bosons in association with two jets was observed in the $W^\pm W^\pm jj$ and $WZjj$ final states. These are amongst the processes with the lowest cross section  measured at the LHC to date. Subsequently, fiducial cross sections have been measured. Results for the combined strong and electroweak $WZjj$ production are presented in a phase space enriched in VBS events, including differential cross sections. Electroweak diboson production was also studied in the semileptonic decay mode and the background-only hypothesis was rejected with a significance of 2.7$\sigma$. 
The reported measurements used a fraction of the full run-2 $pp$ data amounting to $139~\text{fb}^{-1}$. Exploiting the full dataset  will allow for more detailed studies of vector-boson scattering, e.g.\  by making additional, more elusive final states experimentally accessible. 

\section{Recent CMS results in VBS\footnote{speaker: R.~Bellan}}

\subsection{Introduction}

The CMS Collaboration put in place a large effort to cover all channels sensitive to the vector boson scattering (VBS). Arriving to the observation of the VBS is a long journey, but it has finally started and the first milestones of the path have been posed: the first measurement and observation of the electroweak production of vector bosons and jets. However, we just scratched the surface that covers the processes which are sensitive to the most intimate part of the electroweak symmetry breaking (EWSB), the acquisition of the longitudinal degree of polarization of the massive electroweak bosons.

\begin{figure}[th]
  \centerline{
    \raisebox{-0.5\height}{\includegraphics[width=400pt]{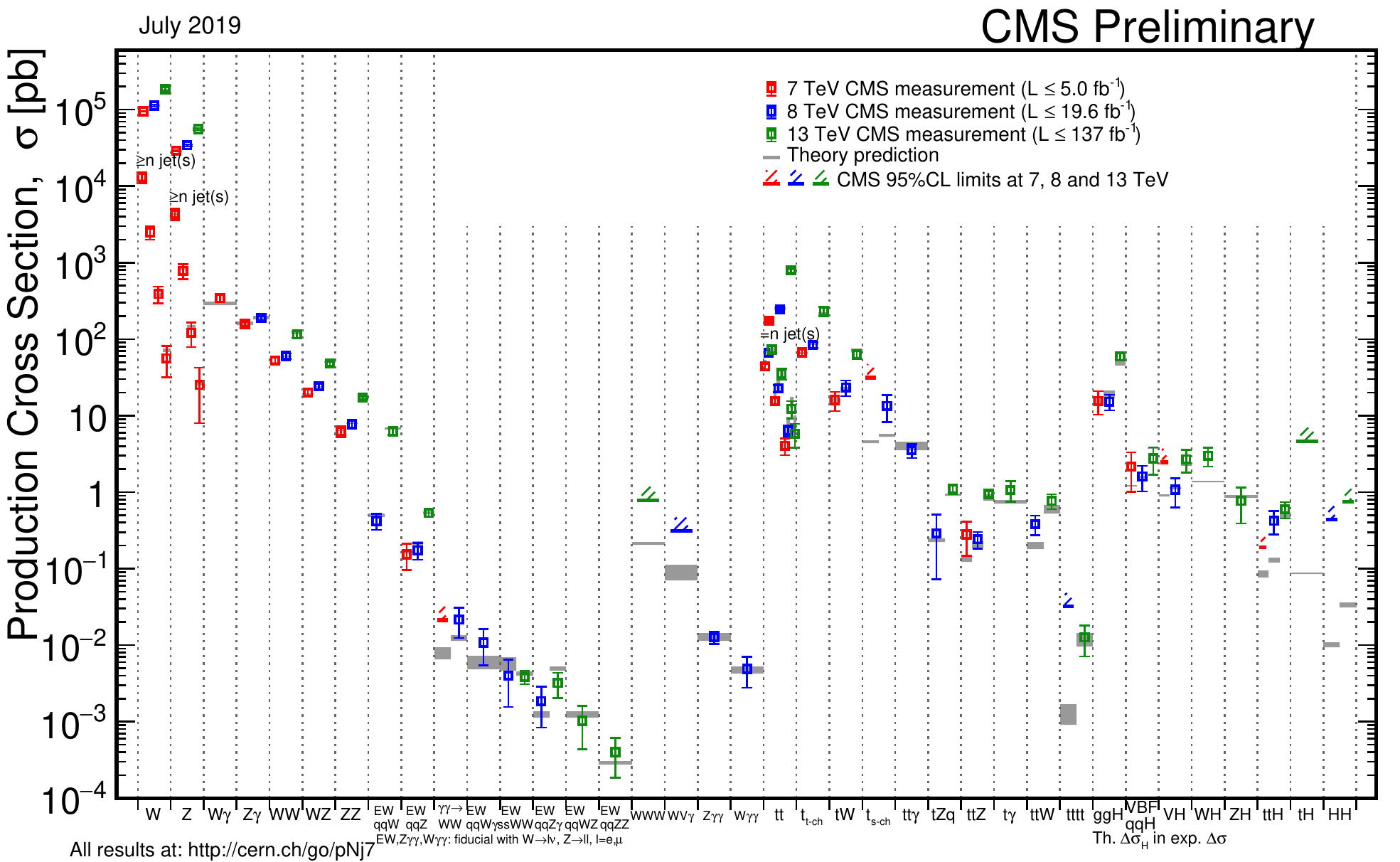}}
  }
  \caption{\label{sigma}
    CMS cross section measurements summary~\cite{SMP}.
  }
\end{figure}

I presented three analyses that targeted the observation of the electroweak production of dibosons in association with jets, pp$\rightarrow$ZZjj$\rightarrow 4\ell$jj~\cite{VBSZZ},
pp$\rightarrow \mathrm{W}^{\pm} \mathrm{W}^{\pm}$jj$\rightarrow \ell^{\pm} \ell^{\pm} \nu \nu$jj~\cite{VBSWW}, and pp$\rightarrow$WZjj$\rightarrow 3\ell \nu$jj~\cite{VBSWZ}, and an analysis searching the evidence of anomalous quartic gauge couplings, using the channels pp$\rightarrow$ZVjj$\rightarrow 2\ell$jjjj and pp$\rightarrow$WVjj$\rightarrow \ell \nu$jjjj~\cite{aQGCZVWV}.
In all cases, $\ell = \mathrm{e}, \mu$ and the data set corresponds to an integrated luminosity of 35.9 fb$^{-1}$ (LHC Run II, 2016 data only).
The measured cross sections are among the smallest measured so far in a collider experiment (Fig.~\ref{sigma}). 
The description of the CMS detector can be found in Ref.~\cite{CMS}.

\subsection{Search for VBS in fully leptonic ZZ+jets final state}

\begin{figure}[th]
  \centerline{
    \raisebox{-0.5\height}{\includegraphics{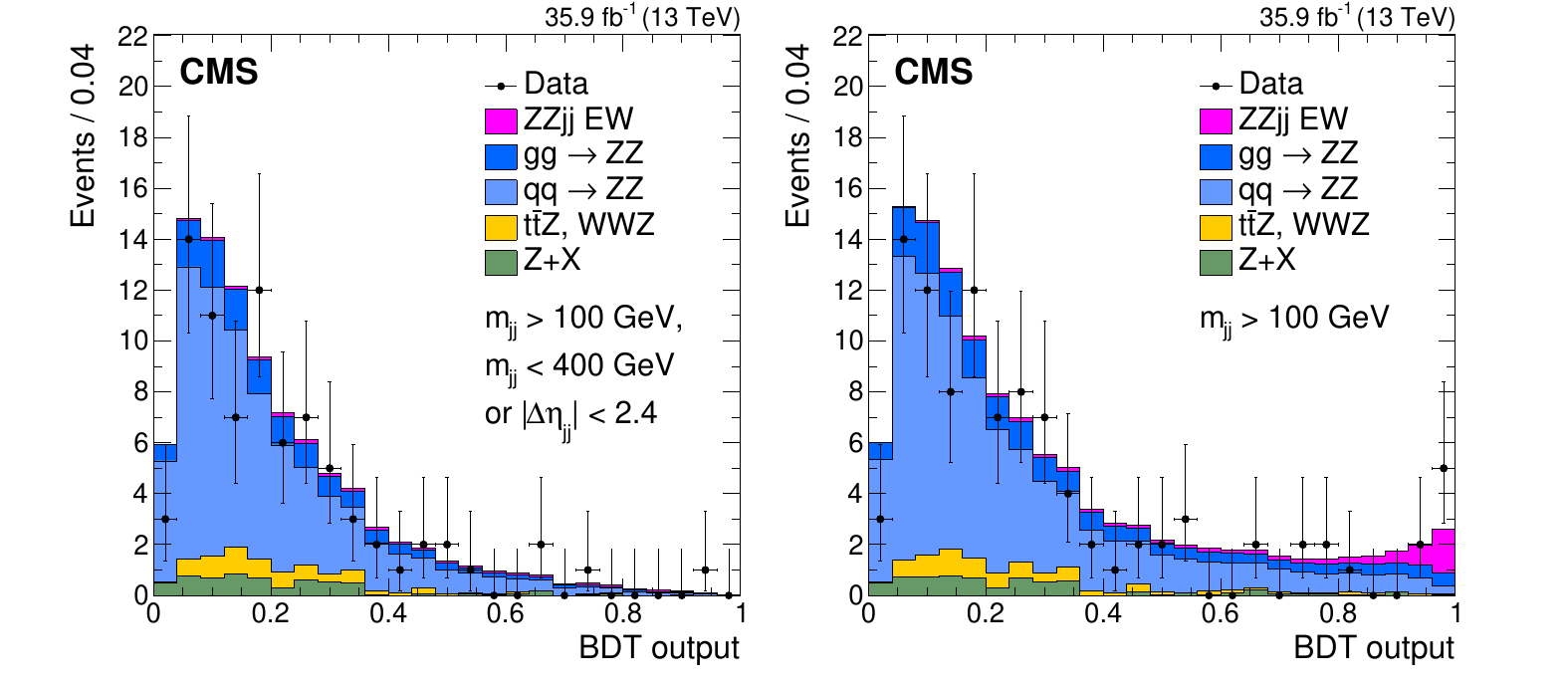}}
  }
  \caption{\label{fig1}
    Distribution of the BDT output in the control region obtained by selecting ZZjj events with $m_{jj}$ < 400 GeV or $|\Delta\eta_{jj}|$ < 2.4 (left) and for the full search region (right)~\cite{VBSZZ}. Points represent the data, filled histograms the expected signal and background contributions. 
}
\end{figure}

In this analysis~\cite{VBSZZ} we search for two pairs of same-flavor opposite-sign leptons that have an invariant mass between 60 and 120 GeV. The leptons must be isolated, prompt and satisfy minimal kinematic cuts: $p_T > 7$~GeV and $|\eta| < 2.5$, or $p_T > 5$~GeV and $|\eta| < 2.4$, for electrons and muons, respectively. Events are considered only if there are at least
two jets (reconstructed with the anti-$k_T$ algorithm~\cite{Cacciari:2008gp}, with a distance parameter of 0.4) with a transverse momentum larger than 30 GeV.
This is the analysis with the cleanest experimental signature of the set: the instrumental background, where jets are misidentified as leptons, is very low, and all kinematic variables can be directly measured. However, the total cross section is very low compared both to the other VBS channels and, most importantly, to the processes that produce two Z bosons and two jets via diagrams of order $\alpha_s^2$. Understanding this background is therefore paramount. CMS developed a dedicated analysis~\cite{ZZjets} to measure differential cross sections in the observables related to the hadronic properties of the process.
To enhance the sensitivity of the analysis, the signal is extracted using a boosted decision trees technique, training it with seven variables. This results in an observed significance of 2.7 standard deviations ($\sigma$), with an expectation of 1.6 $\sigma$. Figure~\ref{fig1} shows the output of the boosted decision tree (BDT) used to extract the signal in a QCD-induced jets control region (left) and in the full search region (right).

\subsection{Search for VBS in fully leptonic W$^\pm$W$^\pm$+jets final state}

\begin{figure}[th]
  \centerline{
    \raisebox{-0.5\height}{\includegraphics[width=540pt]{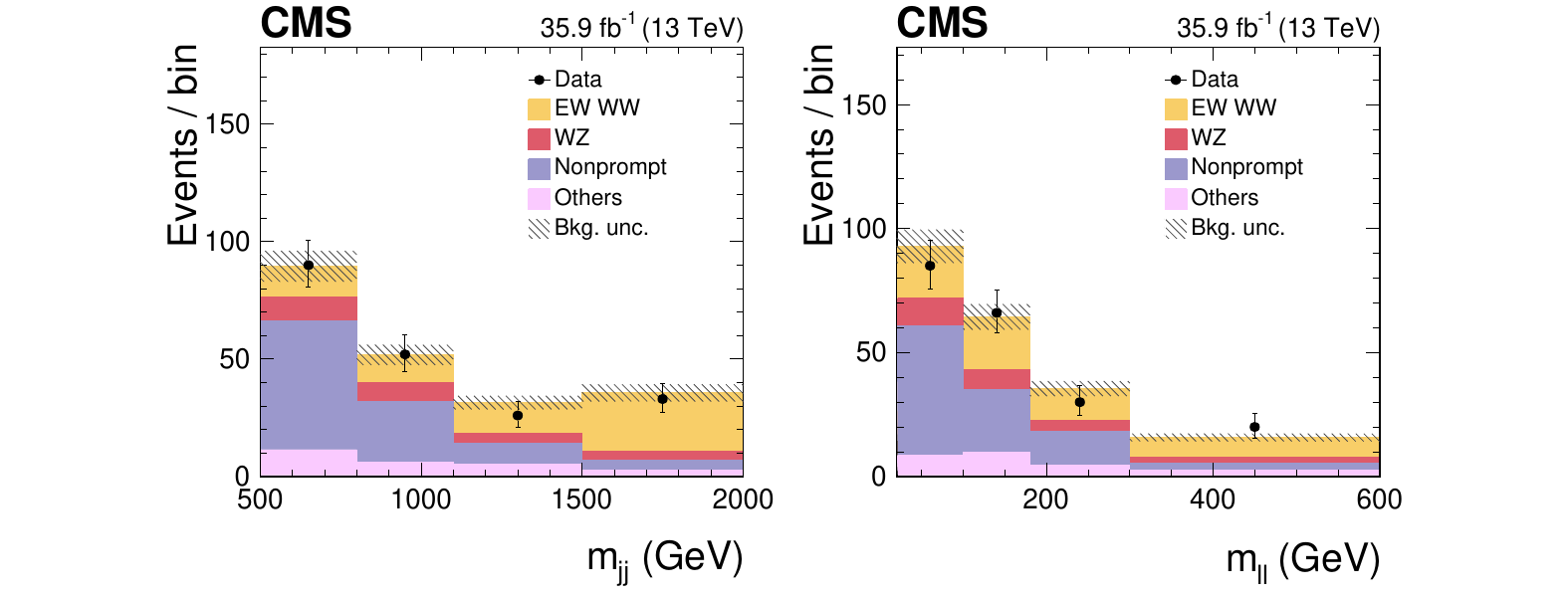}}
  }
  \caption{\label{fig2}
Distributions of $m_{jj}$ (left) and $m_{ll}$ (right) in the signal region of the W$^\pm$W$^\pm$~\cite{VBSWW}. The normalization of the EW W$^\pm$W$^\pm$jj and background distributions corresponds to the result of the fit. The hatched bands include statistical and systematic uncertainties from the predicted yields.
  }
\end{figure}

This is the first analysis that observed the electroweak production of two electroweak vector bosons in association with two jets~\cite{VBSWW}. The analysis strategy is based on the search for two charged leptons with a transverse momentum of at least 25~GeV (leading) and 20 GeV (sub-leading). Events are considered if there are two jets with at least 30~GeV of $p_T$, the two highest-$p_T$ jets form an invariant mass larger than 500~GeV, $\lvert\Delta\eta_{jj}\rvert>2.5$, and for each lepton $Z^*_l = \lvert\eta_l - (\eta_{jet,1} + \eta_{jet,2})/2\rvert/\lvert\Delta\eta_{jj}\rvert < 0.75$. Events with additional leptons are vetoed, as well as events with less than 40~GeV of missing transverse energy.
Despite the presence of neutrinos escaping detection, the channel is very clean, because the final state with two same-sign charged leptons is rare in the standard model. The signal is extracted via a 2D-template fit of the invariant masses of the dijet and the dilepton systems. The final result brought to the observation of the process, with an observed significance of 5.5~$\sigma$ (5.7~$\sigma$ expected). Figure~\ref{fig2} shows the shape of $m_{jj}$ and $m_{ll}$ observables.

\subsection{Search for VBS in fully leptonic WZ+jets final state}

\begin{figure}[b]
  \centerline{
    \raisebox{-0.5\height}{\includegraphics[width=300pt]{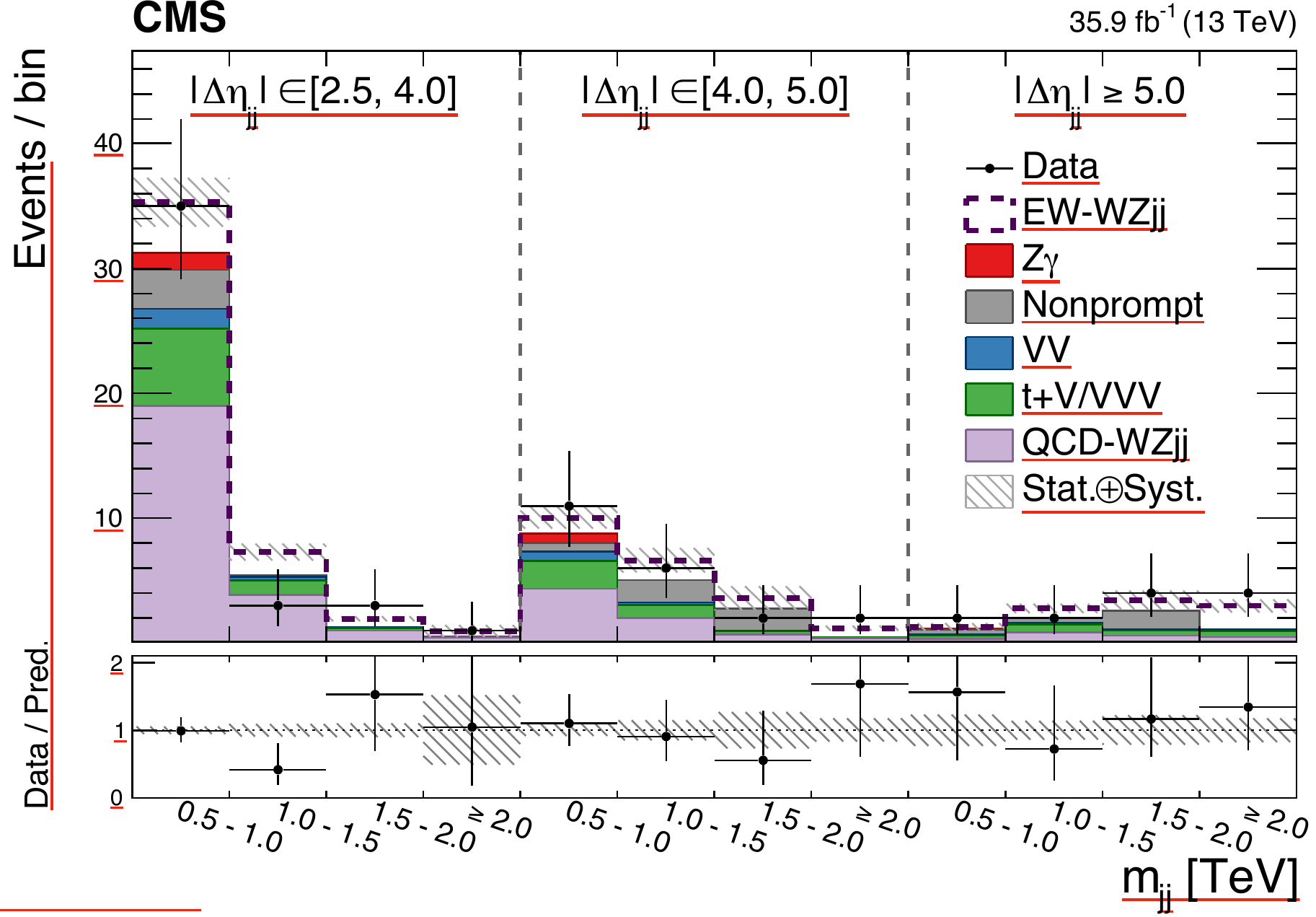}}
  }
  \caption{\label{fig3}
    The one-dimensional representation of the 2D distribution of $m_{jj}$ and $|\Delta\eta_{jj}|$, used for the EW signal extraction~\cite{VBSWZ}. The x axis shows the $m_{jj}$ distribution in the indicated bins,
    split into three bins of $|\Delta\eta_{jj}|$: [2.5, 4], [4, 5], $\geq$ 5. The dashed line represents the EW WZ+jets contribution stacked on top of the backgrounds that are shown as filled histograms. The hatched bands represent the total and relative systematic uncertainties on the predicted yields. The bottom panel shows the ratio of the number of events measured in data to the total number of expected events. The predicted yields are shown with their best-fit normalizations. 
  }
\end{figure}

In this analysis~\cite{VBSWZ} we search for exactly three charged leptons with a $p_T$ larger than 25 and 15 GeV, if the leptons have been associated to the decay of the Z boson, or larger than 20 GeV if the lepton has been judged to come from the W decay. Events are accepted if there are at least two jets with 50 GeV of $p_T$, the two highest-$p_T$ jets form an invariant mass larger than 500 GeV, $|\Delta\eta_{jj}|>2.5$, and $\eta^*_{3l} = |\eta_{3l} - (\eta_{jet,1} + \eta_{jet,2})/2| < 2.5$. In the analysis we both measure the QCD+EW contribution to the WZ+jets cross section and the EW only component, which is approximately 38\% of the previous. The extraction of the EW cross section only is done via a 2D-template fit of $m_{jj}$ and $\Delta\eta_{jj}$. The observed significance is 2.2 $\sigma$, with an expected significance of 2.5 $\sigma$. Figure~\ref{fig3} shows the post-fit yields in the signal region in bins of $m_{jj}$ and $|\Delta\eta_{jj}|$.

\subsection{Search for aQGC in semileptonic ZV and WV + jets final state}

\begin{figure}[th]
  \centerline{
    \raisebox{-0.5\height}{\includegraphics[width=400pt]{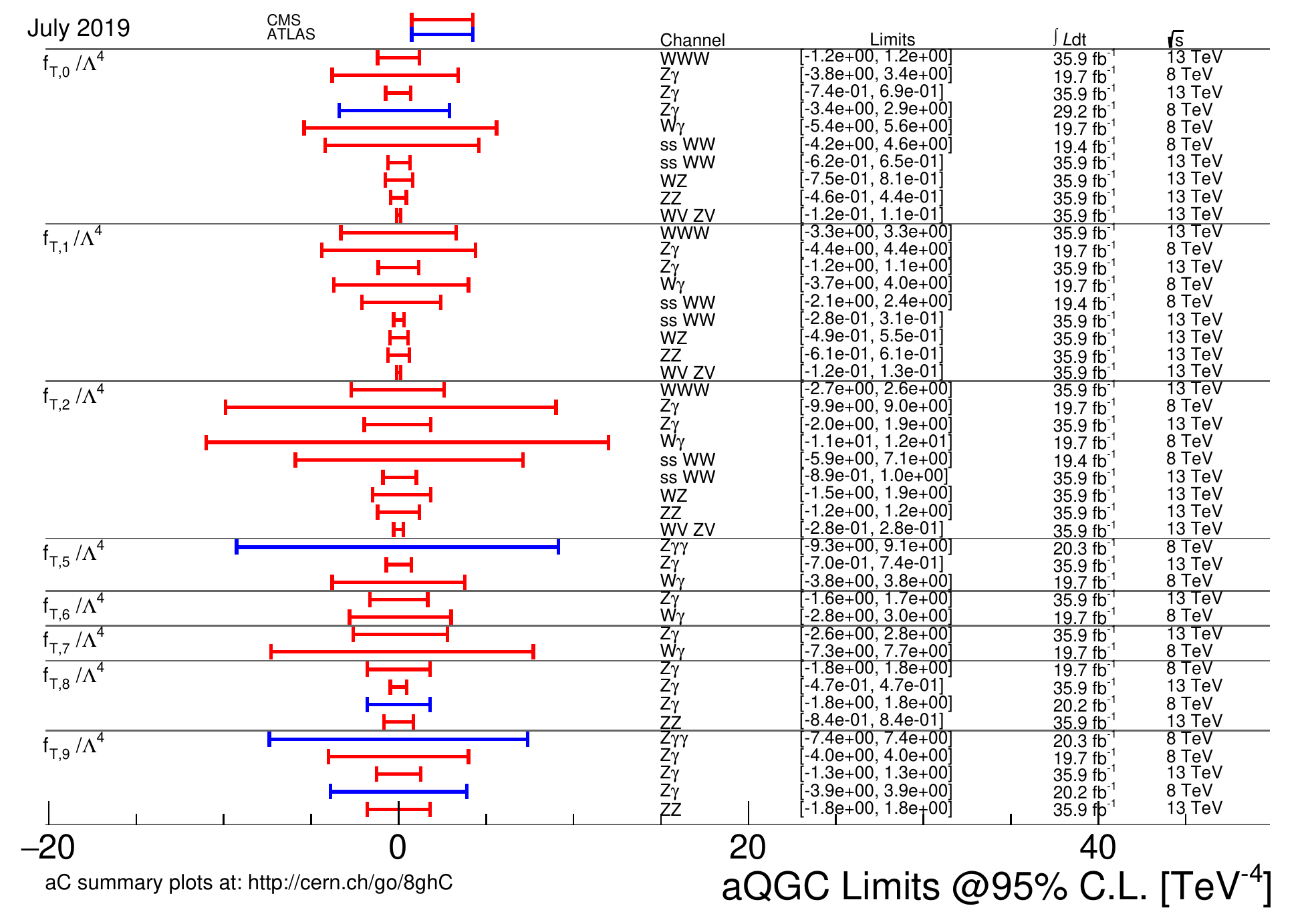}}
  }
  \caption{\label{aQGC}
    95\% confidence level limits on dimension-8 transverse parameters from the ATLAS and CMS Collaborations~\cite{SMP}.
  }
\end{figure}

The CMS Collaboration designed a dedicated analysis~\cite{aQGCZVWV} to search for anomalous quartic gauge couplings. In the analysis we make use of aggressive cuts to highly suppress the background. The hadronically decaying vector boson is reconstructed only in a boosted topology, where the product of the decay are reconstructed as a unique jet, using the anti-$k_T$ algorithm with a distance parameter R=0.8~\cite{aQGCZVWV}. The tag jets, taken as the two which form the dijet system with the highest invariant mass, are required to have $m_{jj}$ larger than 800 GeV, $|\Delta\eta_{jj}|>4$. A detailed description of the cuts can be found in Ref.~\cite{aQGCZVWV}.
The results of the analysis does not show any excess with respect to the standard model prediction, thus the most stringent limits, so far, on dimension-8 operators in an effective field theory framework~\cite{EFT1,EFT2} have been placed.
Figure~\ref{aQGC} shows the summary plots for the anomalous quartic gauge couplings search.

\subsection{Conclusions}

The CMS Collaboration explored several VBS-like final states using 2016 LHC data. So far, we have observed the electroweak production of two same-sign W and two jets,
an hint of the production of the ZZ+jets and WZ+jets through electroweak processes.
The phase space used to  evaluate the fiducial cross sections is not homogeneous among all the analyses; on this point the VBSCan community should give a clear prescription
on how to define the fiducial regions, which should serve as a common ground for all VBS analyses of the ATLAS and CMS Collaborations.

Detailed results on VBS with the CMS detector can be found in Ref.~\cite{SMP}.

\chapter{Experimental Techniques}
\label{WG3}

\section{EFT Combination status report\footnote{speaker: D.~Sampsonidou}}

The Effective Field Theories emerge as the tool to look for deviations
from the Standard Model (SM). They are a low energy parametrization for unknown physics that can become reachable at very high energy. The experimental approach is to associate EFT Dimension-6 and Dimension-8 operators to vertices in form of anomalous couplings, that can be Triple Gauge Couplings (TGCs) or Quartic Gauge Couplings (QGCs).
So far, there is no theoretical model that includes both Dimension-6 and Dimension-8 operators, the aim is to study individually the effect of Dimension-6 and Dimension-8 operators. In both cases, the EFT parameters limits will be combined between various experimental signatures, between various operators and finally between ATLAS and CMS experiments.

In order to extract combined limits, the strategy is to use published data from ATLAS and CMS (HEPData entries) starting from fully leptonic channels, produce EFT Monte Carlo (MC) predictions and proceed to the reinterpretation of the data by setting limits and combining the results.  

\subsection*{Current status of Dimension-6 Operators Combination Plan}
The model that is used for the Dimension-6 operators is SMEFTSim. The  Dimension-6 operators' coefficients can be constrained by measuring diboson and VBS processes. The first relevant operators that are tested are the following: $C_{HD}, C_{H\Box}, C_{W}, C_{HB}, C_{HW}$, $C_{HWB}, C_{ll}, {C_{Hl}}^{(1)},{C_{Hl}}^{(3)}$,${C_{Hq}}^{(1)}$,\\${C_{Hq}}^{(3)}$, $C_{He}, C_{Hu}, C_{Hd}$.
The initial step is to define the experimental signatures that are going to be used. The first candidates are the diboson and VBS channels: ZZjj, WZjj, ssWWjj, osWW. The second step is to define the operators that can be constrained in each process. Thus, sensitivity studies in various experimental signatures and phase spaces need to be performed. Since there are no older limits in the Warsaw basis, the sensitivity should be defined by testing the effect of the change of the EFT parameters on the cross section and by testing the kinematics' shapes sensitivity to the change of the EFT parameters. A test on the WZjj sensitivity has been performed, by generating 1000 events using SMEFTSim, with the U(3)5 flavour symmetric model and the $\alpha$ scheme. In order to test the effect of each EFT parameter on the SM prediction, all operators are set to zero but the one under study that is set to 3.

\begin{figure}[h]
\centering
\includegraphics[scale=0.6]{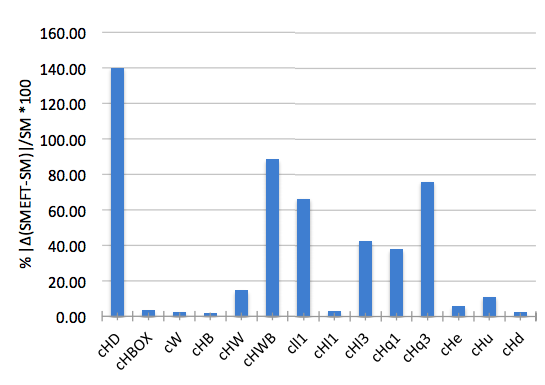}
\caption{Effect of the EFT coefficients to the SM cross section.}
\label{fig-1}       
\end{figure}

\subsection*{Current status of Dimension-8 Operators Combination Plan}
The aQGCs can be parametrized in terms of Dimension-8 operators, by the assumption that the Dimension-6 can already be constrained elsewhere. The EFT parameterization from Eboli, Gonzales-Garcia models. The goal is to combine Dimension-8 EFT parameters limits
in ATLAS and CMS across various VBS channels, by using only published data. In order to
restore unitarity at large $\sqrt{s}$, the clipping method will be used, by setting the anomalous signal contribution to zero for $\sqrt{s}> E_{c}$, while the data and the background contributions remain unchanged. The steps to be taken now are the Monte Carlo production and the limit setting and combination of the results. The Dimension-6 operators' effect on the  Dimension-8 Lagrangian should also be taken into account.
\\\\
In summary, there is an ongoing effort for the EFT parameters limits' combination for Dimension-6 and Dimension-8 operators. The sensitivities of the operators have been tested and the plan is to move on to the EFT MC production and finally the statistical interpretation of the results.

\newcommand{\dimsix}{$\mathrm{dim.}\, 6\,$\,}
\newcommand{\dimeight}{$\mathrm{dim.}\, 8\,$}
\newcommand{\op}{\mathcal{O}}
\newcommand{\amp}{\mathcal{A}}

\section{Dimension-6 EFT for electroweak analyses\footnote{speaker: R.~Gomez-Ambrosio}}

\subsection{Introduction} 
The main goals of the Effective Field Theory (EFT) effort within the WG3 are to: 

\begin{itemize}
    \item Provide dimension-6 (\dimsix) parametrisations for the family of VBS signal and background processes (single boson and diboson productions, vector boson fusion). 
    \item Provide a clear recommendation for other groups to be able to generate and compare with their own productions.
    \item Identify the most interesting operators and bins for the search of \dimsix effects.
    \item Integrate these studies with the LHC electroweak working group (LHC-EWWG). In order to be able to compare these leading order (LO) EFT predictions with the set of next-to-leading (NLO) SM predictions for the electroweak (EW) sector.  
    
\end{itemize}

\subsection{EFT basis} \label{sec:relevantEFToperators}
The basis adopted for this study is the so-called ``Warsaw basis'' \cite{Grzadkowski:2010es}. In particular we adopt the operator classification and labeling used in \cite{Jenkins:2013zja}. 

\begin{itemize}
    \item Operators affecting triple and quartic gauge couplings directly (classes 1 and 4): 
    \begin{align}
        & \mathcal{O}_{W} = \epsilon_{IJK} W_{\mu}^{\nu I} W_{\rho}^{\mu J} W_{\nu}^{\rho K} \nonumber \\
        & \mathcal{O}_{HW} = H^{\dagger}H \, W_{\mu \nu}^{I} W^{\mu \nu I} \nonumber \\ 
        & \mathcal{O}_{HWB} = H^{\dagger} \tau^I H \, W_{\mu \nu}^{I} B^{\mu \nu} \nonumber
    \end{align}
    as well as their CP-odd counterparts, $\lbrace  \mathcal{O}_{\widetilde{W}}, \mathcal{O}_{\widetilde{HW}},  \mathcal{O}_{\widetilde{HWB}},  \mathcal{O}_{\widetilde{HW}} \rbrace $
    \item Operators affecting gauge-quark and gauge-lepton vertices 
     \begin{itemize}
         \item Class 6 (dipole operators): $$ \lbrace \op_{eW}, \op_{eB}, \op_{uW}, \op_{uB}, \op_{dW}, \op_{dB} \rbrace $$
	 They all contribute to the gauge-fermion vertices, but due to their tensor structure they don't interfere with the SM, only with the EFT. These operators should be taken into account when generating \dimsix quadratic terms or comparing with dimension-8 (\dimeight) predictions.
         \item Class 7: 
         
         $$ \lbrace \op_{H \ell}^{(1)} , \op_{H \ell}^{(3)}, \op_{He} , \op_{Hq}^{(1)}, \op_{Hq}^{(3)}, \op_{Hu}, \op_{Hd}, \op_{Hud}  \rbrace $$ 
         
         All of the class-7 operators contribute in one way or another. Either through the gauge-lepton coupling in the final decay process, or through the gauge-quark couplings that lead to the vector boson scattering interaction. 
     \end{itemize}
     \item Four-fermion operators
     \begin{itemize}
         \item  The four-fermion operators are likely to play an important role in this process, since it has two quarks in the initial state and two more in the final state, as well as four leptons.  
     \end{itemize}
\end{itemize}

\begin{figure}[h]
    \centering
    \begin{minipage}{0.33\textwidth}
        \centering
        \includegraphics[width=0.7\textwidth]{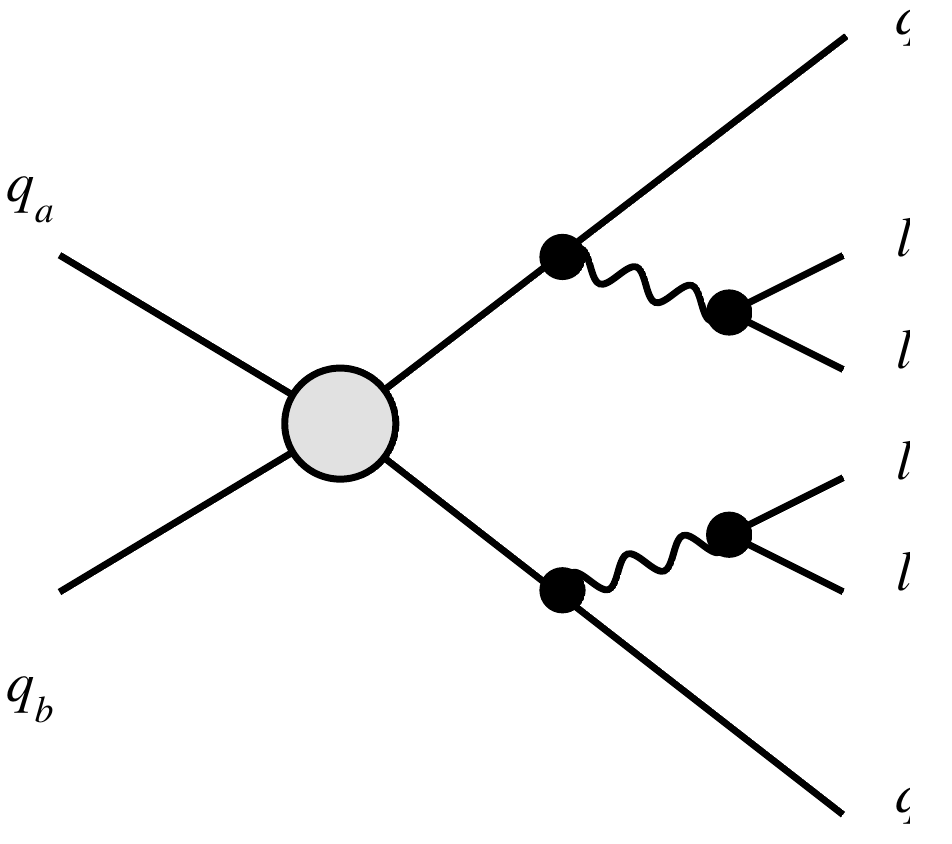} 
    \end{minipage}\hfill
    \begin{minipage}{0.33\textwidth}
        \centering
        \includegraphics[width=0.7\textwidth]{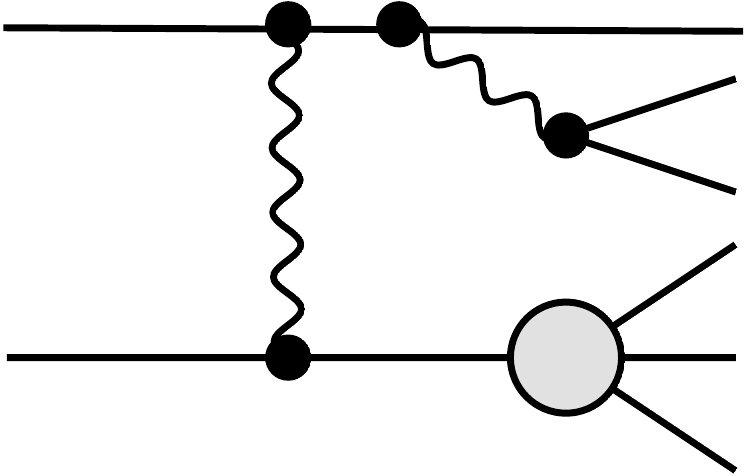} 
    \end{minipage}
        \begin{minipage}{0.33\textwidth}
        \centering
        \includegraphics[width=0.7\textwidth]{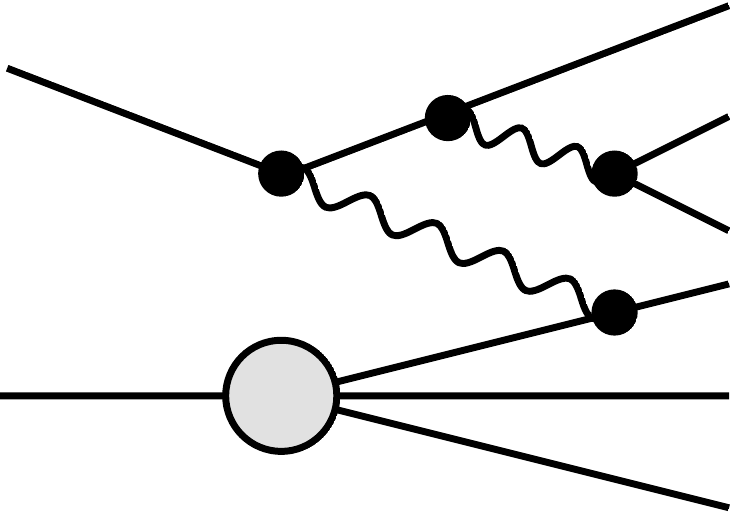} 
    \end{minipage}
\caption{Some examples of four-fermion vertices entering the VBS process.}
\end{figure}


\subsection{Building the amplitude}

Part of the combination effort relies on a precise definition of the EFT amplitudes and cross sections 

\begin{equation}
    \sigma_{EFT} = (\amp_{EFT}^*) \times (\amp_{EFT})
\end{equation}
where $\amp_{EFT}$ has to be defined, for example for two \dimsix operators $\lbrace \op_1^{(6)}, \op_2^{(6)} \rbrace $, 
\begin{equation} \nonumber
    \amp_{EFT}^{(6)} = \amp_{SM} + \frac{c_1^{(6)}}{\Lambda^2} (\amp_{1,6}) +  \frac{c_2^{(6)}}{\Lambda^2} (\amp_{2,6})
\end{equation}
if we further add $\mathrm{dim. 8}$ operators, $\lbrace \op_1^{(8)}, \op_2^{(8)} \rbrace $,
\begin{equation} \nonumber
    \amp_{EFT}^{(8)} = \amp_{EFT}^{(6)} + \frac{c_1^{(8)}}{\Lambda^4} (\amp_{1,8}) +  \frac{c_2^{(8)}}{\Lambda^4} (\amp_{2,8})
\end{equation}
putting all pieces together, 
\begin{align} 
    |\amp_{EFT}|^2 = & |\amp_{SM}|^2 + 
    \frac{c_1^{(6)}}{\Lambda^2} 2  \left( \amp_{1,6}^* \amp_{SM} \right) +
    \frac{c_2^{(6)}}{\Lambda^2} 2  \left( \amp_{2,6}^* \amp_{SM} \right) + \\ &
    + \frac{|c_1^{(6)}|^2}{\Lambda^4} |\amp_{1,6}|^2  + \frac{|c_2^{(6)}|^2}{\Lambda^4} |\amp_{2,6}|^2  + 
    + \frac{c_1^{(8)}}{\Lambda^4} 2\left(  \amp_{1,8}^* \amp_{SM} \right) +  \nonumber \\ & 
  \qquad +       \frac{c_2^{(8)}}{\Lambda^4} 2 \left( \amp_{2,8}^* \amp_{SM} \right) +  \frac{|c_1^{(8)}|^2}{\Lambda^8} |\amp_{1,8}|^2  + \frac{|c_2^{(8)}|^2}{\Lambda^8} |\amp_{2,8}|^2 
       \nonumber
\end{align}
in case the $ \lbrace c^{(6)} \rbrace$ and $\lbrace c^{(8)} \rbrace$ coefficients are of the same order of magnitude, some of the dimension six and eight terms could be of comparable size. 

\subsection{Combining dim.~6 and dim.~8 predictions}

Effective Field Theory parametrisations for VBS are often done in terms of a particular \dimeight basis. This basis, presented in \cite{EFT2} was originally formulated as \textit{``the set of EFT operators that generate quartic gauge couplings, without generating triple gauge couplings''} and it relies on different assumptions to the \dimsix Warsaw basis: it does not assume the SM symmetries and gauge invariance, and hence contains vertices such us $Z_{\mu} Z_{\nu} Z_{\rho} Z_{\sigma}$, not allowed by the SM.

For this reason, it is first necessary to find a complete basis, with both \dimsix and \dimeight \, operators, built on some well defined assumptions. The techniques to build such a basis have been described for example in refs. \cite{Henning:2017fpj,Passarino:2019yjx}.
 

\subsection{EFT for the EW sector}

Until the ``dim. 6+8'' basis is defined, we can work on getting a full parametrisation of the electroweak sector in terms of \dimsix operators. 
We classified our processes in different classes :

\begin{table}[h!]
\begin{center}
\begin{tabular}{|lll |}
\hline
\textbf{Class A:} & Diboson Production  \qquad  & \qquad  \qquad \qquad  \\ \hline 
 A1: WW  & A2: WZ  &  A3: ZZ \\
 A4: W$\gamma$  &  A5: Z($\ell \ell$)$\gamma$ & A6: Z($\nu \nu$)$\gamma$ \\ \hline 
\textbf{Class B:} &  VBS & \\ \hline
B1: ssWW  &  B2: Z$\gamma$jj & B3: ZZjj \\
B4: W$\gamma$jj  & B5: osWW  & B6: $\gamma \gamma$WW \\ \hline
 \textbf{Class C:} &  VBF & \\ \hline 
 C1: Zjj &  C2: Wjj & \\ \hline 
 \end{tabular}
\end{center}
\end{table}

\noindent 
By parametrizing these 12 processes we could set bounds (ideally) to about 12 EFT operators. Moreover, if we identify different observables and phase space regions that have different operator dependencies, we can improve the number of constrains. 
For example, with: 

\begin{equation} \nonumber
  \mu_{\sigma} =  \frac{\sigma_{ZZ,VBS}}{\sigma_{ZZ,SM}} = 1 + a_0 \cdot c_{W} + b_0 \cdot c_{HW} + d_0 \cdot c_{HWB} + \dots
\end{equation}
and,
\begin{equation} \nonumber
    \mu_{p_T} = \frac{p_T (j_1)}{p_T (j_1)_{SM}}\Bigg\vert_{bin 3} = 1 + a_1 \cdot c_{W} + b_1 \cdot c_{HW} + d_1 \cdot c_{HWB} + \dots
\end{equation}
and,
\begin{equation} \nonumber
    \mu_{MM} = \frac{M (Z_1 Z_2)}{M (Z_1 Z_2)_{SM}}\Bigg\vert_{bin 3} = 1 + a_2 \cdot c_{W} + b_2 \cdot c_{HW} + d_2 \cdot c_{HWB} + \dots
\end{equation}
we can solve the equation system:

\begin{equation}
    \begin{pmatrix}
          \mu_{\sigma}  \\
      \mu_{p_T} \\
         \mu_{MM}
\end{pmatrix} = 
\begin{pmatrix}
a_0 & b_0 & d_0 \\
a_1 & b_1 & d_1 \\
a_2 & b_2 & d_2 
\end{pmatrix}
\begin{pmatrix}
c_{W} \\ c_{HW} \\ c_{HW}
\end{pmatrix}
\end{equation}
where $\{a_0, \dots d_2 \} $ are the coefficients we have extracted numerically from our Monte Carlo (MC) studies. The fitting technology should be designed in such a way that is relatively easy to replace those coefficients in the future as the theoretical understanding and the MC productions improve.

\subsection{Generating Numerical Predictions}

For the generation of MC Samples we use \texttt{Madgraph 5} \cite{Frederix:2018nkq}. It would be relevant to repeat this study using \texttt{SHERPA} \cite{Bothmann:2019yzt}, since this is the main generator used for the study of these channels in ATLAS. As a UFO model, we use \texttt{SMEFTsim} \cite{Brivio:2017btx}. We are currently exploring the possibility of migrating to the \texttt{SMEFT@NLO} model, that can also accommodate QCD corrections. 

Further technical details to take into account are the choice of input parameter set (IPS), the choice of lepton and quark masses, and the choice of flavour symmetries:

\begin{enumerate}
    \item We use the IPS with $\lbrace m_w , m_z , G_F  \rbrace$, so-called  $m_W$ scheme, available in both the \texttt{SMEFT@NLO}  and the \texttt{SMEFTsim} packages. The latter also implements the ``alpha scheme'', with  $\lbrace \alpha , m_z , G_F  \rbrace$. A more convenient scheme would be the one with $\lbrace \alpha , m_z , m_w  \rbrace$, since it minimizes the number of operators entering the Lagrangian shifts, see for example ref. \cite{Cullen:2019nnr}. The most relevant scheme to be used should be judged case by case, as the one that matches the available experimental results and ongoing searches. 
    \item We assume the quarks and leptons to be massless, with vanishing Yukawa couplings. The effects of the bottom quark masses in initial state should not be neglected \cite{Krauss:2017wmx}, but that is, for now, beyond our scope.  
    \item We work in the flavour symmetric model, where the CKM matrix is diagonal and no EFT operator connects different generations. 
\end{enumerate}

\subsection{Constraints available from previous fits}

Some bounds from Higgs and diboson production have been derived in previous publications, such as \cite{Ellis:2018gqa}. Triple gauge couplings were broadly studied at LEP experiments and EFT fits of the LEP data are also available \cite{Berthier:2016tkq}. 

When comparing with such results it is important to make sure of the definition of the Wilson coefficients, since different nomenclatures have been used in the literature:

\begin{equation}
    \mathcal{L}_{SMEFT} - \mathcal{L}_{SM} =  \bar{c}_i \op_i \equiv \frac{\tilde{c}_i}{\Lambda^2}  \op_i \equiv {c}_i \frac{v^2 }{\Lambda^2}  \op_i
\end{equation}
The last term is the one implemented in the available UFO models, whereas the first one is the one usually reported in the fits. For example, by generating numerical predictions for Monte Carlo values of $ \lbrace \left[ c= 1- 4 \pi \right] , \left[ \Lambda = 3 - 5 \rm{TeV}  \right]  \rbrace$ we should compare with results reported in the fits as, 

\begin{equation}
    \bar{c} = \frac{c \cdot v^2}{\Lambda^2} = \left\lbrace \frac{1 \cdot 0.06}{9} , \frac{4 \pi \cdot 0.06}{9} ,  \frac{1 \cdot 0.06}{25} , \frac{4 \pi \cdot 0.06}{9} \right\rbrace \approx \{ 2 \cdot 10^{-4} - 0.08 \}
\end{equation}

\subsection{Conclusions}

With this work, we aim to provide a comprehensive recommendation for theorists and experimentalists to be able to generate MC samples to parametrize EFT effects in the different analyses of the LHC Run-2 dataset. 

A precise prediction should parametrize cross-sections and differential distributions, possibly including quadratic and cross-quadratic terms as well as the linear interferences. A \dimeight basis relying on the same assumptions as the \dimsix ones (gauge invariance, SM symmetries) is also desirable.

\phantomsection
\addcontentsline{toc}{part}{Acknowledgements} 

\chapter*{Acknowledgements\markboth{Acknowledgements}{Acknowledgements}}

\indent 
The authors would like to acknowledge the contribution of the COST Action CA16108.
We would like to thank the local organizing committee at the P\^ir\^is Reis University,
and the local secretariat, for their great job and hospitality.

The work of Le~Duc~Ninh is funded by the Vietnam National Foundation for
Science and Technology Development (NAFOSTED) under grant number
103.01-2017.78. He thanks the organizers for the kind invitation.
He also thanks the members of the ATLAS group at LAPP 
for their hospitality and financial support for a research visit via the French ANR grant VBStime.

\renewcommand\leftmark{References}
\renewcommand\rightmark{References}

\bibliographystyle{./StyleFilesMacros/atlasnote}
\cleardoublepage
\phantomsection
\addcontentsline{toc}{part}{References}
\bibliography{IstanbulReport}


\end{document}